\documentclass[onecolumn]{aa}

\usepackage[varg]{txfonts}
\usepackage{hyperref}
\usepackage{color}

\definecolor{CiteColor}{rgb}{0,0,0.35}
\hypersetup{colorlinks,citecolor=CiteColor}

\newcommand{\beq}{\begin{equation}}
\newcommand{\eeq}{\end{equation}}
\newcommand{\bea}{\begin{eqnarray}}
\newcommand{\eea}{\end{eqnarray}}

\newcommand{\ui}{\mathrm{i}}

\newcommand{\SNRyr}{{\mathrm{S/N}_{1\, \rm yr}}}
\DeclareMathOperator{\sinc}{sinc}

\begin{document}


\title{Gravitational waves from bodies orbiting the Galactic Center black hole
and their detectability by LISA}

\author{E.~Gourgoulhon\inst{\ref{luth}}\and A.~Le Tiec\inst{\ref{luth}}\and
F.\,H.~Vincent\inst{\ref{lesia}}\and N.~Warburton\inst{\ref{ucd}}}

\institute{
Laboratoire Univers et Th\'eories, Observatoire de Paris,
Universit\'e PSL, CNRS, Universit\'e Paris Diderot, Sorbonne Paris Cit\'e,
5 place Jules Janssen, 92190 Meudon, France\\
\email{eric.gourgoulhon@obspm.fr}, \email{alexandre.letiec@obspm.fr} \label{luth}
\and
Laboratoire d'\'Etudes Spatiales et d'Instrumentation en Astrophysique,
Observatoire de Paris, Universit\'e PSL, CNRS, Sorbonne Universit\'e,
Universit\'e Paris Diderot, Sorbonne Paris Cit\'e, 5 place Jules Janssen,
92190 Meudon, France \\
\email{frederic.vincent@obspm.fr}\label{lesia}
\and
School of Mathematics and Statistics, University College Dublin, Belfield,
Dublin 4, Ireland\\
\email{niels.warburton@ucd.ie} \label{ucd}
}

\date{Received 6 March 2019; accepted 29 May 2019}

\abstract
{}
{We present the first fully relativistic study of gravitational radiation from
bodies in circular equatorial orbits around the massive black hole at the Galactic Center,
Sgr~A* and we assess the detectability of various kinds of objects by the gravitational wave detector
LISA.}
{Our computations are based on the theory of perturbations of
the Kerr spacetime and take into account the Roche limit induced by tidal forces
in the Kerr metric.  The signal-to-noise
ratio in the LISA detector, as well as the time spent in LISA band, are
evaluated. We have implemented all the computational tools in an open-source SageMath package,
within the Black Hole Perturbation Toolkit framework.}
{We find that white dwarfs, neutrons stars, stellar black holes, primordial black
holes of mass larger than $10^{-4} M_\odot$, main-sequence
stars of mass lower than $\sim 2.5\, M_\odot$, and brown dwarfs orbiting Sgr~A*
are all detectable in one year of LISA data with a signal-to-noise
ratio above 10 for at least $10^5$~years in the slow
inspiral towards either the innermost stable circular orbit (compact objects)
or the Roche limit (main-sequence stars and brown dwarfs). The longest
times in-band, of the order of $10^6$~years, are achieved for primordial black holes
of mass $\sim 10^{-3} M_\odot$ down to $10^{-5} M_\odot$, depending
on the spin of Sgr~A*, as well as for brown dwarfs, just followed by white dwarfs
and low mass main-sequence stars. The long time in-band of these objects makes Sgr~A*
a valuable target for LISA. We also consider bodies on close circular orbits around
the massive black hole in the nucleus of the nearby galaxy M32 and find
that, among them, compact objects and brown dwarfs stay for $10^3$ to $10^4$ years
in LISA band with a one-year signal-to-noise ratio above ten.
}
{}


\maketitle

\section{Introduction} \label{s:intro}

\begin{figure}
\centerline{\includegraphics[width=0.7\textwidth]{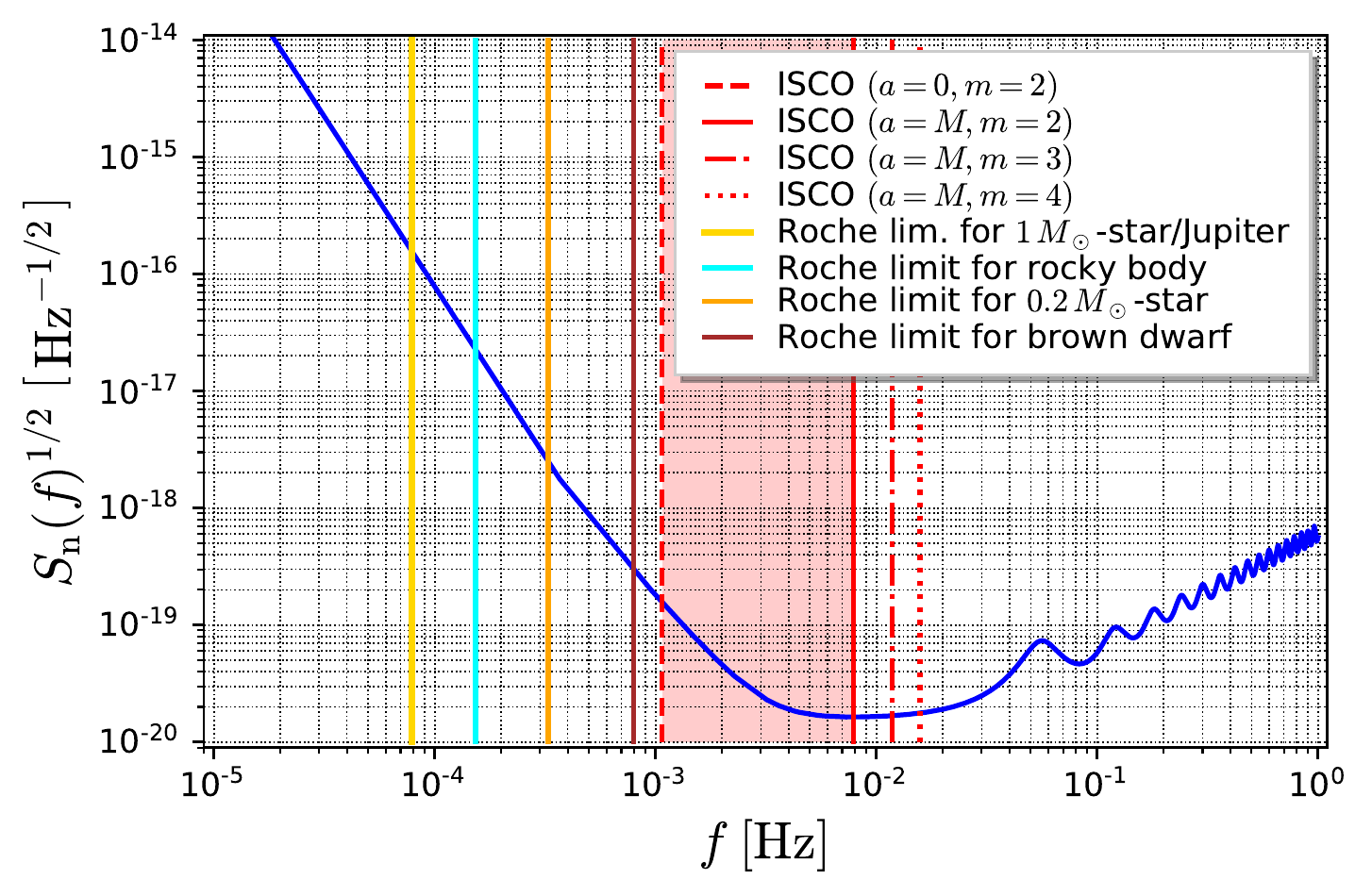}}
\caption{ \label{f:LISA_band}
LISA sensitivity curve \citep{LISA_L3} and various gravitational wave frequencies
from circular orbits around Sgr~A*. The wave frequencies shown above are all
for the dominant $m=2$ mode, except
for the dot-dashed and dotted vertical red lines, which correspond to the
$m=3$ and $m=4$ harmonics of the ISCO of an extreme Kerr BH ($a=M$).
The shaded pink area indicates the location of the frequencies from the ISCO when
$a$ ranges from zero to $M$. The Roche limits are those discussed in
Sect.~\ref{sec:tidal}.}
\end{figure}

The future space-based Laser Interferometer Space Antenna (LISA) \citep{LISA_L3}, selected
as the L3 mission of ESA, will detect
gravitational radiation from various phenomena involving massive black holes (MBHs),
the masses of which range from $10^5$ to $10^7\; M_\odot$ \citep[see e.g.,][and references therein]{Amaro18,Babak_al17}. The mass of the MBH Sgr~A* at the center of our galaxy lies within this range
\citep{gravity18a,gravity18b}:
\beq \label{e:SgrA_mass}
    M_{\rm Sgr\, A^*} = 4.10\pm 0.03 \times 10^6\; M_\odot.
\eeq
More precisely, the angular velocity $\omega_0$ on a circular, equatorial orbit at the Boyer-Lindquist
radial coordinate $r_0$ around a Kerr black hole (BH) is given by the
formula in \citet{BardeenPT72}
\beq \label{e:omega0}
     \omega_0 = \frac{(GM)^{1/2}}{r_0^{3/2} + a (GM)^{1/2}/c} ,
\eeq
where $G$ is the gravitational constant, $c$ the speed of light,
$M$ the BH mass, and $a=J/(cM)$ its reduced spin. Here $J$ is the magnitude of
the BH angular momentum ($a$ has the dimension of a length).
The motion of a particle of mass $\mu\ll M$ on a circular orbit generates some gravitational
radiation with a periodic pattern (the dominant mode of which is
$m=2$) and has the frequency $f_{m=2} = 2 f_0$,
where $f_0 \equiv \omega_0 /(2\pi)$ is the orbital frequency
(details are given
in Sect.~\ref{s:gw_particle}). Combining with Eq.~(\ref{e:omega0}), we obtain
\beq \label{e:f_GW_m2}
     f_{m=2} = \frac{1}{\pi} \frac{(GM)^{1/2}}{r_0^{3/2} + a (GM)^{1/2}/c}.
\eeq
This frequency is maximal at the (prograde) innermost stable circular orbit (ISCO),
which is located at $r_0 = 6GM/c^2$ for
$a=0$ (Schwarzschild BH) and at $r_0 = GM/c^2$ for $a=a_{\rm max}\equiv GM/c^2$ (extreme Kerr
BH). Equation~(\ref{e:f_GW_m2}) leads then to
\beq
     f_{m=2}^{{\rm ISCO},{a=0}} = \frac{c^3}{6^{3/2}\pi G M}
     \quad\mbox{and}\quad
    f_{m=2}^{{\rm ISCO},{a_{\rm max}}} = \frac{c^3}{2\pi G M} .
\eeq
Substituting the mass of Sgr~A* (\ref{e:SgrA_mass}) for $M$, we obtain
\beq \label{e:f_ISCO_SgrA}
     f_{m=2}^{{\rm ISCO},{a=0}} = 1.1  \; {\rm mHz}
     \quad\mbox{and}\quad
    f_{m=2}^{{\rm ISCO},{a_{\rm max}}} = 7.9  \; {\rm mHz} .
\eeq
By convenient coincidence, $f_{m=2}^{{\rm ISCO},{a_{\rm max}}}$
matches almost exactly the frequency of LISA maximal sensitivity,
the latter being $7.86~{\rm mHz}$! (see Fig.~\ref{f:LISA_band}).
The spin of Sgr~A* is currently not known, but it is expected to be quite large,
due to matter accretion since the birth of the MBH. Actually, the tentative measures of
MBH spins in nuclei of other galaxies generally lead to large
values of $a$. See for example Table~3 of the recent review by \citet{NampalliwarB19}, where most
entries have $a > 0.9 \,GM/c^2$.

The adequacy of LISA bandwidth to orbital motions around Sgr~A* was
first stressed by \citet{Freitag03a,Freitag03b}, who estimated the gravitational
radiation from orbiting stars at the (Newtonian) quadrupole order.
By taking into account the tidal forces exerted by the MBH, he showed
that, besides compact objects, low-mass main-sequence stars (mass $\mu \lesssim 0.1\, M_\odot$) can approach
the central MBH sufficiently close to emit gravitational waves in LISA bandwidth.
Via some numerical simulations of the dynamics of the Galactic Center
stellar cluster, he estimated that there could exist a few such stars detectable by LISA, whereas the
probability of observing a compact object was found to be quite low \citep{Freitag03a}.
This study was refined by \citet{BarackC04}, who estimated that the
signal-to-noise ratio (S/N) of a
$\mu=0.06\, M_\odot$ main-sequence star observed $10^6\; {\rm yr}$ before
plunge is of the order eleven in two years of LISA observations. Moreover, they
have shown that the detection of such an event could lead to the spin measurement
of Sgr~A* with an accuracy of $\sim 0.5\%$.
\citet{BerryG13a} investigated the phenomenon of
extreme-mass-ratio burst, which occurs at the periastron passage of
a stellar-mass compact object (mass $\mu$) on a highly eccentric orbit around Sgr~A*.
These authors have shown that LISA can
detect such an event with $\mu=10\;M_\odot$, provided that the periastron distance
is lower than $65 GM/c^2$. The event rate of such bursts could be of the order
of 1 per year \citep{BerryG13c} (see Sect.~7.6 of \citet{Amaro18} for some discussion).
\citet{LinialS17} have computed at the quadrupole order
the gravitational wave emission from orbiting main-sequence stars undergoing Roche lobe overflow,
treated at the Newtonian level. These authors stressed the
detectability by LISA and have showed the possibility of a reverse chirp
signal, the reaction of the accreting system
to the angular momentum loss by gravitational radiation being a widening of the
orbit (outspiral) \citep{DaiB13}. Recently, \citet{KuhnelMSF18}
have computed, still at the quadrupole level, the gravitational wave emission
from an ensemble of macroscopic dark matter candidates  orbiting
Sgr~A*, such as primordial BHs, with masses
in the range $10^{-13} - 10^3\; M_\odot$.

All the studies mentioned above are based on the quadrupole formula
for Newtonian orbits, except that of \citet{BerryG13a}, which is based on the so-called
``kludge approximation''. Now, for orbits close to the ISCO, relativistic effects
are expected to be important. In this article, we present the first study of gravitational waves from stellar objects in close orbits around Sgr~A* in a fully relativistic
framework: Sgr~A* is modeled as a Kerr BH,
gravitational waves are computed via the theory of perturbations of the Kerr metric
\citep{Teukolsky73,Detweiler78,Shibata94,Kennefick98,Hughes00,FinnT00,GlampedakisK02}
and tidal effects are evaluated via the theory of Roche potential
in the Kerr metric developed by \citet{DaiB13}.
Moreover, from the obtained waveforms, we carefully evaluate the signal-to-noise ratio in the LISA detector, taking into account the latest LISA sensitivity curve \citep{RobsonCL18}.
There is another MBH with a mass within the LISA range in the Local Group of galaxies:
the $2.5\times 10^6\, M_\odot$ MBH in the center of the
galaxy M32 \citep{Nguyen_al18}. By
applying the same techniques, we study the detectability by LISA of
bodies in close circular orbit around it.

The plan of the article is as follows. The method employed to compute the
gravitational radiation from a point mass in circular orbit around a Kerr BH
is presented in Sect.~\ref{s:gw_particle}, the open-source code implementing it
being described in Appendix~\ref{s:kerrgeodesic_gw}. The computation of
the signal-to-noise ratio of the obtained waveforms
in the LISA detector is performed in Sect.~\ref{s:SNR}, from which
we can estimate the minimal detectable mass of the orbiting source in terms of
the orbital  radius. Section~\ref{s:orbital_decay} investigates the secular evolution of
a circular orbit under the reaction to gravitational radiation and provides
the frequency change per year and the inspiral time
between two orbits. The potential astrophysical sources are discussed in
Sect.~\ref{s:sources}, taking into account Roche limits for noncompact objects
and estimating the total time spent in LISA band. The case of M32 is treated in
Appendix~\ref{s:M32}.
Finally, the main conclusions are drawn in Sect.~\ref{s:concl}.


\section{Gravitational waves from an orbiting point mass} \label{s:gw_particle}

In this section and the remainder of this article, we use geometrized units,
for which $G=1$ and $c=1$. In addition we systematically use Boyer-Lindquist coordinates $(t,r,\theta,\varphi)$ to describe the Kerr geometry of a rotating BH of mass $M$ and spin parameter $a$, with
$0 \leqslant a < M$.
We consider a particle of mass $\mu \ll M$ on a (stable) prograde circular equatorial orbit of constant coordinate $r = r_0$. Hereafter, we call $r_0$ the \emph{orbital radius}.
The orbital angular velocity $\omega_0$ is given by formula~(\ref{e:omega0}).
In practice this ``particle'' can be any object whose extension is negligible with
respect to the orbital radius. In particular, for Sgr~A*,
it can be an object as large as a solar-type star. Indeed, Sgr~A* mass
(\ref{e:SgrA_mass}) corresponds to a length scale
$M = 6.05 \times 10^6\; {\rm km} \sim 9 R_\odot$, where $R_\odot$ is the
Sun's radius. Moreover, main-sequence stars are centrally condensed objects, so that
their ``effective'' size as gravitational wave generator is smaller
that their actual radius. In addition,
as we shall see in Sect.~\ref{sec:tidal}, their orbital radius must
obey $r_0 > 34 M$ to avoid tidal disruption (Roche limit), so that
$R_\odot / r_0 < 3\times 10^{-3}$.
 Hence,
regarding Sgr~A*,
we may safely describe orbiting stars as point particles.

The gravitational wave emission from a point mass orbiting a Kerr BH
has been computed by many groups, starting from the seminal work of
\citet{Detweiler78}, which is based on the theory of linear perturbations of the Kerr
metric initiated by \citet{Teukolsky73}. The computations have been
extended to eccentric orbits by a number of authors \citep[see e.g.,][]{GlampedakisK02}.
However, in the present study, we limit ourselves to circular orbits, mostly for
simplicity, but also because some of the scenarii discussed in Sect.~\ref{s:sources}
lead naturally to low eccentricity orbits; this involves inspiralling compact
objects that result from the tidal disruption of a binary,
stars formed in an accretion disk, black holes resulting
from the most massive of such stars and a significant proportion
($\sim 1/4$) of the population of brown dwarfs that might be in LISA band.

In Sect.~\ref{subsec:point}, we recall the gravitational waveform obtained from
perturbation analysis of the Kerr metric. It requires the numerical computation of
many mode amplitudes. This is quite technical and we describe the technique
we use to perform the computation
in Sect.~\ref{subsec:modes}. We discuss the limiting case of distant orbits in
Sect.~\ref{s:distant_orbits} and evaluate the Fourier spectrum of the
waveform in Sect.~\ref{s:Fourier_series}, where we present some
specific waveforms.

\subsection{Gravitational waveform}\label{subsec:point}

The gravitational waves generated by the orbital motion of the particle are conveniently encoded in the linear combination $h_+ - \ui h_\times$ of the two polarization states $h_+$ and $h_\times$. A standard result from the theory of linear perturbations of the Kerr BH
\citep{Teukolsky73,Detweiler78,Shibata94,Kennefick98,Hughes00,FinnT00,GlampedakisK02}
yielded the asymptotic waveform as
\beq\label{e:h}
    h_+ - \ui h_\times = \frac{2\mu}{r} \, \sum_{\ell=2}^{+\infty} \sum_{{\scriptstyle m=-\ell\atop \scriptstyle m\not=0}}^\ell \frac{Z^\infty_{\ell m}(r_0)}{(m\omega_0)^2} \, _{-2}S^{am\omega_0}_{\ell m}(\theta,\varphi) \,
    e^{- \ui m (\omega_0 (t-r_*) + \varphi_0)} ,
\eeq
where $(h_+,h_\times)$ are evaluated at the spacetime event of Boyer-Lindquist coordinates $(t,r,\theta,\varphi)$ and $r_*$ is the so-called ``tortoise coordinate'', defined as
\beq
    r_* \equiv r + \frac{2M r_+}{r_+-r_-} \, \ln{\Bigl(\frac{r-r_+}{2M}\Bigr)} - \frac{2Mr_-}{r_+-r_-} \, \ln{\Bigl(\frac{r-r_-}{2M}\Bigr)} \, ,
\eeq
where $r_\pm \equiv M \pm \sqrt{M^2-a^2}$ denote the coordinate locations of the outer ($+$) and inner ($-$) event horizons. The phase $\varphi_0$ in Eq.~(\ref{e:h}) can always be absorbed into a shift of the origin of $t$. The spin-weighted \textit{spheroidal} harmonics $_{-2}S^{am\omega_0}_{\ell m}(\theta,\varphi)$ encode the dependency of the waveform with respect to the polar angles
$(\theta,\varphi)$ of the observer. For each harmonic $(\ell,m)$, they depend on the
(dimensionless) product $a\omega_0$ of the Kerr spin parameter and the orbital angular velocity, and they reduce to the more familiar spin-weighted \emph{spherical} harmonics $_{-2}Y_{\ell m}(\theta,\varphi)$ when $a = 0$.
The coefficients $Z^\infty_{\ell m}(r_0)$ encode the amplitude and phase of each mode. They
depend on $M$ and $a$ and are computed by solving the radial component of the Teukolsky
equation \citep{Teukolsky73};
they satisfy $Z^\infty_{\ell, -m} = (-1)^\ell Z^{\infty*}_{\ell m}$, where the star denotes the complex conjugation.

Given the distance $r=8.12\pm 0.03\; {\rm kpc}$ to Sgr~A* \citep{gravity18a},
the prefactor $\mu/r$ in formula~(\ref{e:h}) takes the following numerical value:
\beq \label{e:mu_ov_r}
    \frac{\mu}{r} = 5.89\times 10^{-18} \left(\frac{\mu}{1\; M_\odot}\right)
    \left(\frac{8.12 \; {\rm kpc}}{r}\right) .
\eeq

\subsection{Mode amplitudes}\label{subsec:modes}

The factor $|Z^\infty_{\ell m}(r_0)|/(m\omega_0)^2$ sets the amplitude of
the mode $(\ell,m)$ of $h_+$ and $h_\times$ according to Eq.~\eqref{e:h}. The complex amplitudes, $Z^\infty_{\ell m}$, are computed by solving the Teukolsky equation \citep{Teukolsky73}, where, typically, the secondary is modeled as a structureless point mass. Generally, the Teukolsky equation is solved in either the time or frequency domain. Time domain calculations are computationally expensive but well suited to modeling a source moving along an arbitrary trajectory. Frequency domain calculations have the advantage that the Teukolsky equation is completely separable in this domain and this reduces the problem from solving partial to ordinary differential equations. This leads to very efficient calculations so long as the Fourier spectrum of the source is sufficiently narrow. Over short timescales\footnote{As discussed further in Sect.~\ref{s:orbital_decay}, an orbiting body's true worldline spirals inwards due to gravitational radiation reaction. A geodesic that is tangent to the worldline at an instance will dephase from the inspiraling worldline on a timescale $\sim M \epsilon^{-1/2}$ where $\epsilon\equiv\mu/M$ is the mass ratio. By approximating the radiation reaction force at each instance by that computed along a tangent geodesic one can compute a worldline that dephases from the true inspiral over the radiation reaction timescale of $\sim M \epsilon^{-1}$ \citep{Hinderer:2008dm}.} the trajectory of a small body with $\mu \ll M$ orbiting a MBH is well approximated by a bound geodesic of the background spacetime. Motion along a bound geodesic is periodic (or bi-periodic \citep{Schmidt:2002qk}) and so the spectrum of the source is discrete. This allows the Teukolsky equation to be solved efficiently in the frequency domain, at least for orbits with up to a moderate eccentricity (for large eccentricities the Fourier spectrum broadens to a point where time domain calculations can be more efficient \citep{Barton:2008eb}). Frequency domain calculations have been carried out for circular \citep{Detweiler78}, spherical \citep{Hughes00}, eccentric equatorial \citep{GlampedakisK02} and generic orbits \citep{Drasco:2005kz,Fujita:2009us,vandeMeent:2017bcc} and we follow this approach in this work.

In the frequency domain the Teukolsky equation separates into spin-weighted spheroidal harmonics and frequency modes. The former can be computed via eigenvalue \citep{Hughes00} or continuous fraction methods \citep{Leaver:1985ax}. The main task is then finding solutions to the Teukolsky radial equation. Typically, this is a two step process whereby one first finds the homogeneous solutions and then computes the inhomogeneous solutions via the method of variation of parameters. Finding the homogeneous solutions is usually done by either numerical integration or via an expansion of the solution in a series of special functions \citep{Sasaki:2003xr}. In this work we make use of both methods as a cross check. Direct numerical integration of the Teukolsky equation is numerically unstable but this can be overcome by transforming the equation to a different form \citep{SasakiN82a,SasakiN82b}. Our implementation is based off the code developed for \citet{Gralla:2015rpa}. For the series method our code is based off of codes used in \citet{Kavanagh:2016idg,Buss:2017vud}. Both of these codes, as well as code to compute spin-weighted spheroidal harmonics, are now publicly available as part of the
Black Hole Perturbation Toolkit\footnote{\url{http://bhptoolkit.org/}}.

The final step is to compute the inhomogeneous radial solutions. In this work we consider circular, equatorial orbits. With a point particle source, this reduces the application of variation of parameters to junction conditions at the particle's radius \citep{Detweiler78}. The asymptotic complex amplitudes, $Z^\infty_{\ell m}$, can then be computed by evaluating the radial solution in the limit $r\rightarrow\infty$.

\begin{figure*}
\begin{tabular}{c@{\!\!}c}
\includegraphics[width=0.51\textwidth]{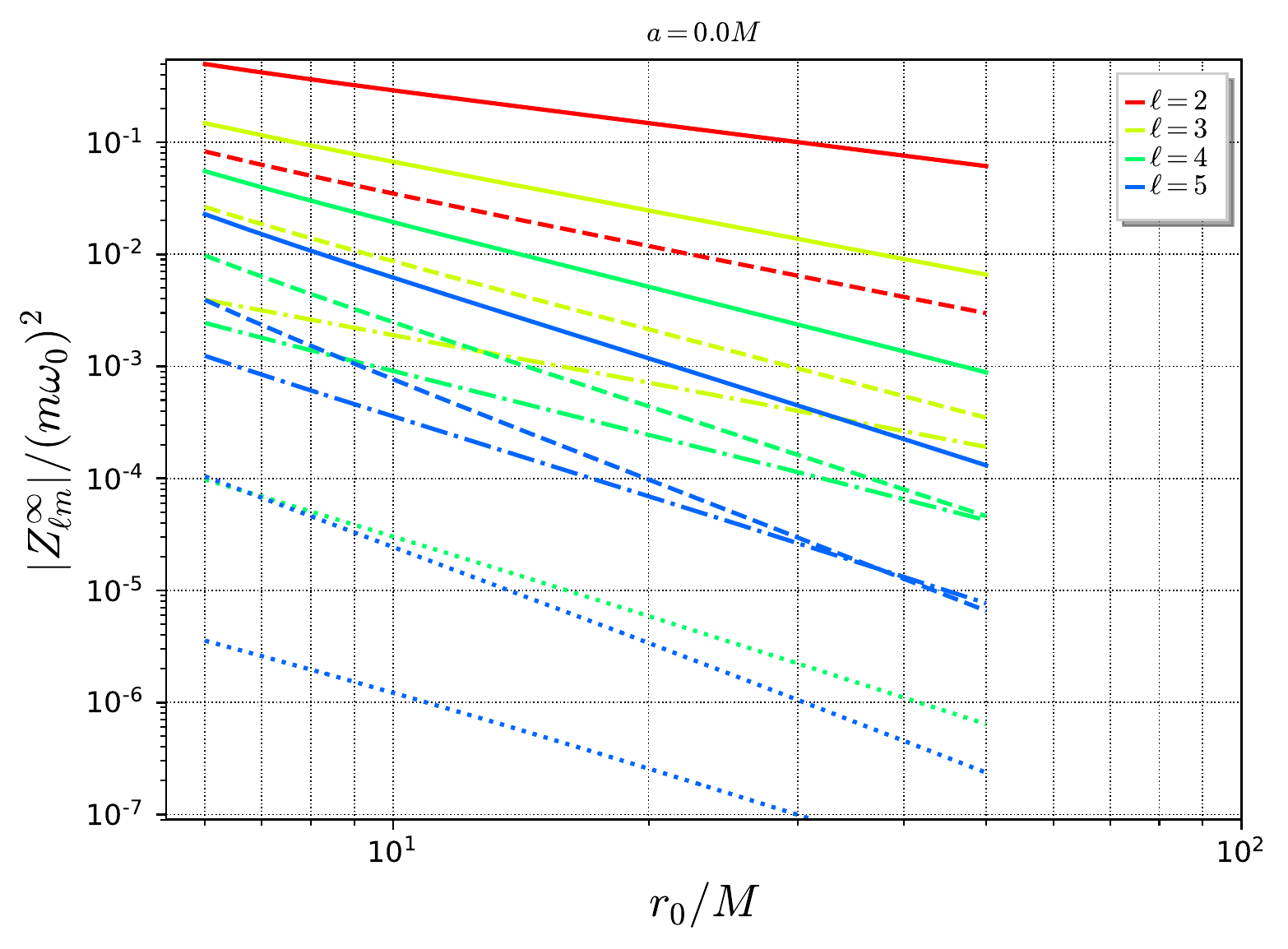} &
\includegraphics[width=0.51\textwidth]{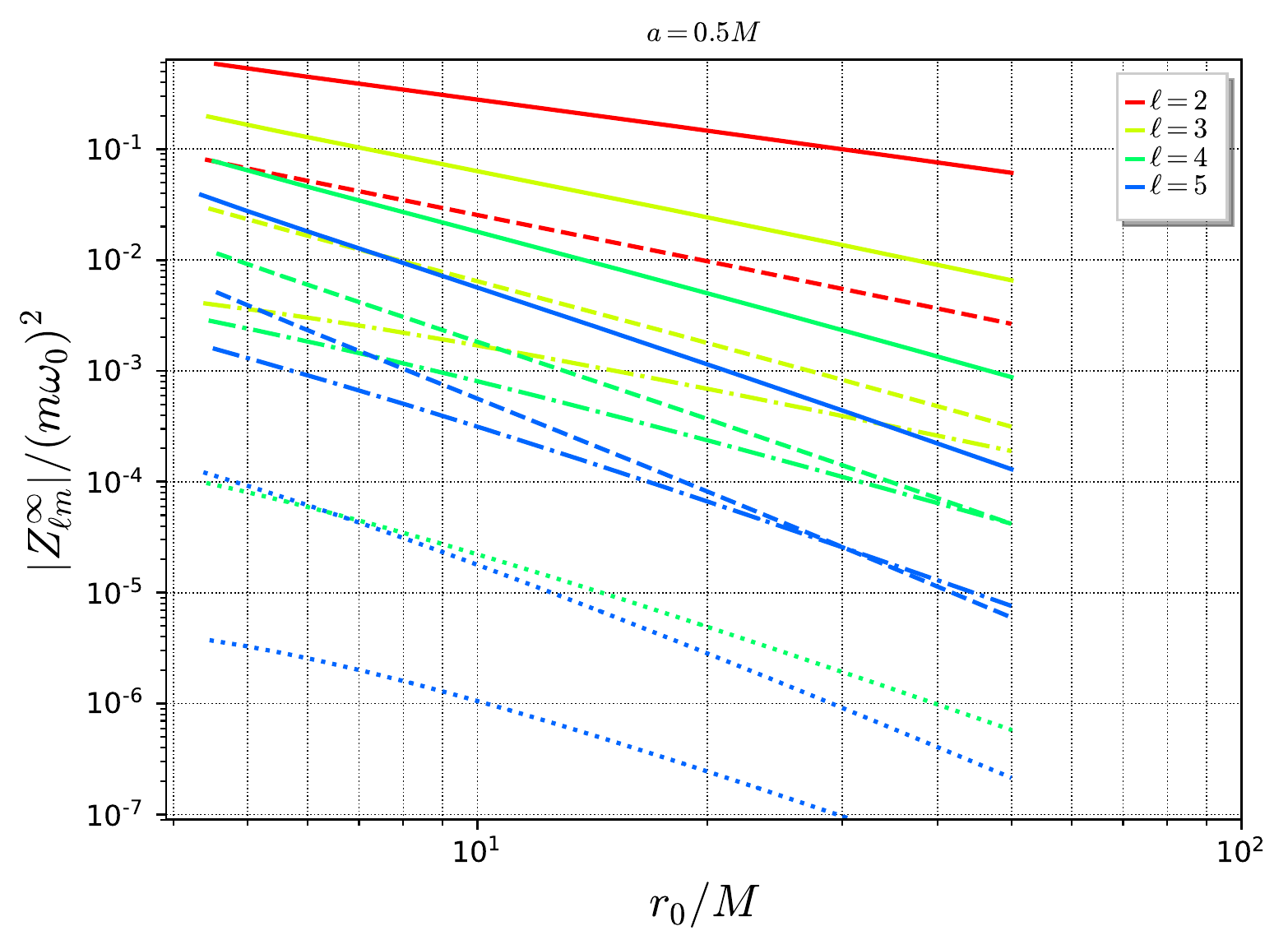} \\
\includegraphics[width=0.51\textwidth]{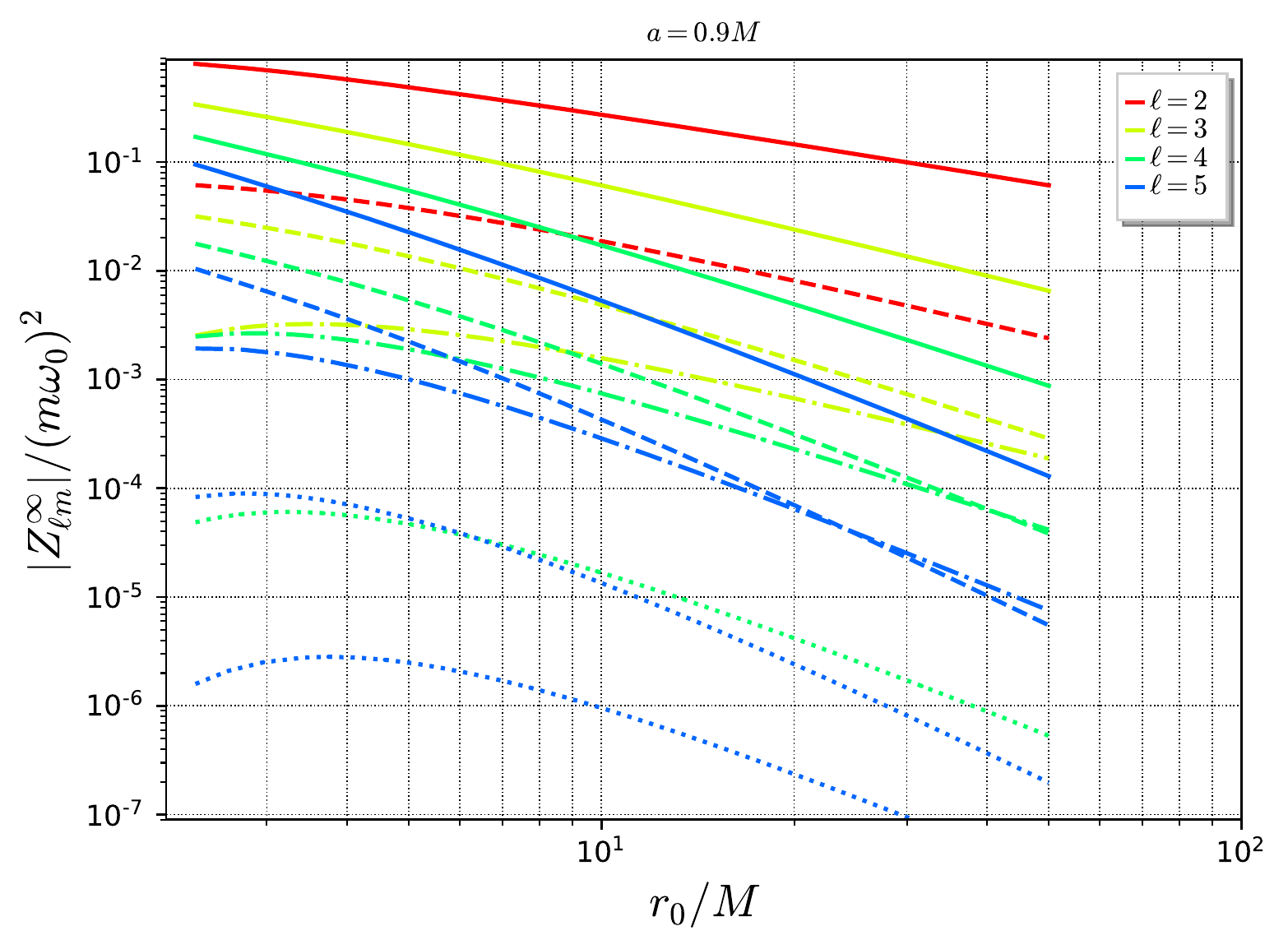} &
\includegraphics[width=0.51\textwidth]{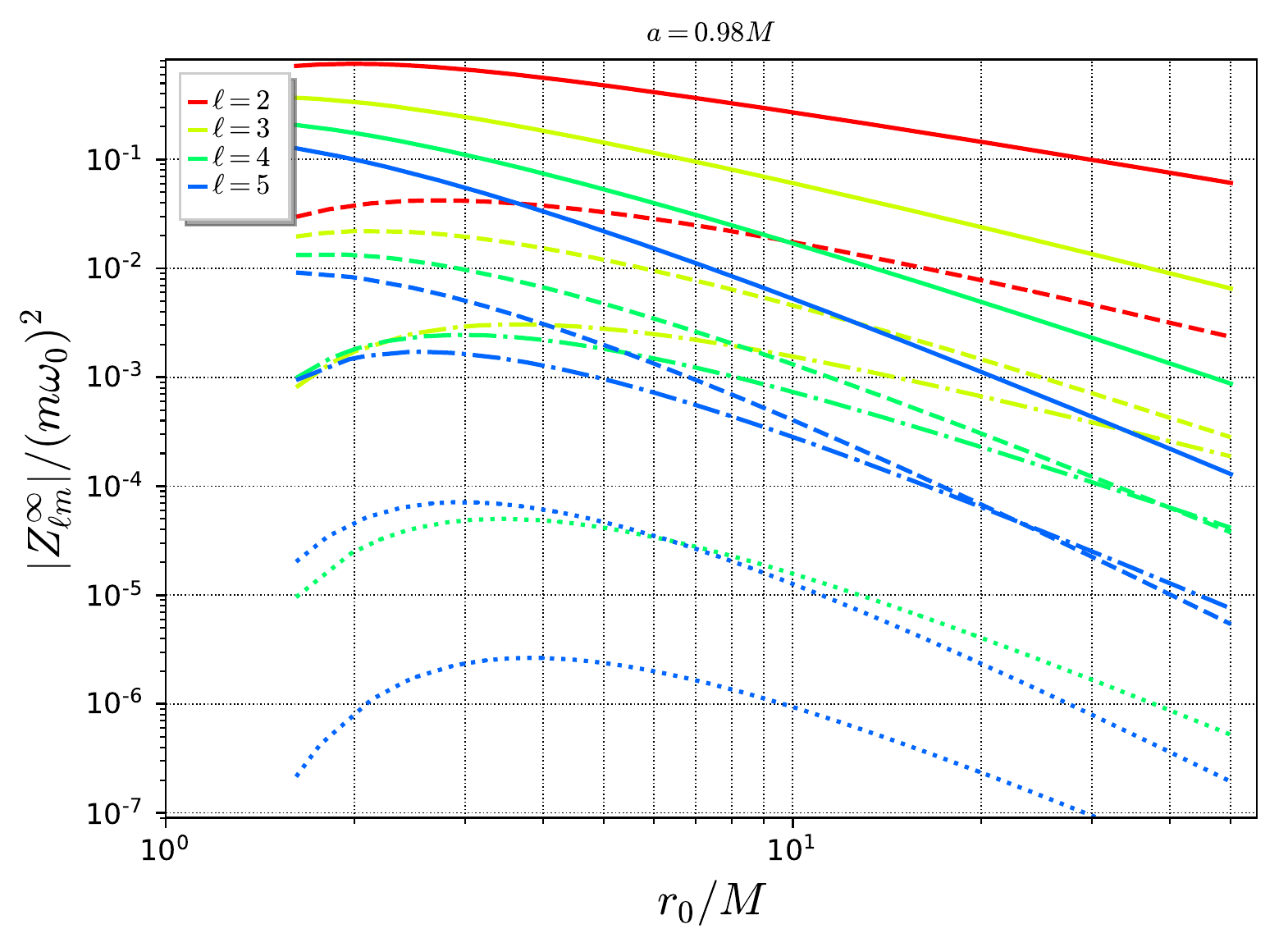}
\end{tabular}
\caption{ \label{f:Zinf_amplitudes}
Amplitude factor $|Z^\infty_{\ell m}(r_0)|/(m\omega_0)^2$ for the harmonic $(\ell,m)$
of the gravitational wave emitted by an orbiting point mass [cf.~Eq.~(\ref{e:h})],
in terms of the orbital radius $r_0$. Each panel corresponds to a given value of the MBH spin:
$a=0$ (Schwarzschild BH), $a=0.5 M$, $a=0.9 M$ and $a=0.98M$. A given color
corresponds to a fixed value of $\ell$ and the line style indicates the
value of $m$: solid: $m=\ell$, dashed: $m=\ell-1$, dot-dashed: $m=\ell-2$,
dotted: $0< m\leqslant \ell-3$.}
\end{figure*}

The mode amplitudes are plotted in Fig.~\ref{f:Zinf_amplitudes}
as functions of the orbital radius $r_0$ for $2 \leqslant \ell \leqslant 5$, $1 \leqslant m \leqslant \ell$ and some selected values of the MBH spin parameter $a$.
Each curve starts at the value of $r_0$ corresponding to the prograde ISCO for the considered $a$.

\subsection{Waveform for distant orbits ($r_0 \gg M$)} \label{s:distant_orbits}

When the orbital radius obeys $r_0 \gg M$, we see from Fig.~\ref{f:Zinf_amplitudes}
that the modes $(\ell,m) = (2, \pm 2)$ dominate the waveform (cf.~the solid
red curves in the four panels of Fig.~\ref{f:Zinf_amplitudes}). Moreover,
for $r_0 \gg M$, the effects of the MBH spin become negligible. This is
also apparent on Fig.~\ref{f:Zinf_amplitudes}: the value of $|Z^\infty_{\ell m}(r_0)|/(m\omega_0)^2$ for $(\ell,m) = (2,2)$ and $r_0 = 50 M$ appears to be independent of $a$,
being equal to roughly $7\times 10^{-2}$ in all the four panels.
The value of $Z^\infty_{2,\pm 2}(r_0)$ at the lowest order in $M/r_0$
is given by e.g., Eq.~(5.6) of \citet{Poisson93a}, and reads\footnote{Our values of
$Z^\infty_{\ell m}(r_0)$ have a sign opposite to those of \citet{Poisson93a}
due to a different choice of metric signature, namely $(+,-,-,-)$ in \citet{Poisson93a} vs.
$(-,+,+,+)$ here, and hence a different sign of $(h_+,h_\times)$.}
\beq \label{e:Zinf_22}
    Z^\infty_{2,\pm 2}(r_0) = 16\sqrt{\frac{\pi}{5}} \frac{M^2}{r_0^4}
        \left[ 1 + O\left(\frac{M}{r_0}\right) \right] .
\eeq
The dependency with respect to $a$ would appear only at the relative order
$(M/r_0)^{3/2}$ (see Eq.~(24) of \citet{Poisson93b}) and can safely be ignored,
as already guessed from Fig.~\ref{f:Zinf_amplitudes}.
Besides, for $r_0 \gg M$, Eq.~(\ref{e:omega0}) reduces to the standard Newtonian expression:
\beq \label{e:omega0_Kepler}
    \omega_0 \simeq \sqrt{\frac{M}{r_0^3}} .
\eeq
Combining with Eq.~(\ref{e:Zinf_22}), we see that the amplitude factor in the waveform (\ref{e:h}) is
\beq \label{e:Zinf_22_omega0}
    \frac{Z^\infty_{2,\pm 2}(r_0)}{(2\omega_0)^2} \simeq
        4 \sqrt{\frac{\pi}{5}} \frac{M}{r_0} .
\eeq
Besides, when $r_0 \gg M$, Eq.~(\ref{e:omega0}) leads to $M\omega_0 \ll 1$ and therefore
to $a\omega_0 \ll 1$ since $|a| \leqslant M$. Accordingly the spheroidal harmonics
$_{-2}S^{am\omega_0}_{\ell m}(\theta,\varphi)$ in Eq.~(\ref{e:h}) can be
approximated by the spherical harmonics $_{-2}Y_{\ell m}(\theta,\varphi)$.
For $(\ell,m)=(2,\pm 2)$, the latter are
\beq \label{e:Y22}
    {}_{-2}Y_{2,\pm 2}(\theta,\varphi) = \frac{1}{8} \sqrt{\frac{5}{\pi}}
        \left(1 \pm \cos\theta \right)^2 e^{\pm 2\ui \varphi} .
\eeq
Keeping only the terms $(\ell,m)=(2,\pm 2)$ in the summations involved in
Eq.~(\ref{e:h}) and substituting expression~(\ref{e:Zinf_22_omega0}) for
the amplitude factor and expression~(\ref{e:Y22}) for
$_{-2}S^{2a\omega_0}_{2,\pm 2}(\theta,\varphi)\simeq {}_{-2}Y_{2,\pm 2}(\theta,\varphi)$,
we get
\beq \label{e:h_22}
    h_+ - \ui h_\times = \frac{\mu}{r} \frac{M}{r_0}
        \left[ (1-\cos\theta)^2 e^{2\ui\psi} + (1+\cos\theta)^2 e^{-2\ui\psi} \right] ,
\eeq
where
\beq \label{e:def_psi}
  \psi \equiv \omega_0 (t - r_*) + \varphi_0  - \varphi .
\eeq
Expanding (\ref{e:h_22}) leads immediately to
\begin{subequations}
\label{e:h_quadrupole}
\begin{align}
h_+(t,r,\theta,\varphi) &= 2\,  \frac{\mu}{r} \frac{M}{r_0} (1+\cos^2\theta)
    \cos\left[2\omega_0 (t - r_*) + 2(\varphi_0-\varphi)\right] , \\
h_\times(t,r,\theta,\varphi) &= 4\, \frac{\mu}{r} \frac{M}{r_0} \cos\theta
    \sin\left[2\omega_0 (t - r_*) + 2(\varphi_0-\varphi)\right] .
\end{align}
\end{subequations}
As expected for $r_0 \gg M$,
we recognize the waveform obtained from the standard quadrupole formula
applied to a point mass $\mu$ on a Newtonian circular orbit around a
mass $M\gg\mu$ (compare with e.g., Eqs.~(3.13)-(3.14) of \citet{Blanchet01}).

\begin{figure*}
\begin{tabular}{c@{\!\!}c}
\includegraphics[width=0.51\textwidth]{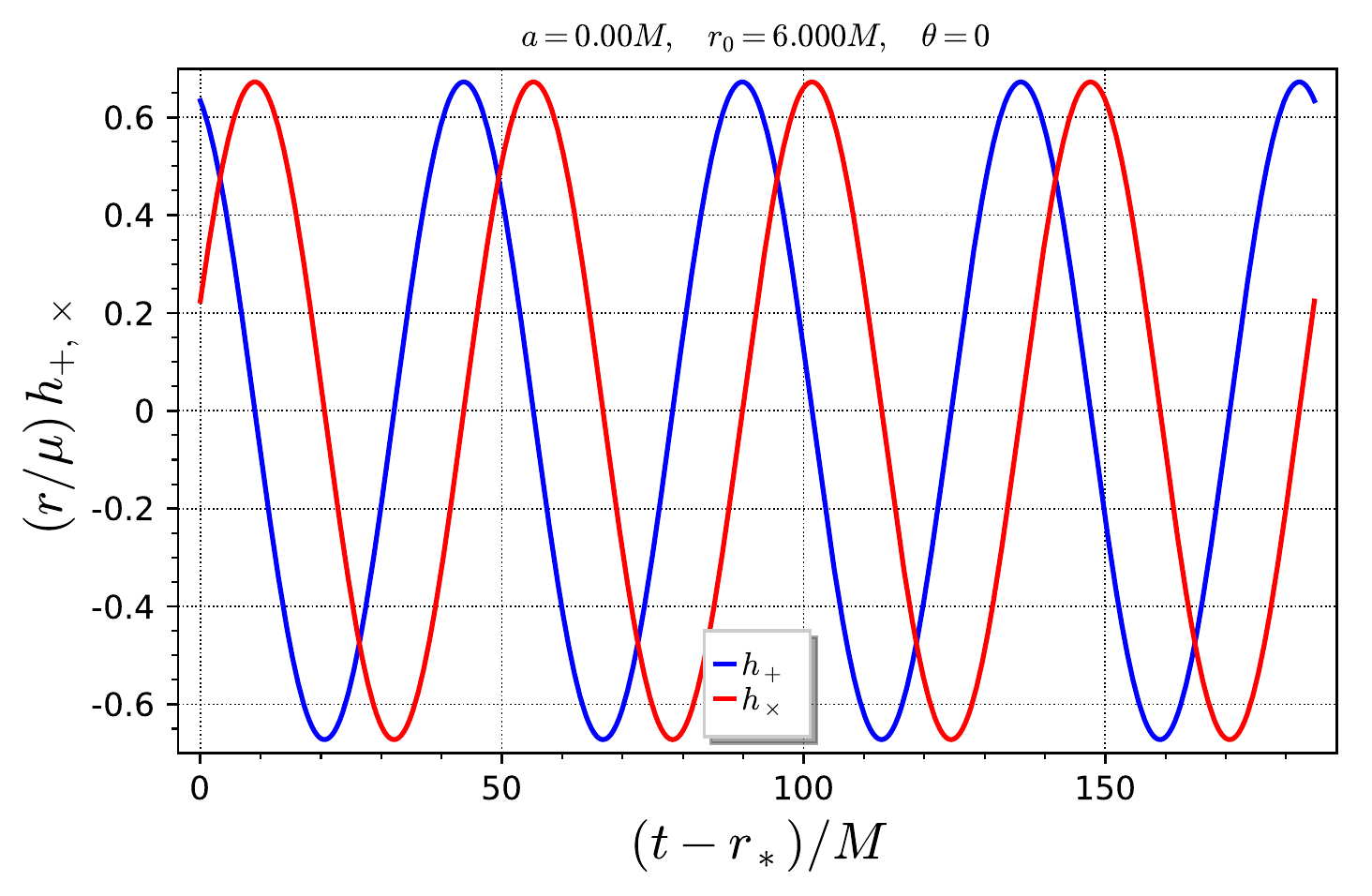} &
\includegraphics[width=0.51\textwidth]{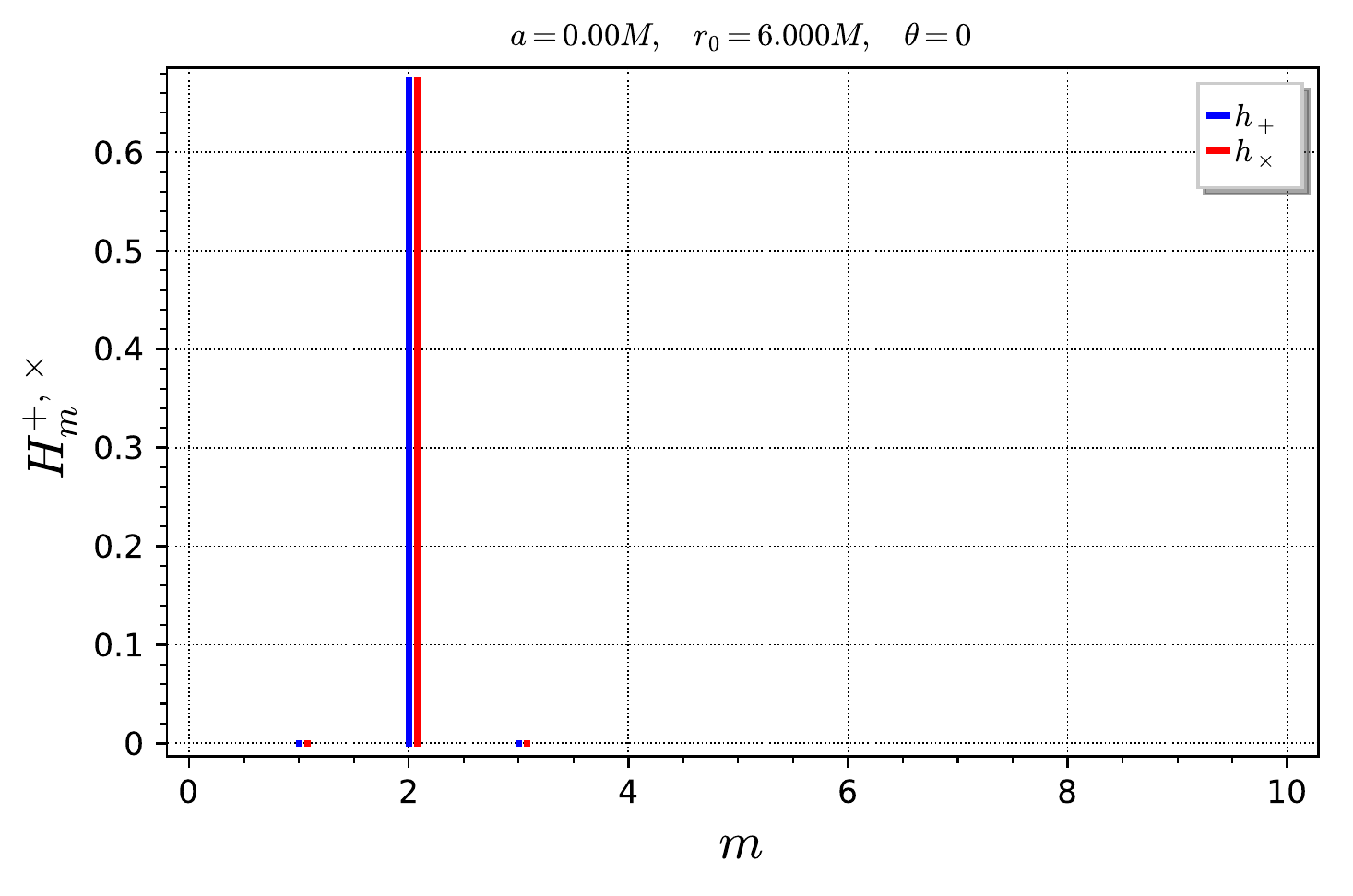} \\
\includegraphics[width=0.51\textwidth]{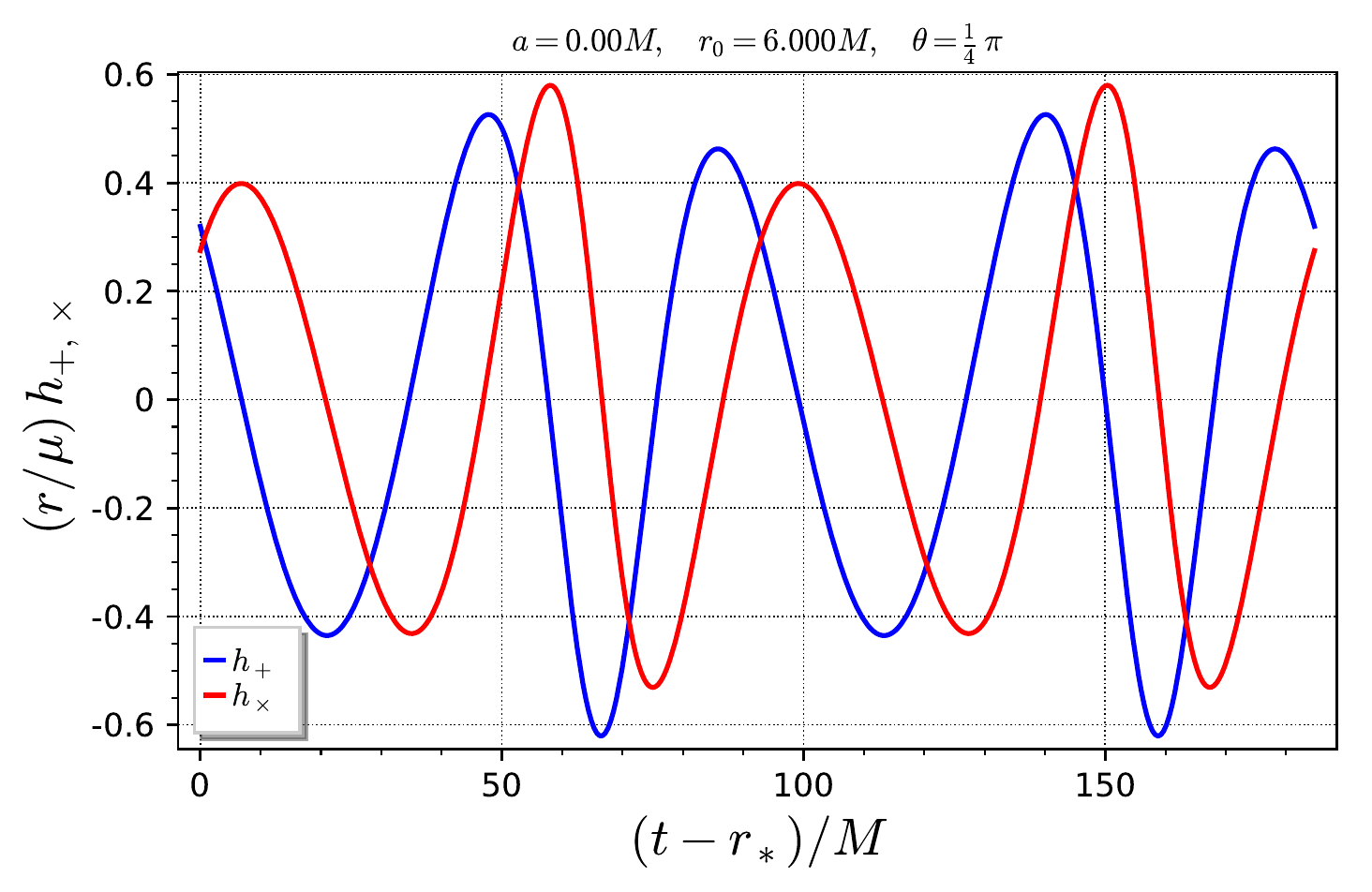} &
\includegraphics[width=0.51\textwidth]{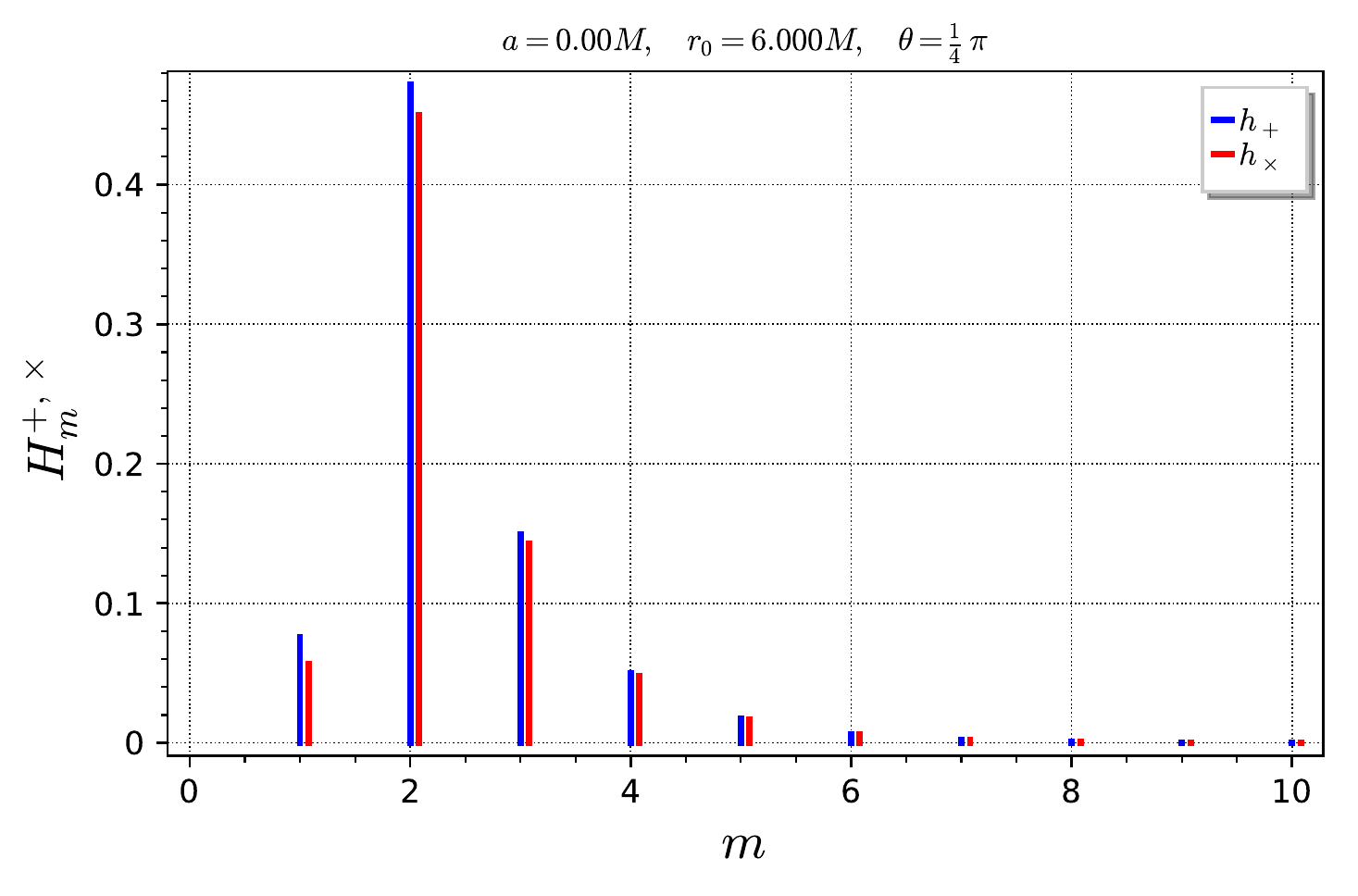} \\
\includegraphics[width=0.51\textwidth]{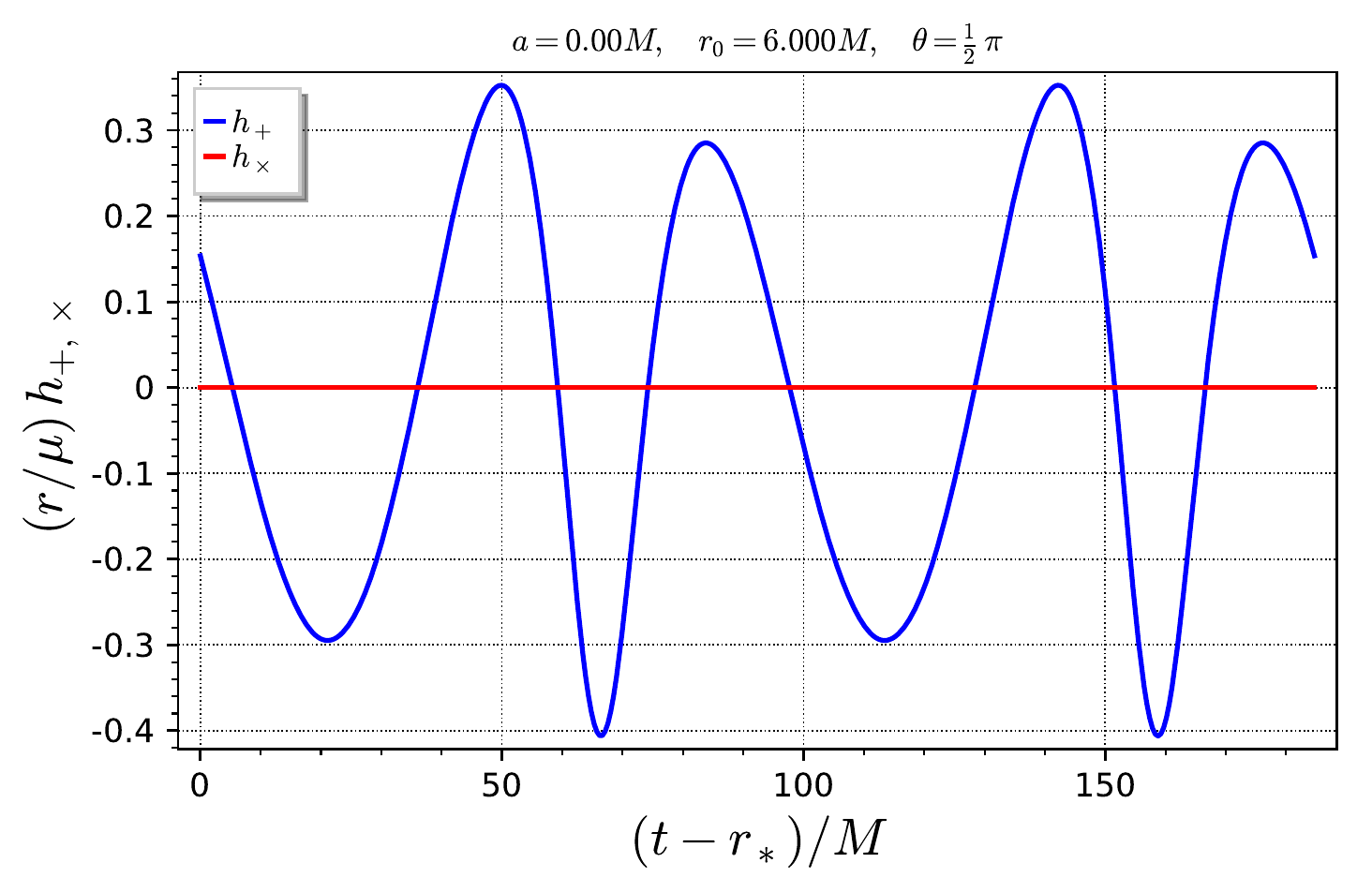} &
\includegraphics[width=0.51\textwidth]{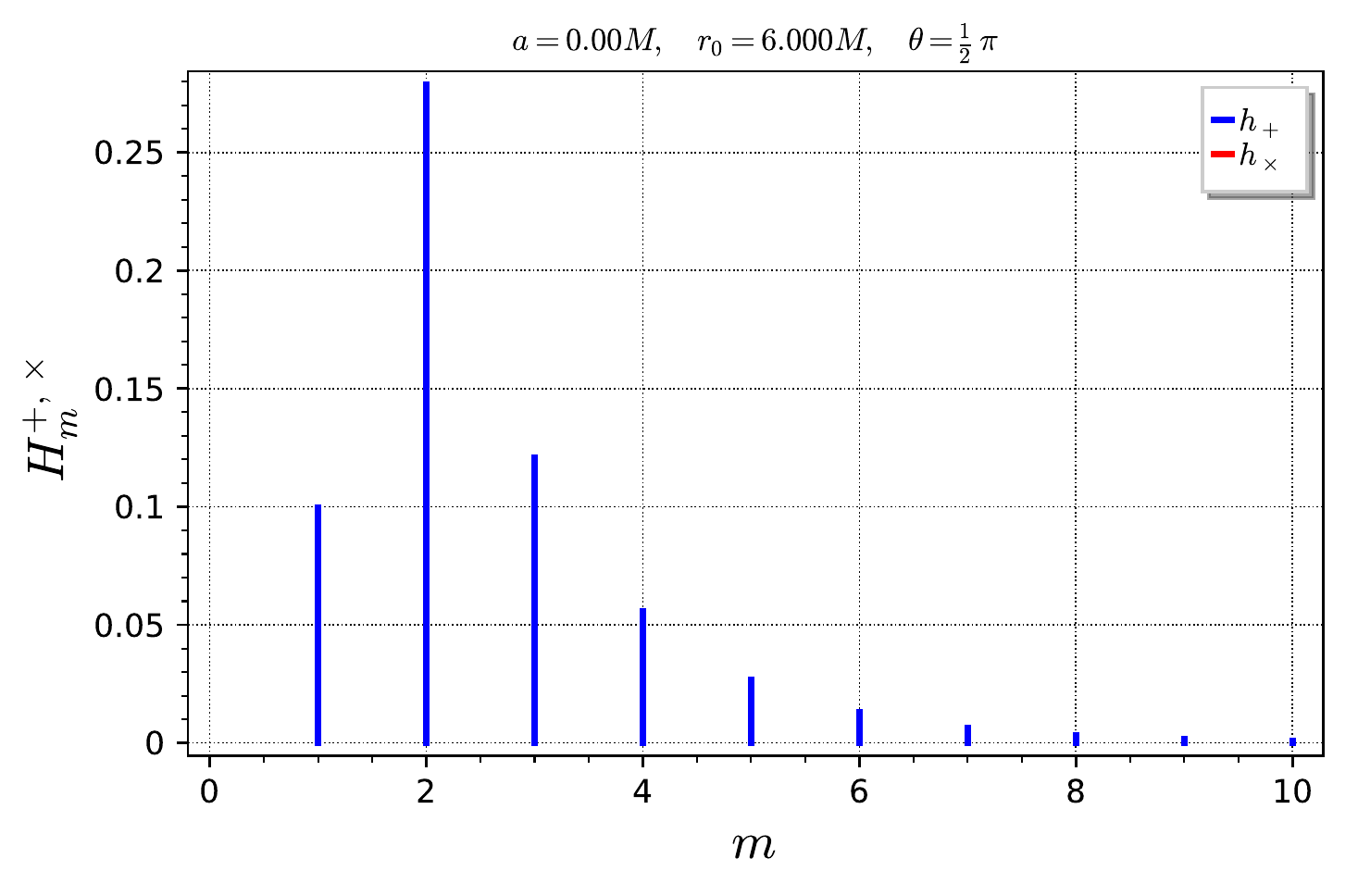}
\end{tabular}
\caption{ \label{f:h_point_mass_a0}
Waveform (left column) and Fourier spectrum (right column)
of gravitational radiation from a point
mass orbiting on the ISCO of a Schwarzschild BH ($a=0$). All amplitudes
are rescaled by $r/\mu$, where $r$ is the Boyer-Lindquist radial coordinate
of the observer and $\mu$ the mass of the orbiting point.
Three values of the colatitude $\theta$ of the observer are considered:
$\theta=0$ (first row), $\theta=\pi/4$ (second row) and $\theta=\pi/2$
(third row).}
\end{figure*}

\begin{figure*}
\begin{tabular}{c@{\!\!}c}
\includegraphics[width=0.51\textwidth]{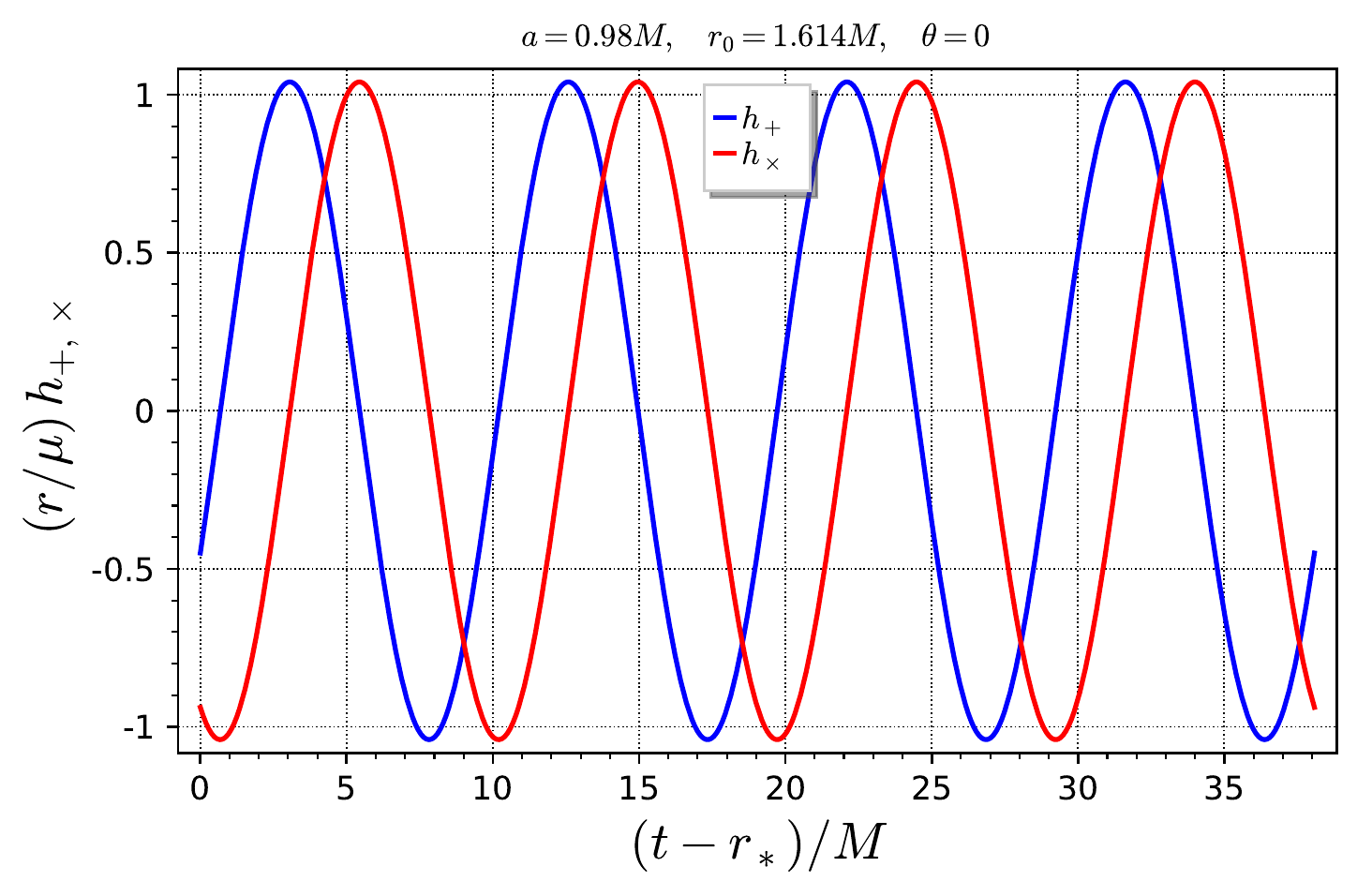} &
\includegraphics[width=0.51\textwidth]{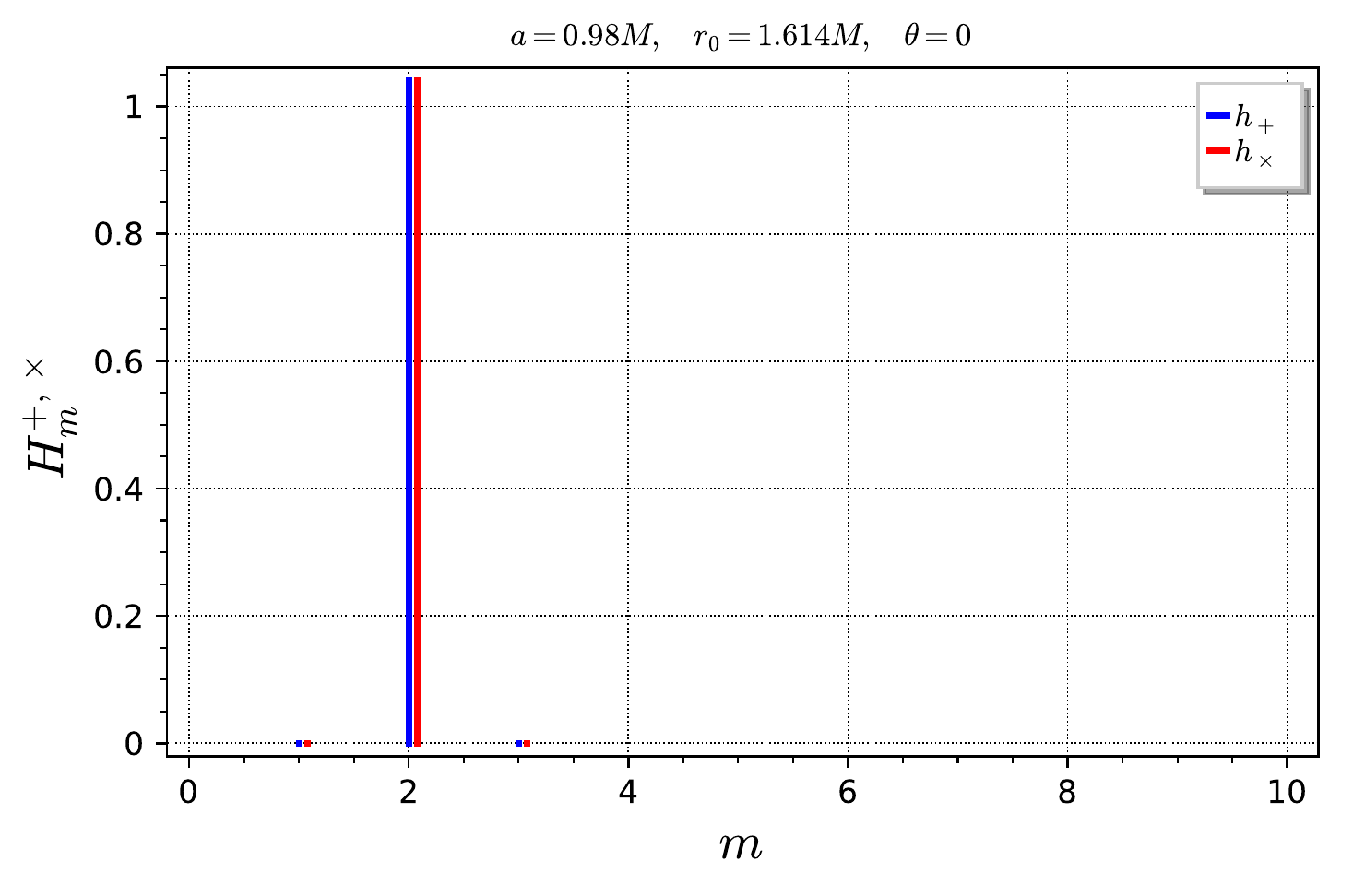} \\
\includegraphics[width=0.51\textwidth]{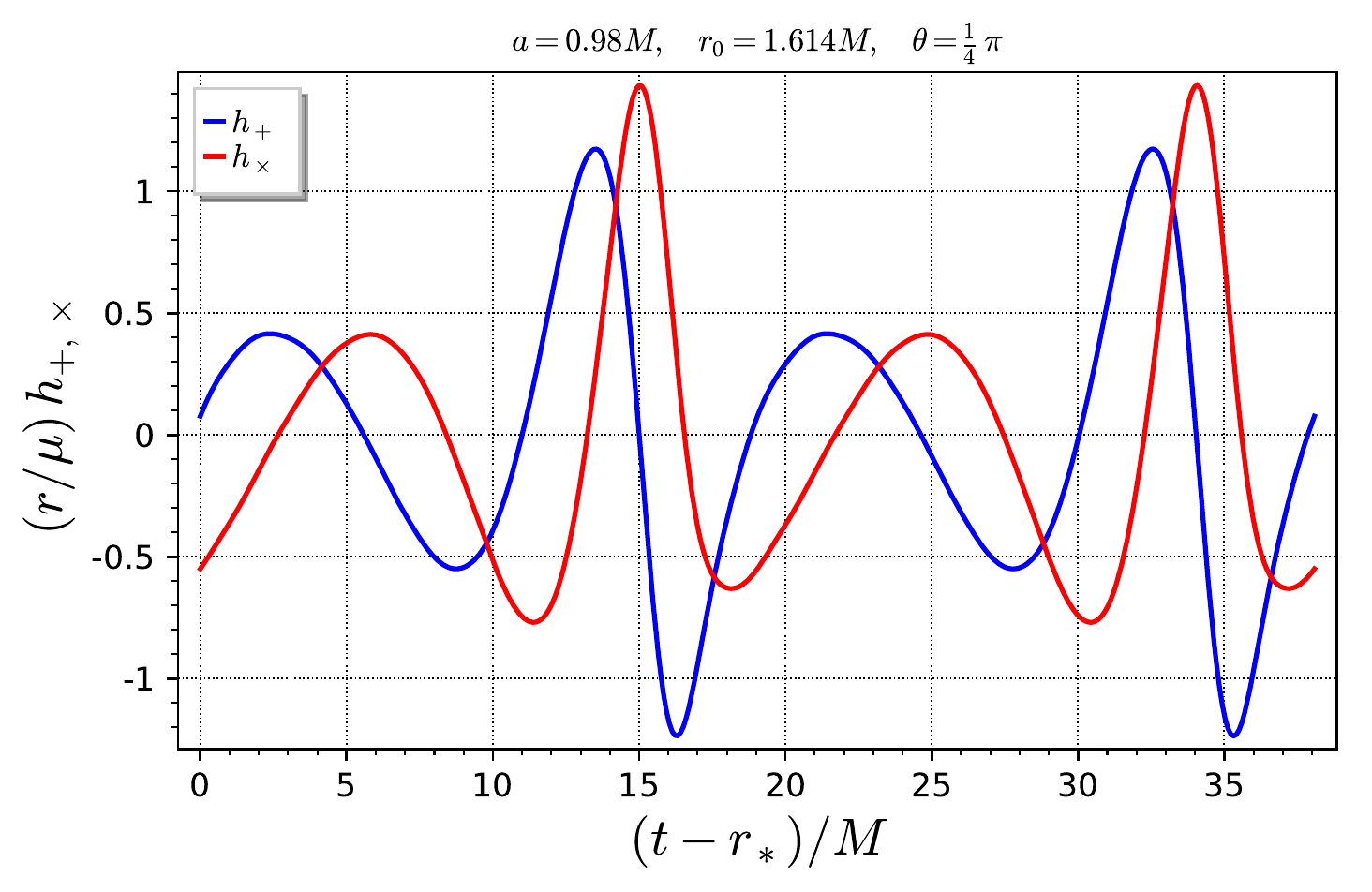} &
\includegraphics[width=0.51\textwidth]{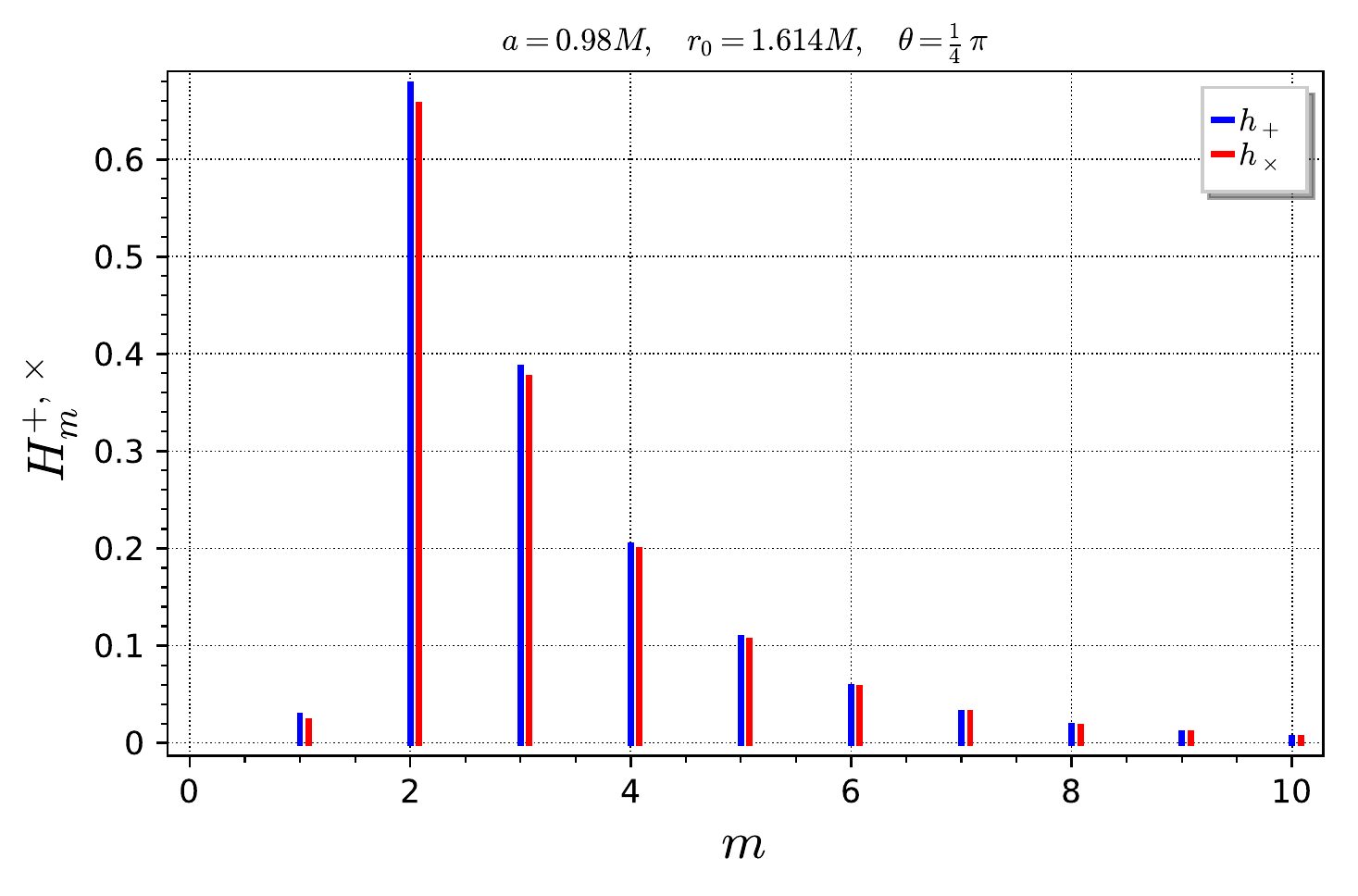} \\
\includegraphics[width=0.51\textwidth]{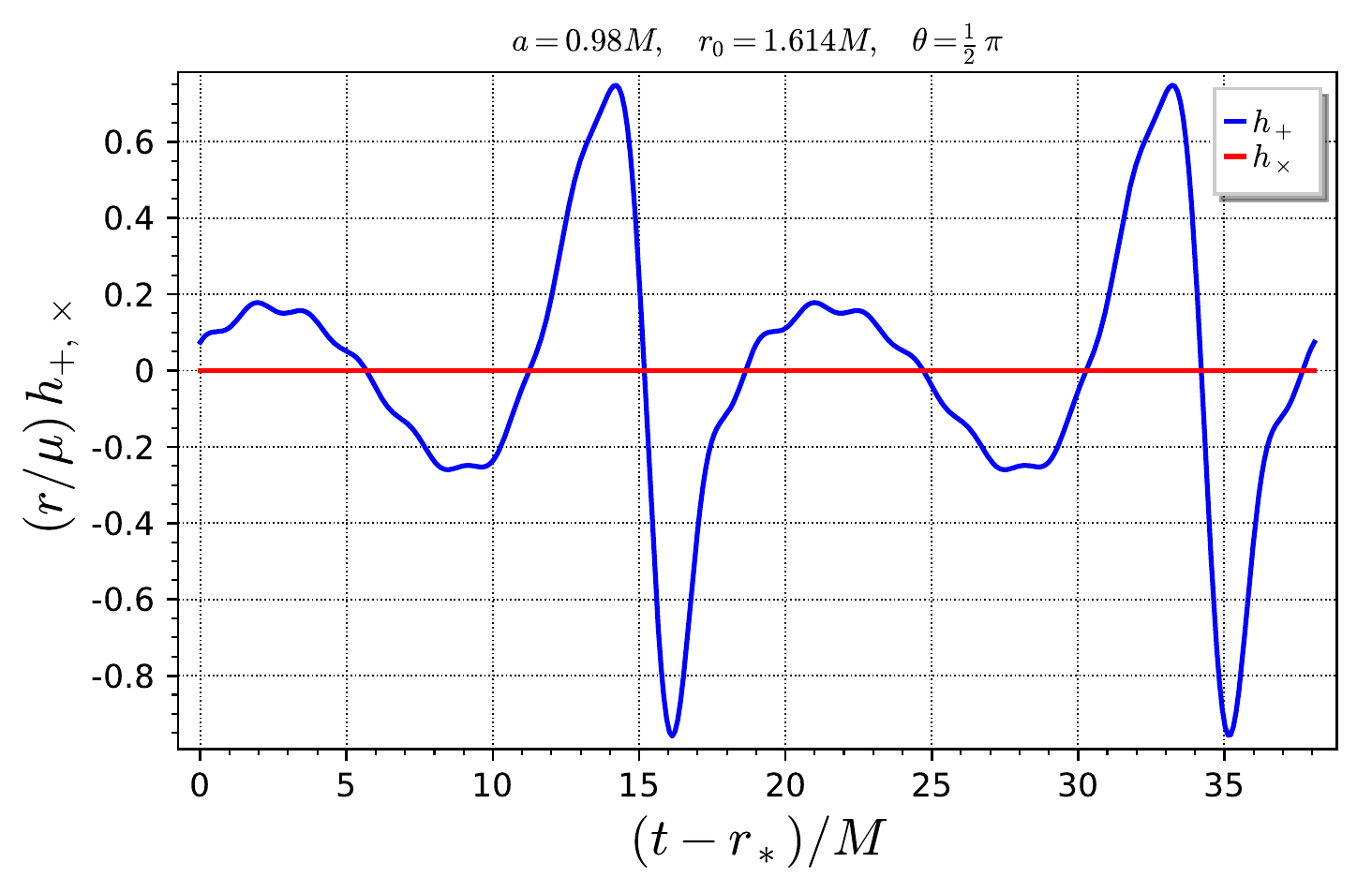} &
\includegraphics[width=0.51\textwidth]{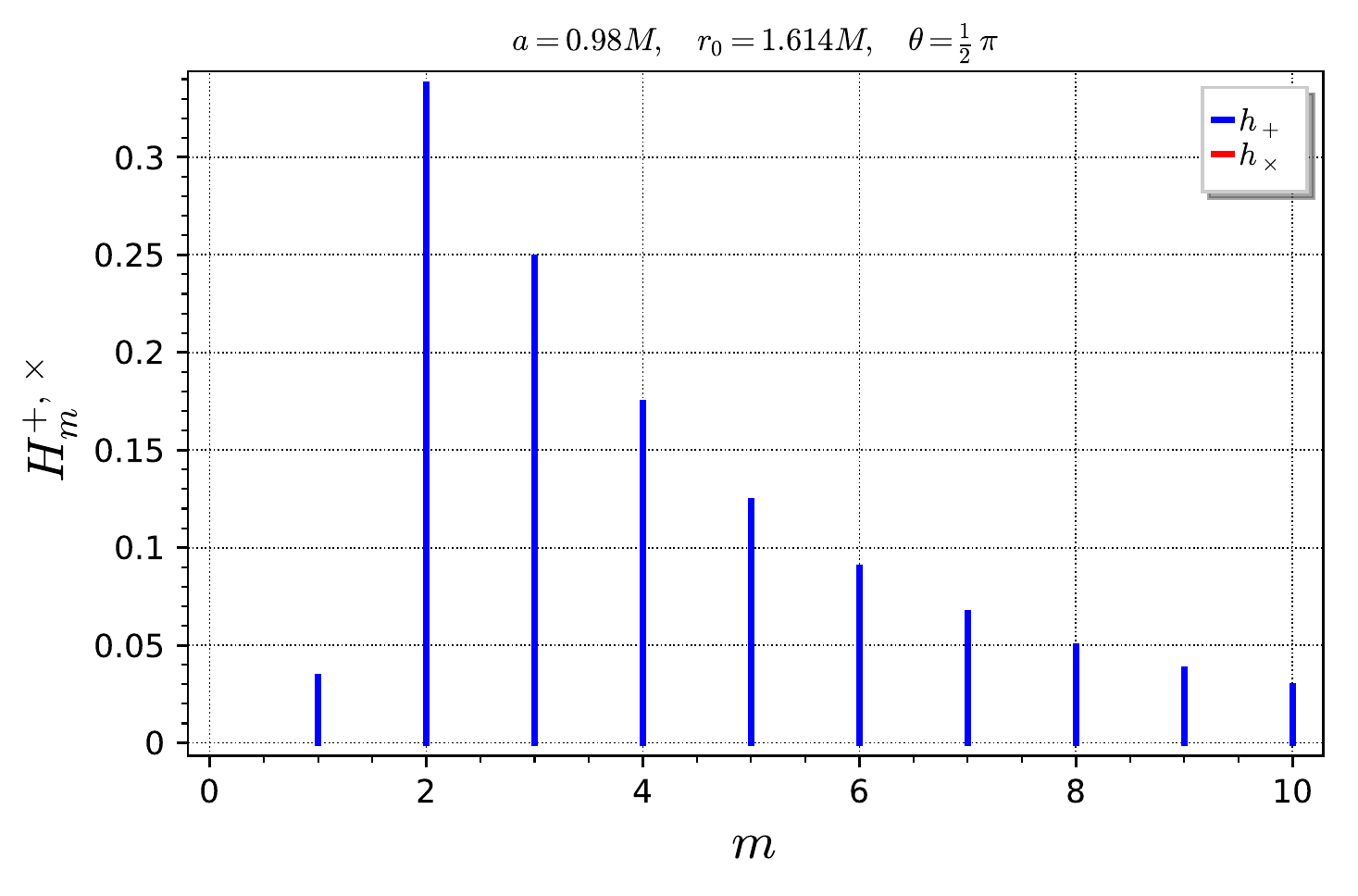}
\end{tabular}
\caption{ \label{f:h_point_mass_a98}
Same as Fig.~\ref{f:h_point_mass_a0}
but for a point mass orbiting on the
prograde ISCO of a Kerr BH with $a=0.98M$.}
\end{figure*}

\subsection{Fourier series expansion} \label{s:Fourier_series}

Observed at a fixed location $(r,\theta,\varphi)$, the waveform $(h_+,h_\times)$
as given by Eq.~(\ref{e:h}) is a periodic function of $t$, or equivalently
of the retarded time $u\equiv t-r_*$, the period being nothing but
the orbital period of the particle:  $T_0 = 2\pi/\omega_0$. It can therefore be
expanded in Fourier series. Noticing that the $\varphi$-dependency of the
spheroidal harmonic $_{-2}S^{am\omega_0}_{\ell m}(\theta,\varphi)$
is simply $e^{\ui m \varphi}$, we may rewrite Eq.~(\ref{e:h}) as an explicit
Fourier series expansion\footnote{The notation $h_{+,\times}$ stands
for either $h_+$ or $h_\times$.}:
\beq \label{e:h_Fourier}
    h_{+,\times} = \frac{\mu}{r} \sum_{m=1}^{+\infty}
        \left[ A_m^{+,\times}(\theta) \cos(m\psi)
         + B_m^{+,\times}(\theta)\sin(m\psi) \right] ,
\eeq
where $\psi$ is given by Eq.~(\ref{e:def_psi})
and $A_m^{+}(\theta)$, $A_m^{\times}(\theta)$, $B_m^{+}(\theta)$ and $B_m^{\times}(\theta)$ are real-valued
functions of $\theta$, involving $M$, $a$ and $r_0$:
\begin{subequations}
\label{e:Fourier_A_B}
\begin{align}
     A_m^+(\theta) & = \frac{2}{(m\omega_0)^2}
        \sum_{\ell=2}^{+\infty}
        \mathrm{Re}\left( Z^\infty_{\ell m}(r_0) \right)
        \left[ (-1)^\ell\,  {}_{-2}S^{-am\omega_0}_{\ell,- m}(\theta,0)
        + {}_{-2}S^{am\omega_0}_{\ell m}(\theta,0) \right] ,\\
    B_m^+(\theta) &= \frac{2}{(m\omega_0)^2}
        \sum_{\ell=2}^{+\infty}
        \mathrm{Im}\left( Z^\infty_{\ell m}(r_0) \right)
        \left[ (-1)^\ell\,  {}_{-2}S^{-am\omega_0}_{\ell,- m}(\theta,0)
        + {}_{-2}S^{am\omega_0}_{\ell m}(\theta,0) \right] ,\\
    A_m^\times(\theta) &= \frac{2}{(m\omega_0)^2}
        \sum_{\ell=2}^{+\infty}
        \mathrm{Im}\left( Z^\infty_{\ell m}(r_0) \right)
        \left[ (-1)^\ell\,  {}_{-2}S^{-am\omega_0}_{\ell,- m}(\theta,0)
        - {}_{-2}S^{am\omega_0}_{\ell m}(\theta,0) \right] ,\\
    B_m^\times(\theta) &= \frac{2}{(m\omega_0)^2}
        \sum_{\ell=2}^{+\infty}
        \mathrm{Re}\left( Z^\infty_{\ell m}(r_0) \right)
        \left[ (-1)^{\ell+1}\,  {}_{-2}S^{-am\omega_0}_{\ell,- m}(\theta,0)
        + {}_{-2}S^{am\omega_0}_{\ell m}(\theta,0) \right] .
\end{align}
\end{subequations}
We then define the spectrum of the gravitational wave at a fixed value
of $\theta$ as the two series
(one per polarization mode):
\beq \label{e:hm_def}
   H_m^{+,\times}(\theta) \equiv \sqrt{\left(A_m^{+,\times}(\theta) \right)^2
        + \left(B_m^{+,\times}(\theta) \right)^2} , \quad
    1 \leqslant m < +\infty .
\eeq

We have developed an open-source SageMath package,
\texttt{kerrgeodesic\_gw} (cf.~Appendix~\ref{s:kerrgeodesic_gw}),
implementing the above formulas, and more generally all the computations
presented in this article, like the signal-to-noise ratio and Roche limit ones
to be discussed below.
The spectrum, as well as the corresponding waveform,
computed via \texttt{kerrgeodesic\_gw},
are depicted in
Figs.~\ref{f:h_point_mass_a0} and
\ref{f:h_point_mass_a98} for $a=0$ and $a=0.98M$ respectively.
In each figure, $\varphi=\varphi_0$ and three values of $\theta$ are selected:
$\theta=0$ (orbit seen face-on), $\pi/4$ and $\pi/2$
(orbit seen edge-on).

We notice that for $\theta=0$, only the Fourier mode $m=2$ is present
and that $h_+$ and $h_\times$ have identical amplitudes and are in quadrature.
This behavior is identical to that given by the large radius (quadrupole-formula)
approximation (\ref{e:h_quadrupole}).
For $\theta>0$, all modes with $m\geqslant 1$ are populated, whereas the
approximation (\ref{e:h_quadrupole}) contains only $m=2$.
For $\theta=\pi/2$, $h_\times$ vanishes identically
and the relative amplitude of the modes $m\not=2$ with respect to the mode $m=2$
is the largest one, reaching $\sim 75\%$ for $m=3$ and $\sim 50\%$
for $m=4$ when $a=0.98 M$.

Some tests of our computations, in particular comparisons with previous
results by \citet{Poisson93a} ($a=0$) and \citet{Detweiler78} ($a=0.5 M$ and
$a=0.9 M$) are presented in
Appendix~\ref{s:kerrgeodesic_gw}.


\section{Signal-to-noise ratio in the LISA detector}
\label{s:SNR}

The results in Sect.~\ref{s:gw_particle} are valid for any BH.
We now specialize them to Sgr~A* and evaluate the signal-to-noise ratio in the
LISA detector, as a function of the mass $\mu$ of the orbiting object, the
orbital radius $r_0$ and the spin parameter $a$ of Sgr~A*.

\subsection{Computation}

Assuming that its noise is stationary and Gaussian, a given detector is characterized
by its one-sided noise power spectral density (PSD) $S_{\rm n}(f)$.
For a gravitational wave search based on the matched filtering technique, the signal-to-noise ratio (S/N) $\rho$ is given by the following formula \citep[see e.g.,][]{JaranowskiK12,MooreCB15}:
\beq \label{e:SNR_matched}
    \rho^2 = 4 \int_0^{+\infty} \frac{|\tilde{h}(f)|^2}{S_{\rm n}(f)}
        \, \mathrm{d} f ,
\eeq
where $\tilde{h}(f)$ is the Fourier transform of the imprint $h(t)$
of the gravitational wave on the detector,
\beq
    \tilde{h}(f) = \int_{-\infty}^{+\infty} h(t)\,
                     \mathrm{e}^{-2\pi \mathrm{i} f t} \, \mathrm{d}t ,
\eeq
$h(t)$ being a linear combination of the two polarization modes $h_+$ and
$h_\times$ at the detector location:
\beq
     h(t) = F_+(\Theta,\Phi,\Psi) \, h_+(t, r, \theta, \varphi)
        + F_\times(\Theta,\Phi,\Psi)\,  h_\times(t, r, \theta, \varphi) .
\eeq
In the above expression, $(t, r, \theta, \varphi)$ are
the Boyer-Lindquist coordinates of the detector (``Sgr~A* frame''),
while $F_+$ and $F_\times$ are the detector beam-pattern coefficients
(or response functions), which depend on the direction $(\Theta,\Phi)$ of the
source with respect to the detector's frame and on the polarization
angle $\Psi$, the latter being
the angle between
the direction of constant azimuth $\Phi$ and
the principal direction ``+'' in the wavefront plane
(i.e. the axis of the $h_+$ mode or equivalently the direction of the semi-major
axis of the orbit viewed as an ellipse in the detector's sky) \citep{ApostolatosCST94}.
For a detector like LISA, where, for high enough frequencies, the gravitational wavelength
can be comparable or smaller than the arm length ($2.5\; {\rm Gm}$), the
response functions $F_+$ and $F_\times$ depend a priori on the gravitational wave
frequency $f$, in addition to $(\Theta,\Phi,\Psi)$ \citep{RobsonCL18}.
However for the gravitational waves considered here, a reasonable upper
bound of the frequency is that of the harmonic $m=4$ (say) of waves from
the prograde ISCO of an extreme Kerr BH (see Fig.~\ref{f:h_point_mass_a98}). From the value given by Eq.~(\ref{e:f_ISCO_SgrA}), this is $f_{\rm max} = 2\times 7.9 \simeq 15.8 \; {\rm mHz}$, the multiplication by $2$ taking into account the transition from $m=2$ to $m=4$.
This value being lower than LISA's transfer frequency $f_* = 19.1 \; {\rm mHz}$
\citep{RobsonCL18}, we may consider that $F_+$ and $F_\times$ do not depend on $f$
(see Fig.~2 in \citet{RobsonCL18}). They are given in terms of
$(\Theta,\Phi,\Psi)$ by Eq.~(3.12) of \citet{Cutler98} (with the prefactor
$\sqrt{3}/2$ appearing in Eq.~(3.11) included in them).

Generally, the function $S_{\rm n}(f)$ considered in the LISA literature,
and in particular to present the LISA sensitivity curve, is not the true
noise PSD of the instrument, $P_{\rm n}(f)$ say, but rather $P_{\rm n}(f)/R(f)$, where
$R(f)$ is the average over the sky (angles
$(\Theta,\Phi)$) and over the polarization (angle $\Psi$) of the square of
the response functions $F_+$ and $F_\times$, so that
Eq.~(\ref{e:SNR_matched}) yields directly
the sky and polarization average S/N by substituting
$|\tilde{h}_+(f)|^2 + |\tilde{h}_\times(f)|^2$ for
$|\tilde{h}(f)|^2$ (see \citet{RobsonCL18} for details).
With Sgr~A* as a target, the direction angles $(\Theta,\Phi)$ are of course
known and, for a short observation time (1 day say), they are approximately
constant. However, on longer observation times, theses angles varies due to the
motion of LISA spacecrafts on their orbits around the Sun. Moreover, the polarization angle
$\Psi$ is not known at all, since it depends on the orientation of the orbital
plane around the MBH, which is assumed to
be the equatorial plane, the latter being currently unknown. For these
reasons, we consider the standard sky and polarization average sensitivity
of LISA, $S_{\rm n}(f) = P_{\rm n}(f) / R(f)$, as given e.g., by Eq.~(13)
of \citet{RobsonCL18}, and define the \emph{effective signal-to-noise ratio} $\rho$
by
\beq \label{e:SNR_eff}
    \rho^2 = 4 \int_0^{+\infty}
    \frac{|\tilde{h}_+(f)|^2+|\tilde{h}_\times(f)|^2}{S_{\rm n}(f)}
        \, \mathrm{d} f ,
\eeq
where $\tilde{h}_+(f)$ and $\tilde{h}_\times(f)$ are the Fourier transforms
of the two gravitational wave signals $h_+(t)$ and $h_\times(t)$, as given
by Eq.~(\ref{e:h}) or Eq.~(\ref{e:h_Fourier}), over some observation time $T$:
\beq \label{e:FT_trunc}
    \tilde{h}_{+,\times}(f) = \int_{-T/2}^{T/2} h_{+,\times}(t)\,
                     \mathrm{e}^{-2\pi \mathrm{i} f t} \, \mathrm{d}t .
\eeq
As shown in Appendix~\ref{s:comput_SNR}, plugging the expressions (\ref{e:h_Fourier})
for $h_+(t)$ and $h_\times(t)$ into Eqs.~(\ref{e:SNR_eff})-(\ref{e:FT_trunc})
leads to the following S/N value:
\beq \label{e:SNR_sum_m}
    \rho = \frac{\mu}{r} \sqrt{T} \left( \sum_{m=1}^{+\infty}
        \frac{\left(H_m^+(\theta)\right) ^2 +
            \left(H_m^{\times}(\theta)\right) ^2}{S_{\rm n}(m f_0)}
        \right) ^{1/2}
         \quad\mbox{for}\quad f_0 T \gg 1,
\eeq
where the coefficients $H_m^+(\theta)$ and $H_m^{\times}(\theta)$
are defined by Eq.~(\ref{e:hm_def}) and $f_0=\omega_0/(2\pi)$ is the orbital
frequency, $\omega_0$ being the function of $M$, $a$ and $r_0$ given by
Eq.~(\ref{e:omega0}).

\begin{figure}
\centerline{\includegraphics[width=0.7\textwidth]{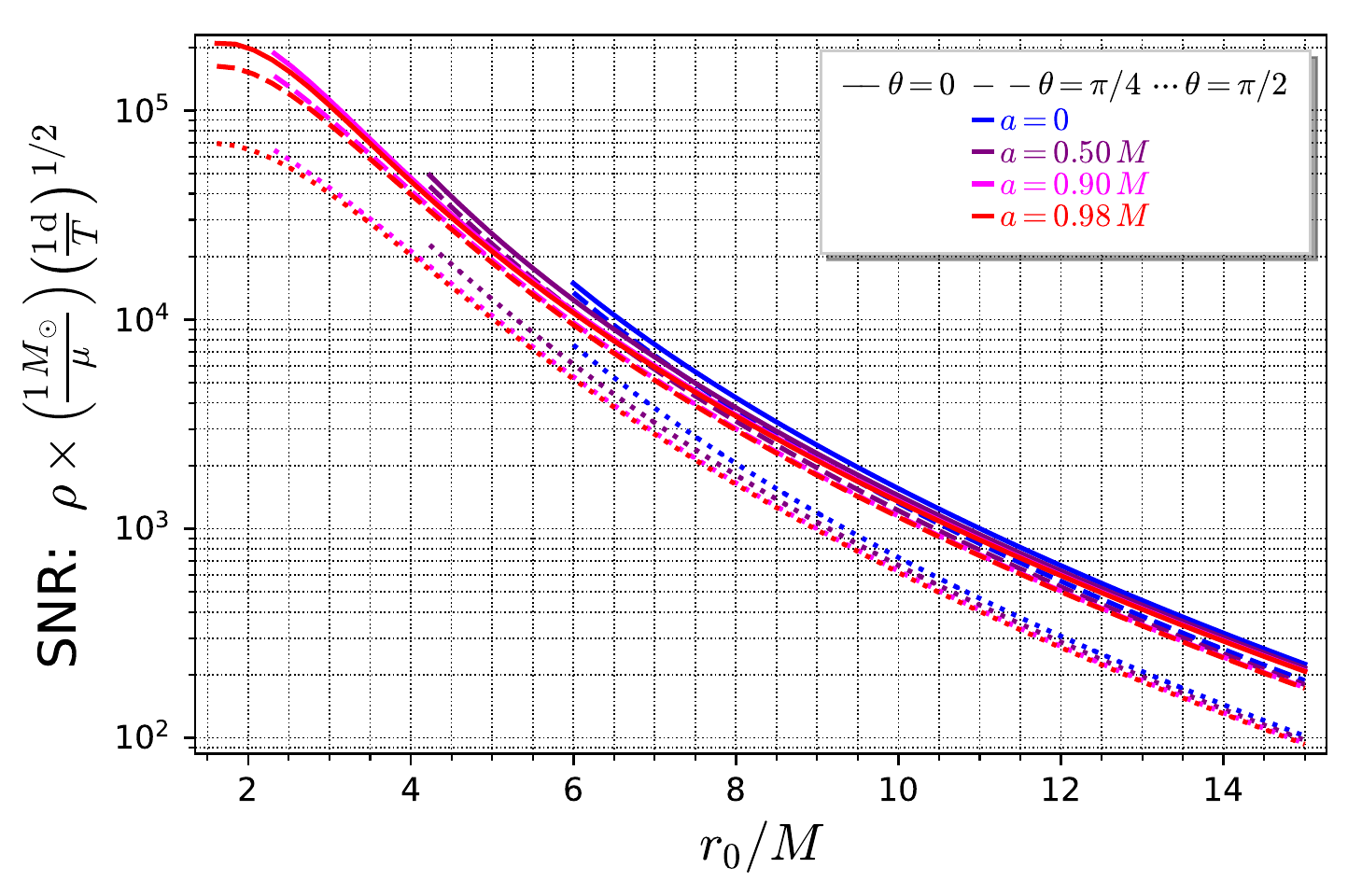}}
\caption{ \label{f:snr_radius_15}
Effective (direction and polarization averaged) S/N in LISA
for a 1-day observation of an
object of mass $\mu$ orbiting Sgr~A*, as a function of the orbital radius $r_0$ and for
selected values of the Sgr~A*'s spin parameter $a$ and well
as selected values of the inclination angle $\theta$. Each curve starts at the
ISCO radius of the corresponding value of $a$.}
\end{figure}

\begin{figure}
\centerline{\includegraphics[width=0.7\textwidth]{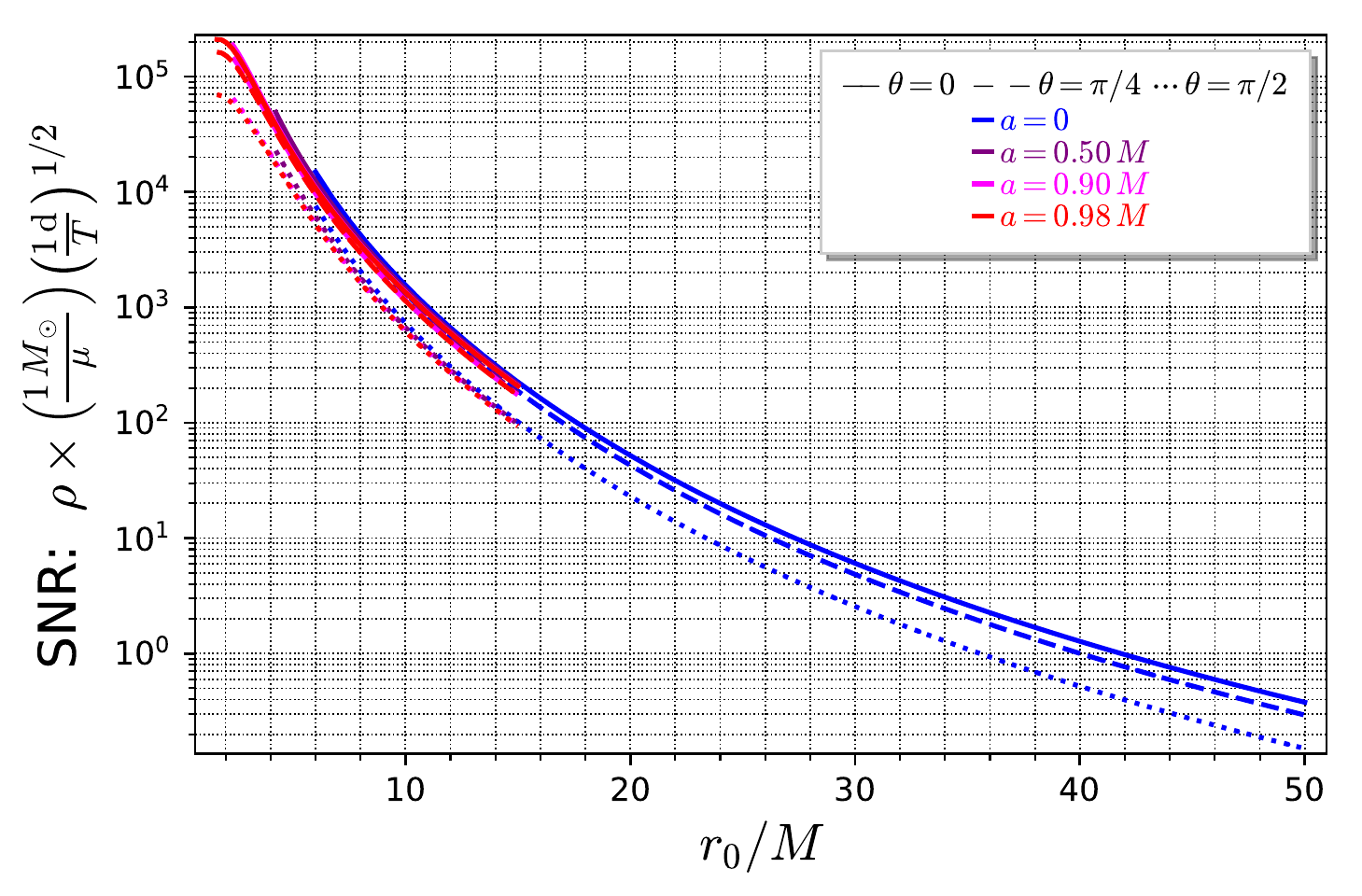}}
\caption{ \label{f:snr_radius_50}
Same as Fig.~\ref{f:snr_radius_15}, except for $r_0$ ranging up to $50 M$.
For $r_0  > 15 M$, only the $a=0$ curves are plotted, since
the MBH spin plays a negligible role at large distance.
}
\end{figure}

The effective S/N resulting from Eq.~(\ref{e:SNR_sum_m})
is shown in Figs.~\ref{f:snr_radius_15}--\ref{f:snr_radius_50}.
We use the value (\ref{e:mu_ov_r}) for $\mu/r$ and
the analytic model of \citet{RobsonCL18} (their Eq.~(13))
for LISA sky and polarization average sensitivity $S_{\rm n}(f)$.
We notice that for a given value of the orbital radius $r_0$
and a given MBH spin $a$, the S/N is maximum for
the inclination angle $\theta=0$
and minimal for $\theta=\pi/2$, the ratio between the two values varying
from $\sim 2$ for $a=0$ to $\sim 3$ for $a=0.98 M$. This behavior was
already present in the waveform amplitudes displayed in
Figs.~\ref{f:h_point_mass_a0}--\ref{f:h_point_mass_a98}.

Another feature apparent from Figs.~\ref{f:snr_radius_15} and \ref{f:snr_radius_50}
is that at fixed orbital radius $r_0$, the S/N is a decaying function of $a$.
This results from the fact that the orbital frequency $f_0$ is a decaying
function of $a$ [cf.~Eq.~(\ref{e:omega0})], which both reduces the gravitational
wave amplitude and displaces the wave frequency to less favorable parts of
LISA's sensitivity curve.

At the ISCO, the S/N for $\theta=0$ is
\beq \label{e:SNR_ISCO}
    \rho_{\rm ISCO} = \alpha 10^5 \left( \frac{\mu}{1\; M_\odot} \right)
    \left( \frac{T}{1\; {\rm d}} \right) ^{1/2} ,
\eeq
with the coefficient $\alpha$ given in Table~\ref{t:SNR_ISCO}.
It should be noted that if the observation time is one year, then the factor
$(T/1 {\rm d})^{1/2}$ is $\sqrt{365.25} \simeq 19.1$.

\begin{table}
\caption{
Coefficient $\alpha$ in formula (\ref{e:SNR_ISCO}) for the S/N from the
ISCO.}
\label{t:SNR_ISCO}
\centering
\begin{tabular}{lcccc}
\hline\hline
$a/M$ & 0 & 0.50 & 0.90 & 0.98\\
\hline
$\alpha$ & 0.149 & 0.490 & 1.87 & 2.09\\
\hline
\end{tabular}
\end{table}

\begin{figure}
\centerline{\includegraphics[width=0.7\textwidth]{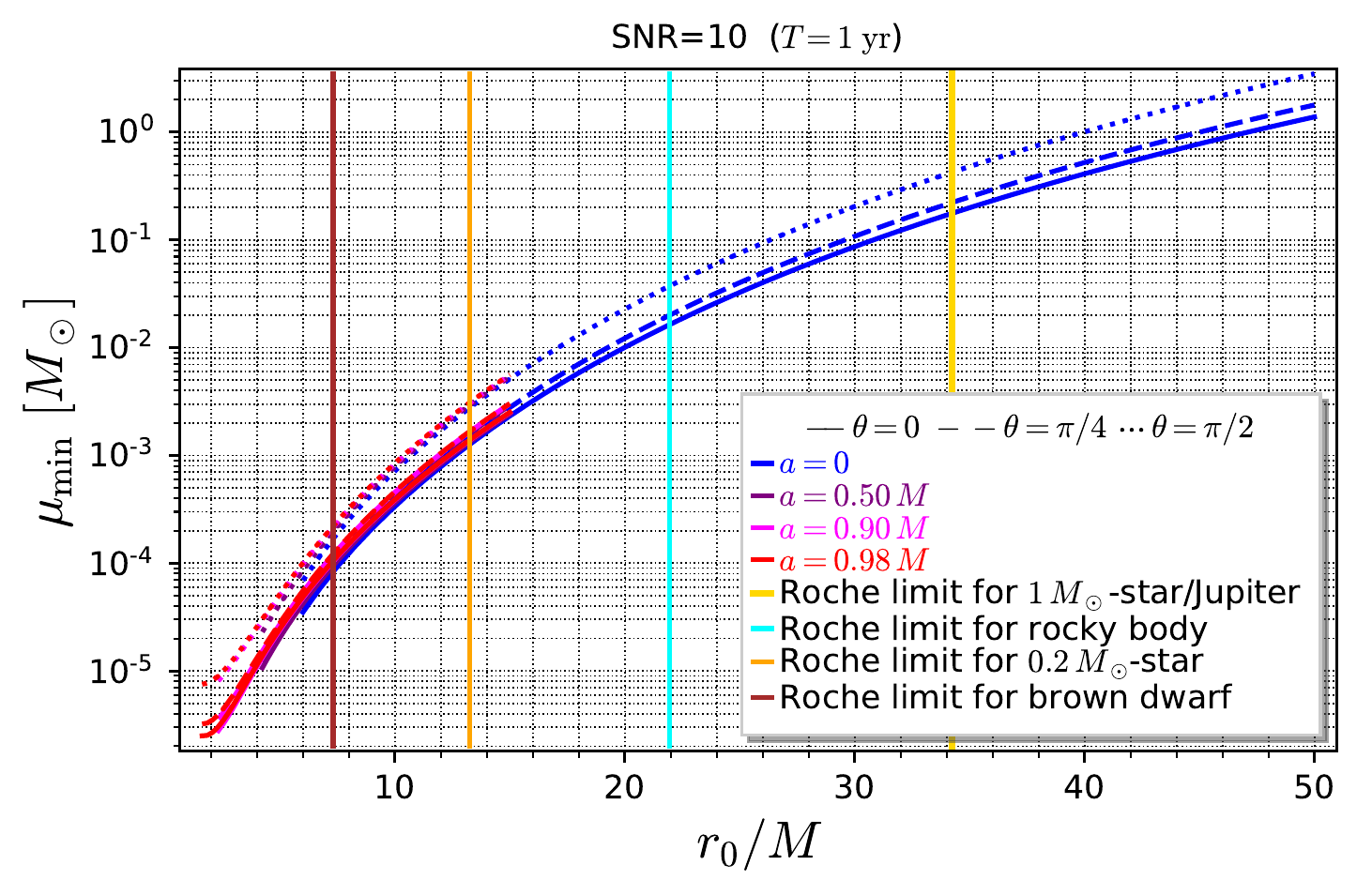}}
\caption{ \label{f:min_detect_mass}
Minimal detectable mass with a S/N larger than 10 in one year of LISA observation,
as a function of the orbital radius $r_0$. The various Roche limits are
those considered in Sect.~\ref{sec:tidal}. As in Fig.~\ref{f:snr_radius_50},
for $r_0  > 15 M$, only the $a=0$ curves are shown,
the MBH spin playing a negligible role at large distance.
}
\end{figure}

\subsection{Minimal detectable mass}

As clear from Eq.~(\ref{e:SNR_sum_m}), the S/N $\rho$ is proportional to the
mass $\mu$ of the orbiting body and to the square root of the
observing time $T$.
It is then easy to evaluate the minimal
mass $\mu_{\rm min}$ that can be detected by analyzing one year of LISA data, setting
the detection threshold to
\beq \label{e:SNR_threshold}
    \SNRyr = 10 ,
\eeq
where $\SNRyr$ stands for the value of $\rho$ for $T = 1\; {\rm yr}$.
The result is shown in Fig.~\ref{f:min_detect_mass}.
If one does not take into account any Roche limit, it is worth noticing
that the minimal detectable mass is quite small: $\mu_{\rm min} \simeq 3\times 10^{-5}\, M_\odot$
at the ISCO of a Schwarzschild BH ($a=0$), down to
$\mu_{\rm min} \simeq 2 \times 10^{-6}\, M_\odot$ (the Earth mass) at the ISCO of a rapidly rotating
Kerr BH ($a\geqslant 0.90M$).


\section{Radiated energy and orbital decay}
\label{s:orbital_decay}

In the above sections, we have assumed that the orbits are exactly circular,
i.e. we have neglected the reaction to gravitational radiation. We now take
it into account and discuss the resulting secular evolution of the orbits.

\subsection{Total radiated power} \label{s:radiated_power}

The total power (luminosity) emitted via gravitational radiation is given by
\citep{Detweiler78}:
\beq
    L = \lim_{r\to+\infty} \frac{r^2}{16\pi} \oint_{\mathcal{S}_r}
        \bigl| \dot{h}_+ - \ui \dot{h}_\times \bigr|^2
            \sin\theta\, \mathrm{d}\theta\, \mathrm{d}\varphi ,
\eeq
where $\mathcal{S}_r$ is the sphere of constant value of $r$ and an overdot stands for the partial derivative with respect to the time coordinate $t$, i.e. $\dot{h}_{+,\times} \!\equiv\! \partial h_{+,\times} / \partial t$.
Substituting the waveform (\ref{e:h}) into this expression
leads to
\beq
    L = \lim_{r\to+\infty} \frac{\mu^2}{4\pi} \oint_{\mathcal{S}_r}
    \Bigg|
    \sum_{\ell=2}^{+\infty} \sum_{{\scriptstyle m=-\ell\atop \scriptstyle m\not=0}}^\ell
    \frac{Z^\infty_{\ell m}(r_0)}{m\omega_0} \, _{-2}S^{am\omega_0}_{\ell m}(\theta,\varphi) \,
    e^{- \ui m (\omega_0 (t-r_*) + \varphi_0)}
     \Bigg| ^2
         \, \sin\theta\, \mathrm{d}\theta\, \mathrm{d}\varphi .
\eeq
Thanks to the orthonormality property of the spin-weighted spheroidal harmonics,
\beq
    \oint_{\mathbb{S}^2} {}_{-2}S^{am\omega_0}_{\ell m}(\theta,\varphi)
        \; {}_{-2}S^{am'\omega_0}_{\ell' m'}(\theta,\varphi)^*
        \, \sin\theta\, \mathrm{d}\theta\, \mathrm{d}\varphi  =
        \delta_{\ell\ell'} \delta_{m m'} ,
\eeq
the above expression simplifies to
\beq \label{e:luminosity}
    L = \left( \frac{\mu}{M}\right)^2 \tilde{L}\left(\frac{r_0}{M}\right)
    \quad\mbox{with}\quad
     \tilde{L}\left(\frac{r_0}{M}\right) \equiv
    \frac{M^2}{4\pi}
    \sum_{\ell=2}^{\infty} \sum_{{\scriptstyle m=-\ell\atop \scriptstyle m\not=0}}^\ell
    \frac{\left| Z^\infty_{\ell m}(r_0) \right| ^2}{(m\omega_0)^2} .
\eeq
It should be noted that $\tilde{L}$ is a dimensionless function of $x\equiv r_0/M$, the
dimension of $Z^\infty_{\ell m}$ being an inverse squared length (see e.g.,
Eq.~(\ref{e:Zinf_22})) and $\omega_0$ being the function of $r_0/M$
given by Eq.~(\ref{e:omega0}).
Moreover, the function $\tilde{L}(x)$ depends only on
the parameter $a/M$ of the MBH.

As a check of Eq.~(\ref{e:luminosity}),
let us consider the limit of large orbital radius:
$r_0 \gg M$. As discussed in Sect.~\ref{s:distant_orbits}, only the terms
 $(\ell,m) = (2, \pm 2)$ are pertinent in this case, with
 $Z^\infty_{2,\pm 2}(r_0)$ given by Eq.~(\ref{e:Zinf_22})
 and $\omega_0$ related to $r_0$ by Eq.~(\ref{e:omega0_Kepler}).
Equation~(\ref{e:luminosity}) reduces then to
\beq \label{e:L_quadrupole}
    L \simeq \frac{32}{5} \left(\frac{\mu}{M}\right)^2
        \left(\frac{M}{r_0}\right)^5 \quad (r_0 \gg M)
    \quad\iff\quad
    \tilde{L}(x) \simeq \frac{32}{5 x^5} \quad (x \gg 1).
\eeq
We recognize the standard result from the
quadrupole formula at Newtonian order
\citep{LandauL71} (see also the lowest order of formula~(314) in the review by
\citet{Blanchet14}).

\begin{figure}
\centerline{\includegraphics[width=0.7\textwidth]{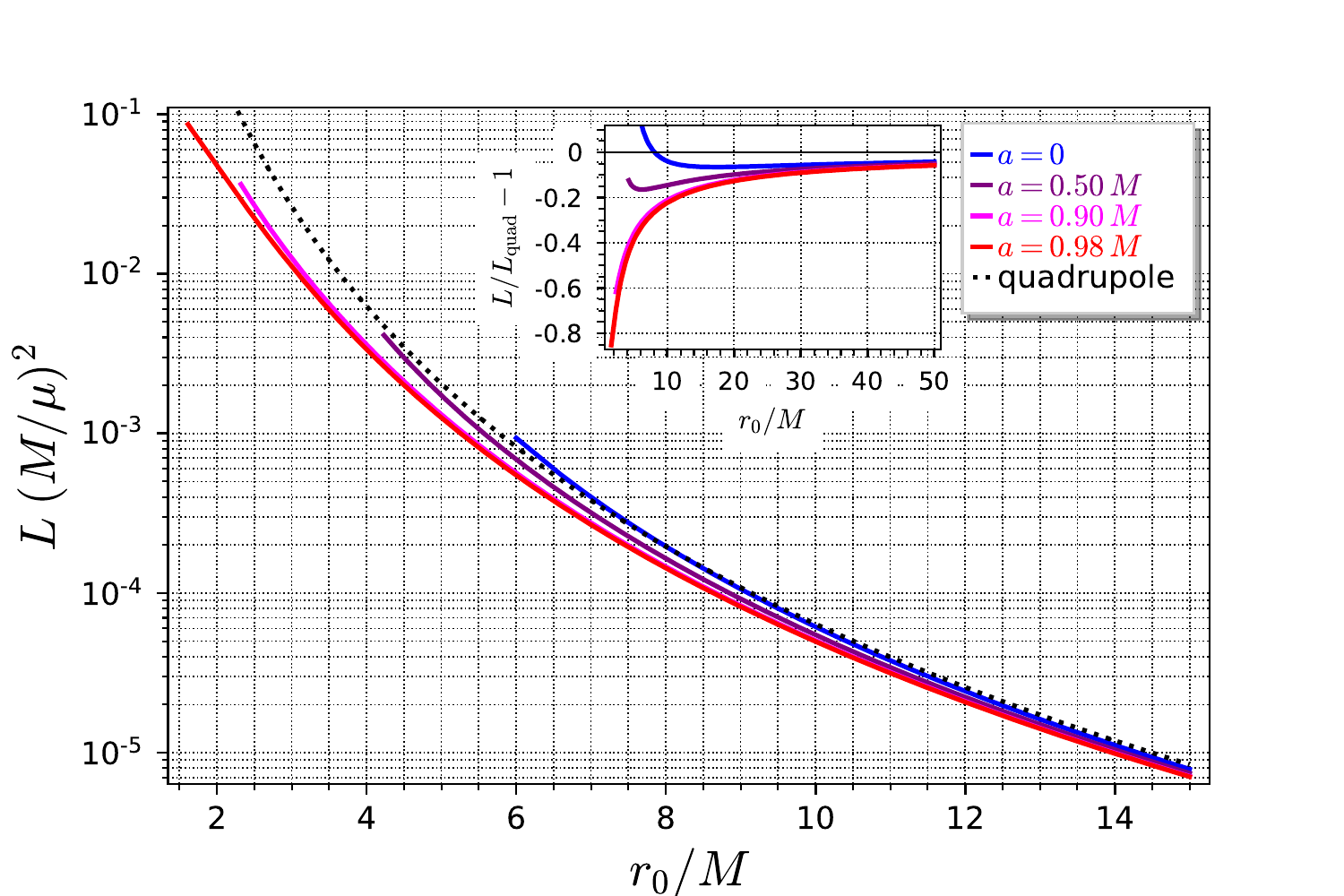}}
\caption{ \label{f:gw_power}
Gravitational wave luminosity $L$ for an object of mass $\mu$
in circular equatorial orbit around a
Kerr BH of mass $M$ and spin parameter $a$,
as a function of the orbital radius $r_0$.
Each curve starts at the prograde ISCO radius of the corresponding value of $a$.
The dotted curve corresponds to the quadrupole approximation as
given by Eq.~(\ref{e:L_quadrupole}). The inset shows the relative difference
with respect to the quadrupole formula (\ref{e:L_quadrupole})
up to $r_0 = 50 M$.}
\end{figure}

The total emitted power $L$ (actually the function $\tilde{L}(r_0/M)$)
is depicted in Fig.~\ref{f:gw_power}.
A test of our computations
is provided by the comparison with Figs.~6 and 7 of \citet{Detweiler78}'s study.
At the naked eye, the agreement is quite good, in
particular for the values of $L$ at the ISCO's.
Moreover, for large values of $r_0$ all curves converge towards the curve
of the quadrupole formula (\ref{e:L_quadrupole}) (dotted curve), as they
should.
However, as the inset of Fig.~\ref{f:gw_power} reveals, the relative deviation from the
quadrupole formula is still $\sim 5\%$ for orbital radii as large as
$r_0 \sim 50 M$. This is not negligibly small and justifies the fully relativistic
approach that we have adopted.

\subsection{Secular evolution of the orbit}

For a particle moving along any geodesic in Kerr spacetime, in particular
along a circular orbit, the \emph{conserved energy} is $E \equiv - p_a \xi^a$, where $p_a$ is the
particle's 4-momentum 1-form and $\xi^a$ the Killing vector associated
with the pseudo-stationarity of Kerr spacetime ($\xi = \partial/\partial t$
in Boyer-Lindquist coordinates). Far from the MBH, $E$
coincides with the particle's energy as an inertial observer at rest with
respect to the MBH would measure. For a circular orbit of radius
$r_0$ in the equatorial plane of a Kerr BH
of mass $M$ and spin parameter $a$, the
expression of $E$ is \citep{BardeenPT72}
\beq \label{e:cons_ener}
    E = \mu
    \frac{1 - 2M/r_0 + a M^{1/2}/r_0^{3/2}}{\left(1-3M/r_0 + 2a M^{1/2}/r_0^{3/2}\right) ^{1/2}} ,
\eeq
where $\mu \equiv (-p_a p^a)^{1/2}$ is the particle's rest mass.

Due to the reaction to gravitational radiation, the particle's worldline
is actually not a true timelike geodesic of Kerr spacetime, but is slowly
inspiralling towards the central MBH. In particular, $E$ is not truly
constant. Its secular evolution is governed by the balance
law \citep{FinnT00,BarackP19,Isoyama_al19}
\beq \label{e:balance_law}
   \dot{E} = - L - L_{\rm H},
\eeq
where $\dot{E}\equiv \mathrm{d}E/\mathrm{d}t$,
$L$ is the gravitational wave luminosity evaluated in Sect.~\ref{s:radiated_power}
and $L_{\rm H}$ is the power radiated down to the event horizon of the
MBH. It turns out that in practice, $L_{\rm H}$ is quite small compared to $L$.
From Table~VII of \citet{FinnT00}, we notice that for $a=0$, one has
always $|L_{\rm H}/\dot{E}| < 4\times 10^{-3}$ and for $a=0.99M$, one has
$|L_{\rm H}/\dot{E}| < 9.5\times 10^{-2}$, with
$|L_{\rm H}/\dot{E}| < 10^{-2}$ as soon as $r_0 > 7.3 M$. In the following,
we will neglect the term $L_{\rm H}$ in our numerical evaluations
of $\dot{E}$.

\begin{figure}
\centerline{\includegraphics[width=0.7\textwidth]{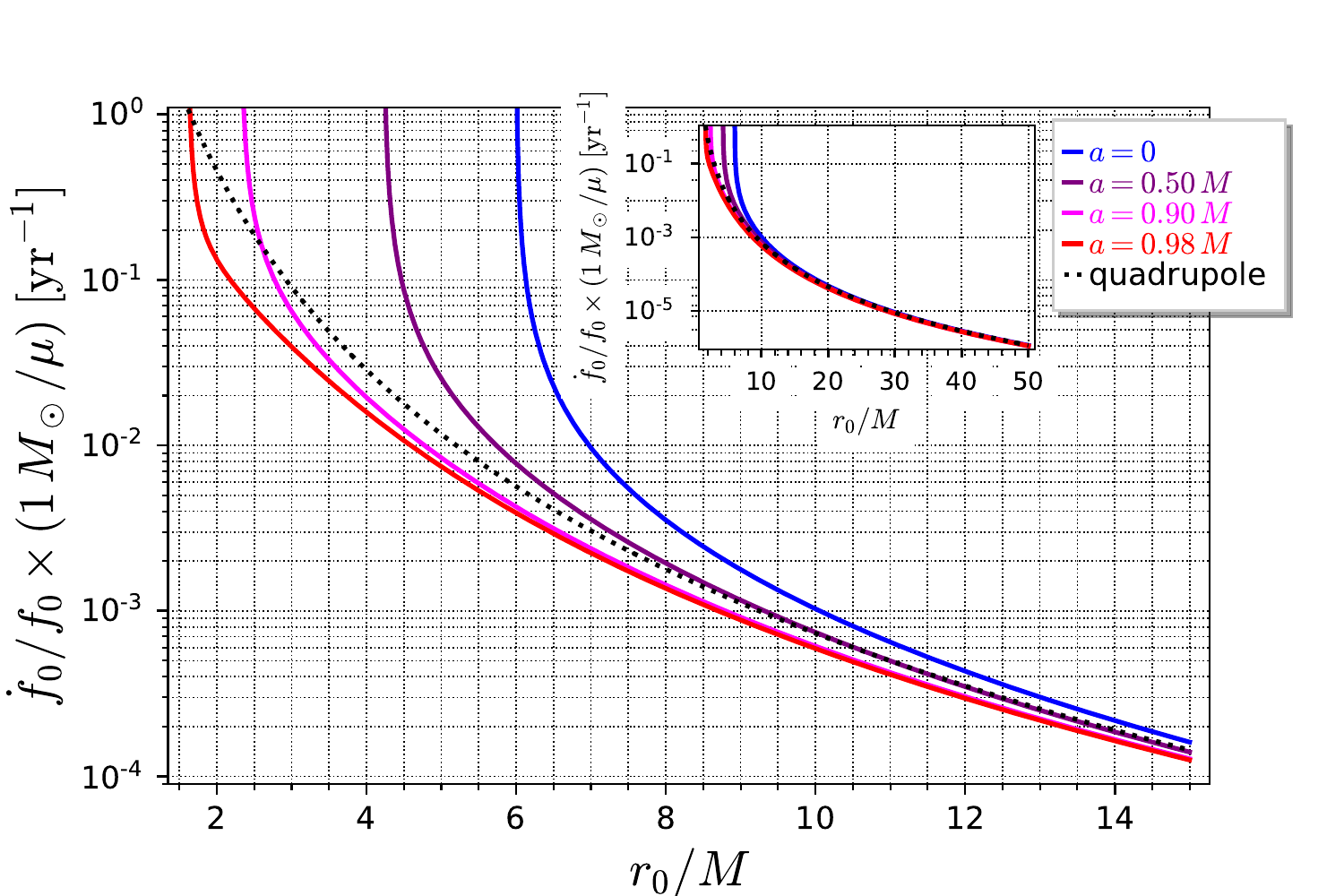}}
\caption{ \label{f:df0_over_f0}
Relative change in orbital frequency $\dot{f}_0/f_0$
induced by the reaction to gravitational radiation for an object of mass
$\mu$
in circular equatorial orbit around
Kerr BH of mass $M$ equal to that of Sgr~A*
as a function of the orbital radius $r_0$ [Eq.~(\ref{e:dot_f0})].
Each curve has a vertical asymptote at the ISCO radius for the
corresponding value of $a$.
The dotted curve corresponds to the quadrupole approximation and
quasi-Newtonian orbits. The inset shows the curves extended
up to $r_0 = 50 M$.}
\end{figure}

From Eq.~(\ref{e:cons_ener}), we have
\beq \label{e:dE_dr0}
    \dot{E} =
        \frac{\mu M}{2r_0^2} \;
        \frac{1 - 6M/r_0 + 8a M^{1/2}/r_0^{3/2}
    - 3 a^2/r_0^2}{\left(1-3M/r_0 + 2aM^{1/2}/r_0^{3/2}\right) ^{3/2}}
    \; \dot{r}_0.
\eeq
In view of Eq.~(\ref{e:omega0}), the
secular evolution of the orbital frequency $f_0=\omega_0/(2\pi)$
is related to
$\dot{r}_0$ by
\beq \label{e:df0_dr0}
   \frac{\dot{f}_0}{f_0} \ =
   - \frac{3}{2} \; \frac{1}{1 + a M^{1/2}/r_0^{3/2}}\;  \frac{\dot{r}_0}{r_0} \, .
\eeq
By combining successively Eqs.~(\ref{e:df0_dr0}), (\ref{e:dE_dr0}), (\ref{e:balance_law})
and (\ref{e:luminosity}), we get
\beq \label{e:dot_f0}
    \frac{\dot{f}_0}{f_0} = 3 \frac{\mu}{M^2} \left[
        \, \tilde{L}\left(\frac{r_0}{M}\right) + \tilde{L}_{\rm H}\left(\frac{r_0}{M}\right) \right]
        \frac{r_0/M}{1 + aM^{1/2}/r_0^{3/2}} \;
        \frac{\left(1-3M/r_0 + 2aM^{1/2}/r_0^{3/2}\right) ^{3/2}}{1 - 6M/r_0
        + 8a M^{1/2}/r_0^{3/2} - 3 a^2/r_0^2} ,
\eeq
where we have introduced the rescaled horizon flux function $\tilde{L}_{\rm H}$,
such that
\beq \label{e:def_dot_tEH}
  L_{\rm H} = \left(\frac{\mu}{M} \right)^2 \tilde{L}_{\rm H}\left(\frac{r_0}{M}\right) .
\eeq

This relative change in orbital frequency is depicted in
Fig.~\ref{f:df0_over_f0}, with a $y$-axis scaled to the mass (\ref{e:SgrA_mass})
of Sgr~A* for $M$ and to $\mu=1\, M_\odot$.
One can note that $\dot{f}_0$ diverges at the
ISCO. This is due to the fact that $E$ is minimal at the ISCO, so that
$\mathrm{d}E/\mathrm{d} r_0 = 0$ there. At this point,
a loss of energy cannot be compensated by a slight decrease of the orbit.

\begin{figure}
\centerline{\includegraphics[width=0.7\textwidth]{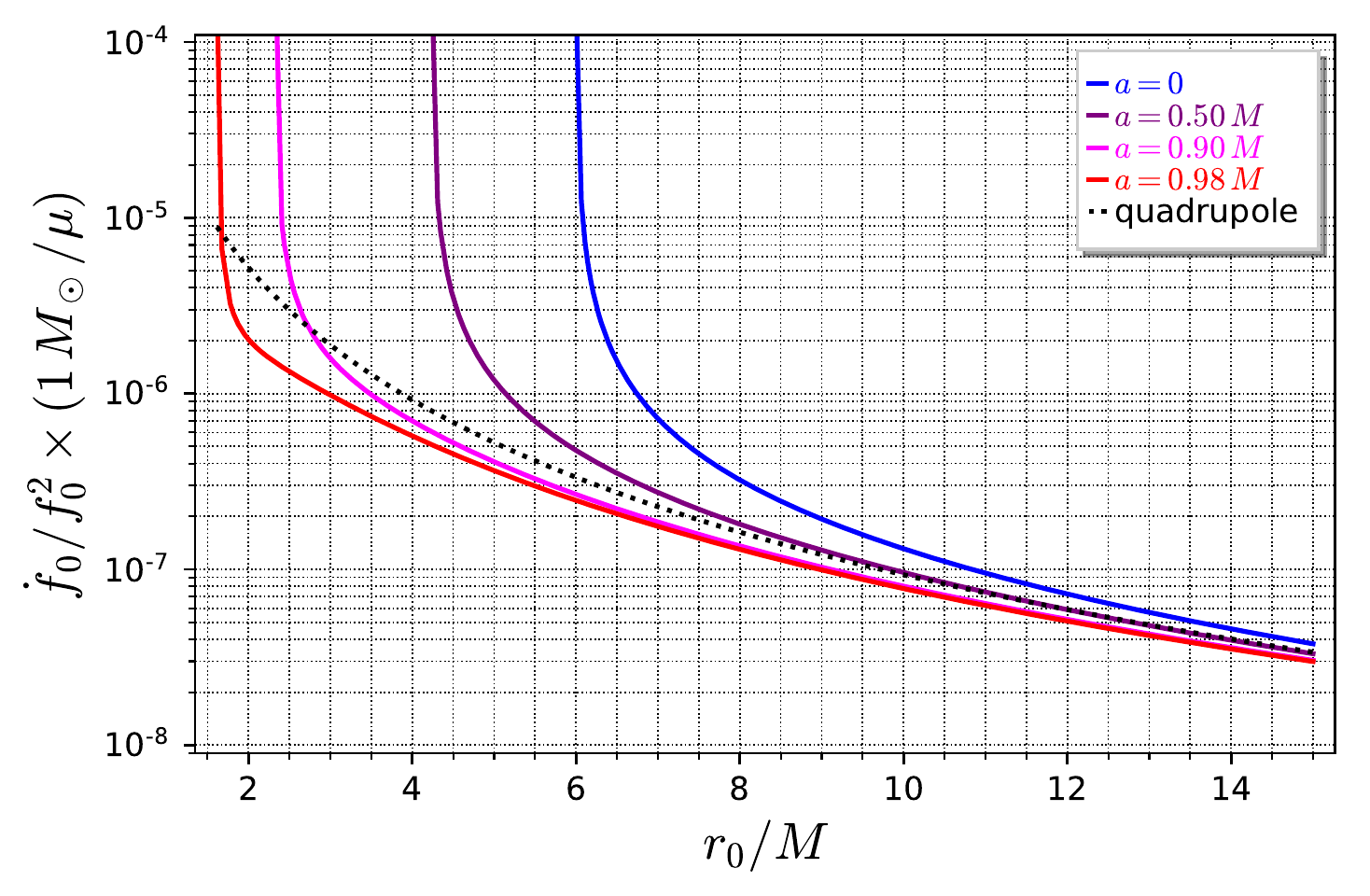}}
\caption{ \label{f:adiab_param}
Adiabaticity parameter $\dot{f}_0/f_0^2$
as a function of the orbital radius $r_0$.
The dotted curve corresponds to the quadrupole approximation and
quasi-Newtonian orbits.}
\end{figure}

Another representation of the orbital frequency evolution, via the
adiabaticity parameter $\varepsilon\equiv \dot{f}_0/f_0^2$,
is shown in Fig.~\ref{f:adiab_param}. The adiabaticity parameter $\varepsilon$ is a dimensionless
quantity, the smallness of which guarantees the validity of
approximating the inspiral trajectory by a succession of circular orbits
of slowly shrinking radii. As we can see on Fig.~\ref{f:adiab_param},
$\varepsilon < 10^{-4}$ except very near the ISCO, where
$\dot{f}_0$ diverges.

\subsection{Inspiral time}

By combining Eqs.~(\ref{e:dE_dr0}), (\ref{e:balance_law}), (\ref{e:def_dot_tEH})
and (\ref{e:luminosity}), we get an expression for
$\dot{r}_0^{-1} = {\mathrm{d}t}/{\mathrm{d}r_0}$ as a function of $r_0$.
Once integrated, this leads to the time required for the orbit to shrink from
an initial radius $r_0$ to a given radius $r_1 < r_0$:
\beq \label{e:inspiral_time}
    T_{\rm ins}(r_0,r_1) = \frac{M^2}{2\mu}
    \int_{r_1/M}^{r_0/M}
    \frac{1 - 6/x + 8 \bar{a}/x^{3/2}
    - 3 \bar{a}^2/x^2}{\left(1-3/x + 2\bar{a}/x^{3/2}\right) ^{3/2}}
    \frac{\mathrm{d}x}{x^2 (\tilde{L}(x) + \tilde{L}_{\rm H}(x))} \, ,
\eeq
where $\bar{a} \equiv a / M = J/M^2$ is the dimensionless Kerr parameter.
We shall call $T_{\rm ins}(r_0,r_1)$ the \emph{inspiral time} from $r_0$ to
$r_1$. For an object whose evolution is only driven by the reaction to
gravitation radiation (e.g.,~a compact object, cf.~Sect.~\ref{sec:compact}),
we define then the \emph{life time} from the orbit $r_0$ as
\beq \label{e:life_time}
    T_{\rm life}(r_0) \equiv T_{\rm ins}(r_0,r_{\rm ISCO}) .
\eeq
Indeed, once the ISCO is reached, the plunge into the MBH is very fast,
so that $T_{\rm life}(r_0)$ is very close to the actual life time outside
the MBH, starting from the orbit of radius $r_0$.

\begin{figure}
\centerline{\includegraphics[width=0.7\textwidth]{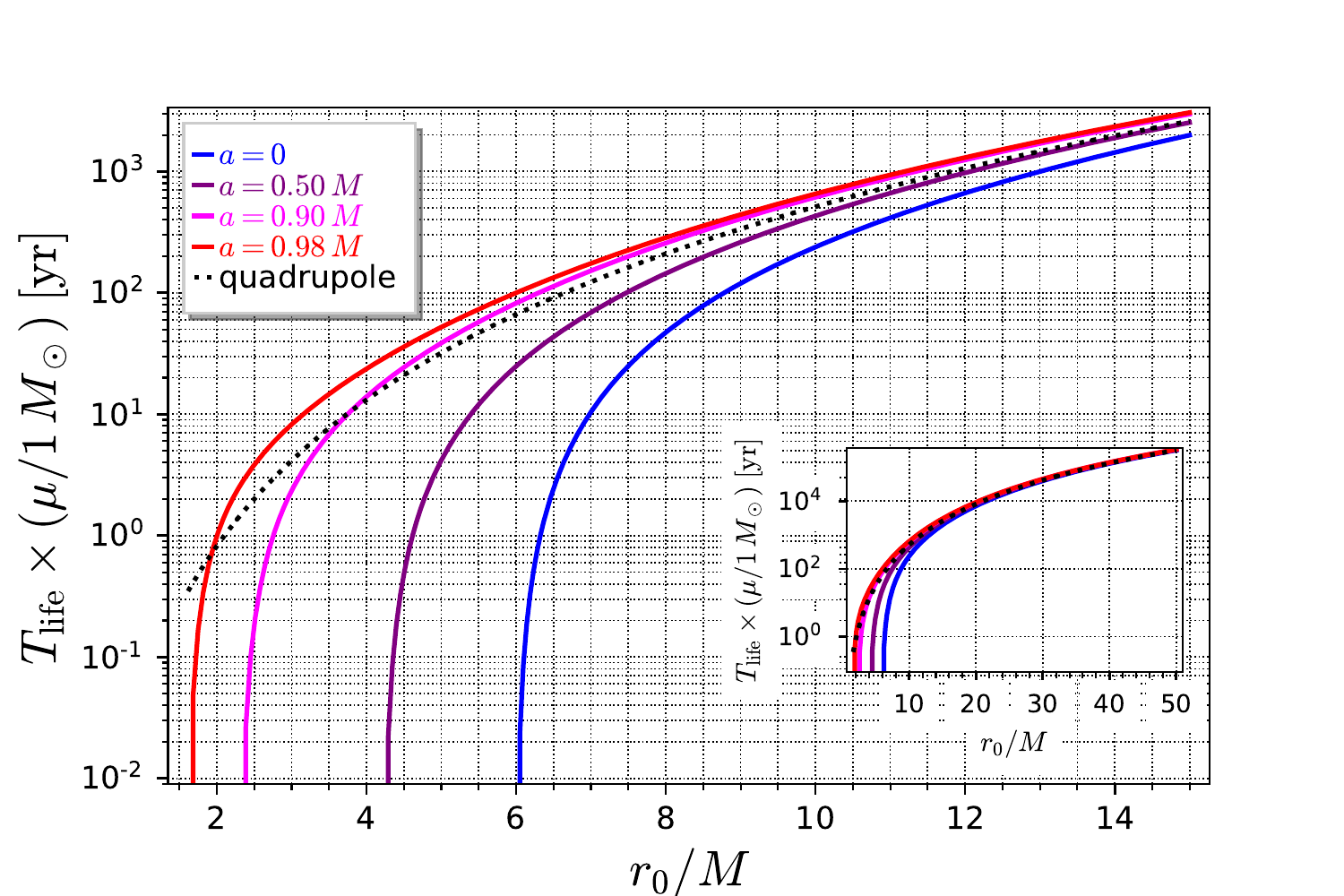}}
\caption{ \label{f:life_time_scale}
Life time of a (compact) object of mass $\mu$ in circular equatorial orbit around
a Kerr BH with a mass $M$ equal to that of Sgr~A* as a function of the
orbital radius $r_0$ [Eq.~(\ref{e:life_time})]. The inset shows the curves extended
up to $r_0 = 50 M$.}
\end{figure}

\begin{figure}
\centerline{\includegraphics[width=0.7\textwidth]{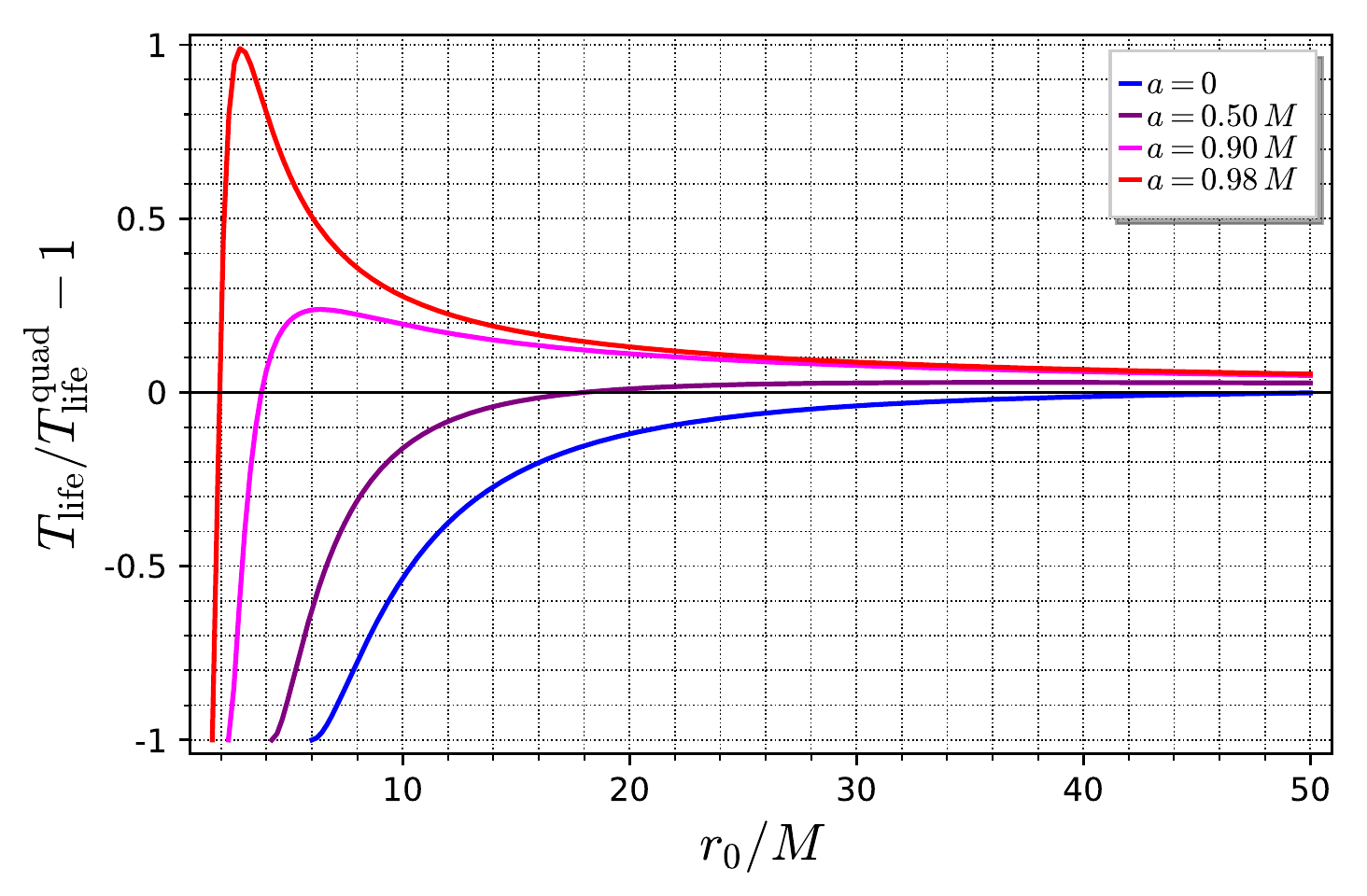}}
\caption{ \label{f:life_time_diff}
Relative difference between the life time given by Eq.~(\ref{e:life_time}) and
the value given the quadrupole formula, Eq.~(\ref{e:life_time_quadrupole}),
as a function of the orbital radius $r_0$.}
\end{figure}

The life time is depicted in Fig.~\ref{f:life_time_scale},
which is drawn for $M=M_{\rm Sgr\, A^*}$.
It appears from Fig.~\ref{f:life_time_scale} that the life time near the ISCO is
pretty short; for instance, for $a=0$ and a solar-mass object, it is only 34 days at $r_0 = 6.1 M$. Far from the ISCO, it is much larger and reaches
$\sim 3\times 10^5 \; {\rm yr}$ at $r_0 = 50 M$ (still for $\mu= 1 \, M_\odot$).
The dotted curve in Fig.~\ref{f:life_time_scale} corresponds to the value obtained for Newtonian orbits and the quadrupole formula (\ref{e:L_quadrupole}):
$T_{\rm life} = 5/256\, (M^2/\mu) (r_0/M)^4$ \citep{Peters64}, a value which can be recovered
by taking the limit $x \to +\infty$ in Eqs.~(\ref{e:inspiral_time})-(\ref{e:life_time}) and using expression~(\ref{e:L_quadrupole}) for $\tilde{L}(x)$, as well as
$\tilde{L}_{\rm H}(x)=0$. For $M=M_{\rm Sgr\, A^*}$, the quadrupole formula
becomes
\beq \label{e:life_time_quadrupole}
    T_{\rm life}^{\rm\, quad} \simeq 4.2\times 10^4 \, \left( \frac{1 \, M_\odot}{\mu} \right)
        \left( \frac{r_0}{30 M} \right) ^4 \; {\rm yr} .
\eeq
The relative difference between the exact formula (\ref{e:life_time})
and the quadrupole approximation (\ref{e:life_time_quadrupole}) is plotted
in Fig.~\ref{f:life_time_diff}. Not surprisingly, the difference is very large in the
strong field region, reaching $100\%$ close to the ISCO. For $r_0 = 20 M$,
it is still $\sim 10\%$. Even for $r_0 = 50 M$, it is as large as
$3$ to $5\%$
for $a\geqslant 0.5 M$ and $0.1\%$ for $a=0$.

\section{Potential sources}
\label{s:sources}

Having established the signal properties and detectability by LISA, let us
now discuss astrophysical candidates for the orbiting object.
A preliminary required for the discussion is the evaluation of the tidal effects
exerted by  Sgr~A* on the orbiting body, since this can make the innermost orbit
to be significantly larger than the ISCO. We thus start by investigating
the tidal limits in Sect.~\ref{sec:tidal}. Then, in
Sect.~\ref{sec:formation}, we review the scenarios which might lead to
the presence of stellar
objects in circular orbits close to Sgr~A*.
The various categories of sources are then discussed in the remaining
subsections: compact objects (Sect.~\ref{sec:compact}), main-sequence
stars (\ref{sec:stars}), brown dwarfs (Sect.~\ref{s:brown_dwarfs}),
accretion flow (\ref{sec:accretion}), dark matter (Sect.~\ref{sec:DM})
and artificial sources (Sect.~\ref{sec:alien}). As it will appear in
the discussion, not all these sources are on the same footing regarding the
probability of detection by LISA.

\subsection{Tidal radius and Roche radius}
\label{sec:tidal}

In Sects.~\ref{s:gw_particle}-\ref{s:orbital_decay}, we have considered
an idealized point mass. When the orbiting object has some extension, a
natural question is whether the object integrity can be maintained in
presence of the tidal forces exerted by the central MBH. This leads to the
concept of \emph{tidal radius} $r_{\rm T}$, defined as the
minimal orbital radius for which the tidal forces cannot disrupt the orbiting body. In other words, the considered object
cannot move on an orbit with $r_0 < r_{\rm T}$.
The tidal radius is given by the formula
\beq \label{e:def_r_tidal}
    r_{\rm T} = \alpha \left( \frac{M}{\rho} \right) ^{1/3} ,
\eeq
where $M$ is the mass of the MBH, $\rho$ the mean density of the
orbiting object and $\alpha$ is a coefficient of order 1, the value of which
depends on the object internal structure and rotational state.
From the naive argument of equating the self-gravity and the tidal
force at the surface of a spherical Newtonian body, one gets
$\alpha = (3/(2\pi))^{1/3} = 0.78$. If one further assumes that the object
is corotating, i.e. is in synchronous rotation with respect to the
orbital motion, then one gets
$\alpha = (9/(4\pi))^{1/3} = 0.89$. \citet{Hills75} uses $\alpha=(6/\pi)^{1/3}=1.24$,
while \citet{Rees88} uses $\alpha=(3/(4\pi))^{1/3}=0.62$.
For a Newtonian incompressible fluid ellipsoid in
synchronous rotation, $\alpha = 1.51$ \citep{Chandrasekhar69}.
This result has been generalized
by \citet{Fishbone73} to incompressible fluid ellipsoids in the Kerr metric:
$\alpha$ increases then from $1.51$ for $r\gg M$ to $1.60$ (resp. $1.56$)
for $r=10 M$ and $a=0$ (resp. $a=0.99M$)
(cf.~Fig.~5 of \citet{Fishbone73}, which displays $1/(\pi\alpha^3)$). Taking into account the compressibility
decreases $\alpha$: $\alpha=1.34$ for a polytrope of index $n=1.5$ \citep{IshiiSM05}.

For a stellar type object on a circular orbit, a more relevant quantity
is the \emph{Roche radius}, which marks the onset of tidal stripping near
the surface of the star, leading to some steady accretion to the MBH (Roche lobe
overflow) without the total disruption of the star \citep{DaiBE13,DaiB13}.
For centrally condensed bodies, like main-sequence stars, the Roche radius is given by the condition that the stellar material fills
the Roche lobe.
In the Kerr metric, the volume $V_{\rm R}$
of the Roche lobe generated by a mass $\mu$
on a circular orbit of radius $r_0$ has been evaluated by
\citet{DaiB13}, yielding to the approximate formula\footnote{See Eqs.~(10), (26) and (27)
of \citet{DaiB13}.} $V_{\rm R} \simeq \mu M^2 \mathcal{V}_{\rm R}$, with
\beq \label{e:Roche_volume}
    \mathcal{V}_{\rm R} \equiv \left(\frac{r_0}{M}\right) ^3
    \left[ \frac{0.683}{1 + \frac{\chi}{2.78}}
    + \frac{\frac{0.456}{1 + \frac{\chi}{4.09}}
             -\frac{0.683}{1 + \frac{\chi}{2.78}}}{
    \sqrt{\frac{r_0}{r_{\rm ISCO}}}
    + F(a,\chi) \left(\frac{r_0}{r_{\rm ISCO}(a)} - 1 \right)}
    \right] ,
\eeq
where $r_{\rm ISCO}$ is the radius of the prograde ISCO,
$\chi \equiv \Omega / \omega_0$ is the ratio between the angular velocity
$\Omega$ of the star (assumed to be a rigid rotator) with respect to some
inertial frame to the orbital angular velocity $\omega_0$ and $F(a,\chi)$
is the function defined by
\beq
    F(a,\chi) \equiv - 23.3 + \frac{13.9}{2.8 + \chi}
        + \left( 23.8 - \frac{14.8}{2.8 + \chi} \right) (1-a)^{0.02}
        + \left( 0.9 - \frac{0.4}{2.6 + \chi} \right) (1 - a)^{-0.16} .
\eeq
It should be noted that $\chi=1$ for a corotating star.
The Roche limit is reached when the actual volume of the star equals the volume
of the Roche lobe. If $\rho$ stands for the mean mass density of the star, this
corresponds to the condition $\mu = \rho V_{\rm R}$, or equivalently
\beq \label{e:Roche_radius_eq}
    \rho M^2 \mathcal{V}_{\rm R} - 1 = 0.
\eeq
Solving this equation for $r_0$ leads to the orbital radius $r_{\rm R}$ at
the Roche limit, i.e. the Roche radius.
The mass $\mu$ has disappeared from
Eq.~(\ref{e:Roche_radius_eq}), so that  $r_{\rm R}$
depends only on the mean density $\rho$
and the rotational parameter $\chi$.
For $r_0 \gg r_{\rm ISCO}$, we can neglect the second term in the square
brackets in Eq.~(\ref{e:Roche_volume}) and obtain an explicit expression:
\beq \label{e:Roche_radius_far}
    r_{\rm R} \simeq 1.14 \left( 1 + \frac{\chi}{2.78}
        \right)^{1/3}  \left( \frac{M}{\rho} \right) ^{1/3}
    \quad \mbox{for}\quad r_{\rm R} \gg M .
\eeq
This equation has
the same shape as the tidal radius formula (\ref{e:def_r_tidal}).
Using Sgr~A* value (\ref{e:SgrA_mass}) for $M$, we may rewrite the
above formula as
\beq \label{e:Roche_radius_far_sol}
    \frac{r_{\rm R}}{M} \simeq 33.8 \left( 1 + \frac{\chi}{2.78}
        \right)^{1/3}  \left( \frac{\rho_\odot}{\rho} \right) ^{1/3}
        \quad \mbox{for}\quad r_{\rm R} \gg M ,
\eeq
where $\rho_\odot \equiv 1.41\times 10^3 \; {\rm kg \cdot m}^{-3}$ is the mean
density of the Sun.

\begin{figure}
\centerline{\includegraphics[width=0.7\textwidth]{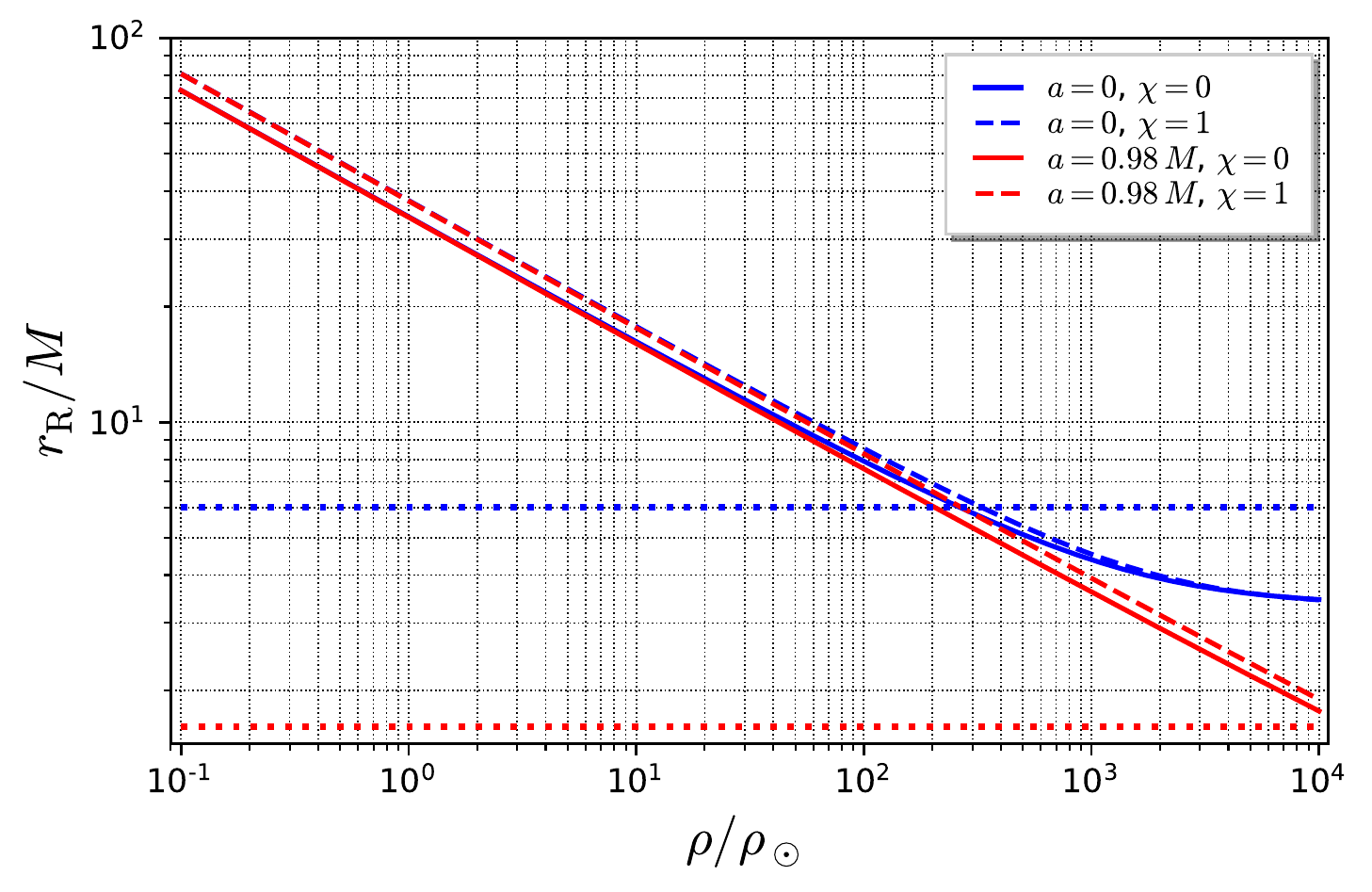}}
\caption{ \label{f:roche_limit}
Roche radius $r_{\rm R}$ as a function of the mean
density $\rho$ of the star (in solar units), for two values of the
MBH spin $a$ and two rotational states of the star: irrotational ($\chi=0$)
and corotating ($\chi=1$). The blue (resp. red) dotted horizontal line marks
the ISCO radius for $a=0$ (resp. $a=0.98\, M$).}
\end{figure}

\begin{table*}
\caption{
Roche radius $r_{\rm R}$ for different types of objects orbiting Sgr~A*.
The first three lines give the mass $\mu$,
the mean radius $R$ and the mean mass density $\rho$, all in solar units.
$\chi=0$ stands for an irrotational body and $\chi=1$ for a corotating one.
See the text for the chosen characteristics of the red dwarf and the brown dwarf.}
\label{t:Roche_radius}
\centering
\begin{tabular}{lcccccc}
\hline\hline
  & Jupiter & Sun & Earth & red dwarf & brown dwarf & white dwarf \\
\hline
$\mu / M_\odot$   & $9.55\times 10^{-4}$ & $1$ & $3.0\times 10^{-6}$ & $0.20$ & $0.062$ & $0.80$ \\
$R / R_\odot$   & $0.10$ & $1$ & $9.17\times 10^{-3}$ & $0.22$ & $0.078$ & $5.58\times 10^{-3}$ \\
$\rho/\rho_\odot$ & $0.94$  & $1$ & $3.91$& $18.8$    & $131.$      &  $1.10\times 10^6$ \\
$r_{\rm R}/M$ ($a=0$, $\chi=0$) & $34.9$ & $34.2$ & $21.9$ & $13.3$ & $7.31$ & $0.28$ \\
$r_{\rm R}/M$ ($a=0$, $\chi=1$) & $38.5$ & $37.7$ & $24.1$ & $14.5$ & $7.86$ & $0.32$ \\
$r_{\rm R}/M$ ($a=0.98M$, $\chi=0$) & $34.8$ & $34.1$ & $21.8$ & $13.0$ & $6.93$ & $0.52$ \\
$r_{\rm R}/M$ ($a=0.98M$, $\chi=1$) & $38.4$ & $37.6$ & $24.0$ & $14.3$ & $7.57$ & $0.52$ \\
\hline
\end{tabular}
\end{table*}

The numerical resolution of Eq.~(\ref{e:Roche_radius_eq}) for $r_{\rm R}$
has been implemented
in the \texttt{kerrgeodesic\_gw} package (cf.~Appendix~\ref{s:kerrgeodesic_gw})
and the results are shown in Fig.~\ref{f:roche_limit} and Table~\ref{t:Roche_radius}.
The straight line behavior in the left part of Fig.~\ref{f:roche_limit}
corresponds to the power law $r_{\rm R} \propto \rho^{-1/3}$
in the asymptotic formula (\ref{e:Roche_radius_far_sol}).
In Table~\ref{t:Roche_radius}, the characteristics of the red dwarf star
are taken from Fig. 1 of \citet{ChabrierGB07} --- it corresponds to a
main-sequence star of spectral type M4V. The brown dwarf model of Table~\ref{t:Roche_radius} is
the model of minimal
radius along the 5~Gyr isochrone in Fig.~1 of \citet{Chabrier_al09}.
This brown dwarf is close to the hydrogen burning limit and to the maximum
mean mass density $\rho$ among brown dwarfs and main-sequence stars.
We note from Table~\ref{t:Roche_radius} that it has
a Roche radius very close to the Schwarzschild ISCO.
We note as well that $r_{\rm R} < M$ for a white dwarf.
This means that such
a star is never tidally disrupted above Sgr ~A*'s event horizon.
A fortiori, neutron stars share the same property.

\subsection{Presence of stellar objects in the vicinity of Sgr~A*}
\label{sec:formation}

The Galactic Center is undoubtably a very crowded region. For instance, it
is estimated that there are $\sim 2\times 10^4$ stellar BHs
in the central parsec, a tenth of which are located
within $0.1\; {\rm pc}$ of Sgr~A* \citep{FreitagAK06}.
The recent detection of a dozen of
X-ray binaries in the central parsec \citep{Hailey_al18} supports these
theoretical predictions.
The two-body relaxation
in the central cluster causes some mass segregation: massive
stars lose energy to lighter ones and drift to the center
\citep{hopman05,FreitagAK06}.
Accordingly BHs are expected to dominate the \emph{mass} density
within $0.2\; {\rm pc}$. However, they do not dominate the \emph{number} density,
main-sequence stars being more numerous than BHs \citep{FreitagAK06,Amaro18}.
The number of stars or stellar BHs very close to Sgr~A* (i.e. located at $r < 100 M$)
is expected to be quite small though. Indeed the central parsec region is
very extended in terms of Sgr~A*'s length scale:
$1\; {\rm pc} = 5.1\times 10^6 \, M$, where $M$ is Sgr~A*'s mass.
At the moment, the closest known stellar object orbiting Sgr~A* is the star S2,
the periastron of which is located at $r_{\rm p} = 120 {\rm\; au} \simeq 3 \times 10^3 M$
\citep{gravity18a}.

The most discussed process for populating the vicinity of the central MBH is
the extreme mass ratio inspiral (EMRI) of a (compact) star or
stellar BH \citep{Amaro_al07,Amaro18}. In the standard scenario (see e.g., \citet{Amaro18} for a review), the inspiralling object originates from the
two-body scattering by other stars in the Galactic Center cluster.
It keeps a very high eccentricity until the final plunge in the MBH,
despite the circularization effect of gravitational radiation
\citep{hopman05}. Such an EMRI is thus not an eligible source for the
process considered in the present article, which is limited to circular orbits.

Another kind of EMRI results from the tidal separation of a binary by the MBH \citep{MillerFHL05}.
In such a process, a member of the binary is ejected at high speed while the
other one is captured by the MBH and inspirals towards it, on an initially
low eccentricity orbit. Gravitational radiation is then efficient
in circularizing the orbit, making it almost circular when it enters LISA band.
Such an EMRI is thus fully relevant to
the study presented here. The rate of formation of these zero-eccentricity EMRIs is very low,
being comparable to those of high-eccentricities EMRIs \citep{MillerFHL05}, which is
probably below $10^{-6}\; {\rm yr}^{-1}$ \citep{Amaro18,hopman05}. However,
as discussed in Sect.~\ref{sec:compact}, due to their long life time ($> 10^5\;{\rm yr}$)
in the LISA band,
the probability of detection of these EMRIs is not negligibly small.

Another process discussed in the literature and leading to objects on
almost circular  orbits is the formation of stars in an accretion disk
surrounding the MBH
\citep[see e.g.,][and references therein]{collin99,nayakshin07,collin08}.
Actually, it was particularly surprising to find in the inner parsec of the Galaxy
a population of massive (few $10\,M_\odot$) young
stars, that were formed $\approx 6$~Myr ago~\citep{genzel10}.
Indeed, forming stars in the extreme environment of a MBH is
not obvious because of the strong tidal forces that would break typical
molecular clouds.
A few scenarios were proposed to account for this young stellar population;
see \citet{mapelli16} for a recent dedicated review.
Among these, \textit{in situ} formation might take place
in a geometrically thin Keplerian (circularly orbiting)
accretion disk surrounding the
MBH~\citep{collin99,nayakshin07,collin08}.
Such an accretion disk is not presently detected,
and would have existed in past periods of AGN activity
at the Galactic Center~\citep{ponti13,Ponti_al14}.

Stellar formation in a disk is supported by the fact that
the massive young stellar population proper motion
was found to be consistent
with rotational motion in a disk~\citep{paumard06}.
It is interesting to note that the on-sky orientation of this
stellar disk is similar to the orientation of the
orbital plane of a recently detected flare of Sgr~A*~\citep{gravity18b}.
However, such a scenario suffers from the fact that the young stars
observed have a median eccentricity of $0.36 \pm 0.06$~\citep{bartko09},
while formation in a Keplerian disk leads
to circular orbits.
On the other side, the recently detected X-ray binaries \citep{Hailey_al18}
mentioned above are most probably quiescent
BH binaries. These BHs are likely to have formed
\textit{in situ} in a disk~\citep{GenerozovSMO18}, giving more support
to the scenario discussed here.

A population of stellar-mass BHs
will form after the death of the most massive stars born in the accretion disk.
These would be good candidates for the scenario discussed here,
provided the initially circular orbit is maintained after supernova explosion.
The recent study of \citet{bortolas17} shows that BHs
formed from the supernova explosion of one of the members
of a massive binary keep their initial orbit without noticeable
kink from the supernova explosion. Given that a large fraction
(tens of percent)
of the Galactic Center
massive young stars are likely to be binaries~\citep{sana11},
this shows that circular-orbiting BHs are likely
to exist within the framework of the Keplerian \textit{in-situ}
star formation model.
This scenario was already advocated by \citet{levin07},
which considers the fragmentation of a self-gravitating thin
accretion disk that forms massive stars, leading to the formation
of BHs that inspiral in towards Sgr~A*, following quasi-circular orbits,
in a typical time of $\approx 10$~Myr.


\subsection{Compact objects}
\label{sec:compact}

As discussed in Sect.~\ref{sec:tidal}, compact objects --- BHs, neutron
stars and white dwarfs --- do not suffer any tidal disruption above the
event horizon of Sgr~A*. Their evolution around Sgr~A* is thus entirely
given by the reaction to gravitational radiation with the timescale shown
by Fig.~\ref{f:life_time_scale}.

\begin{figure}
\centerline{\includegraphics[width=0.7\textwidth]{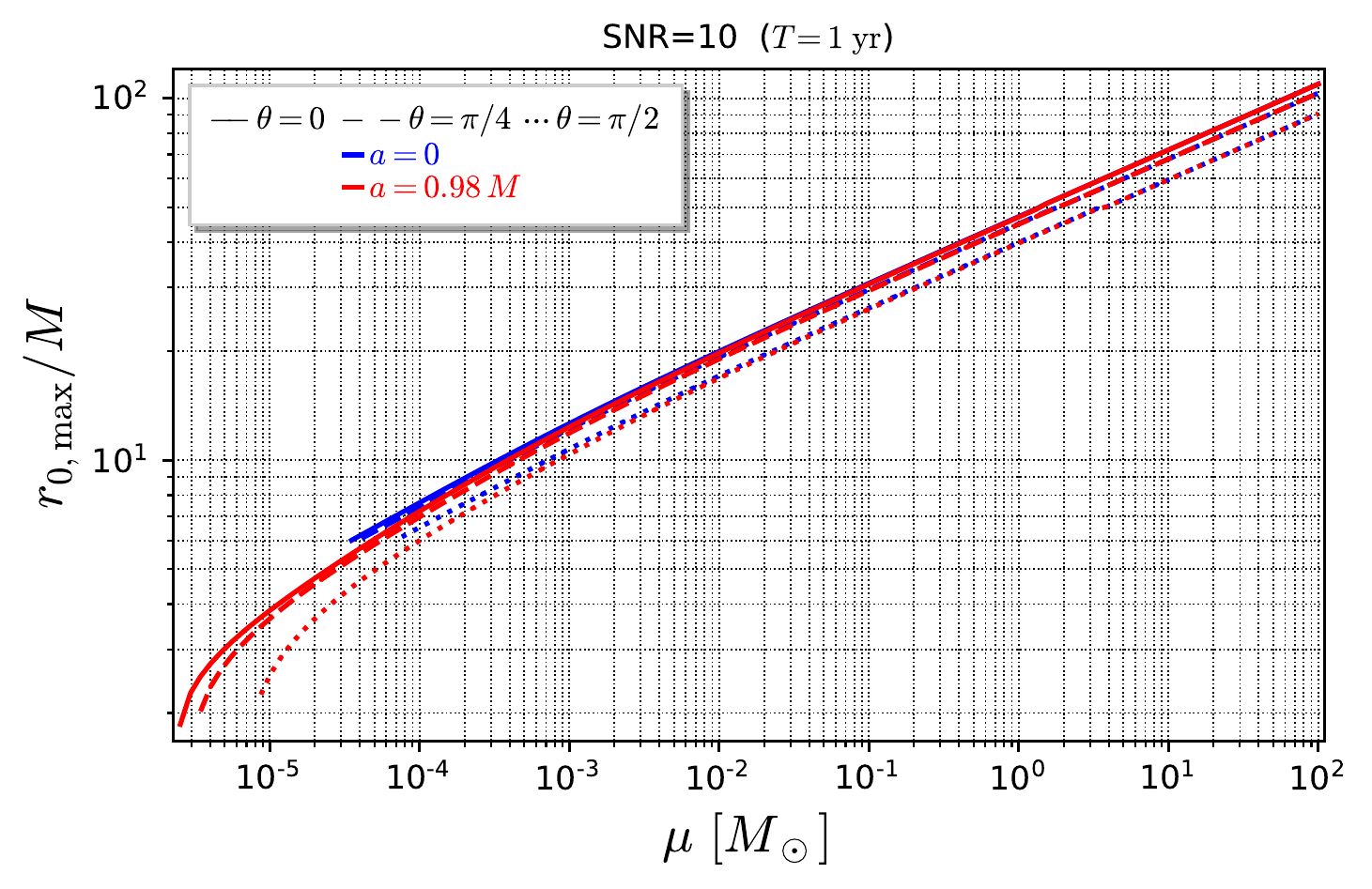}}
\caption{ \label{f:max_detect_radius}
Maximum orbital radius $r_{0,\rm max}$ for a $\text{S/N}=10$ detection by LISA in
one year of data, as a function of the mass $\mu$ of the object
orbiting around Sgr~A*.
}
\end{figure}

Let us define the \emph{entry in LISA band}
as the moment in the slow
inspiral when $\SNRyr$ reaches 10, which is the
threshold we adopt for a positive detection [Eq.~(\ref{e:SNR_threshold})].
The orbital radius at the
entry in LISA band in plotted in Fig.~\ref{f:max_detect_radius} as a function
of the mass $\mu$ of the inspiralling object. It is denoted by $r_{0,\rm max}$
since it is the maximum radius at which the detection is possible.
Some selected values are displayed in Table~\ref{t:compact_obj}.
The mass of the primordial BH has been chosen arbitrarily to be the
mass of Jupiter ($10^{-3} \, M_\odot$), as a representative of a low mass
compact object.

\begin{figure}
\centerline{\includegraphics[width=0.7\textwidth]{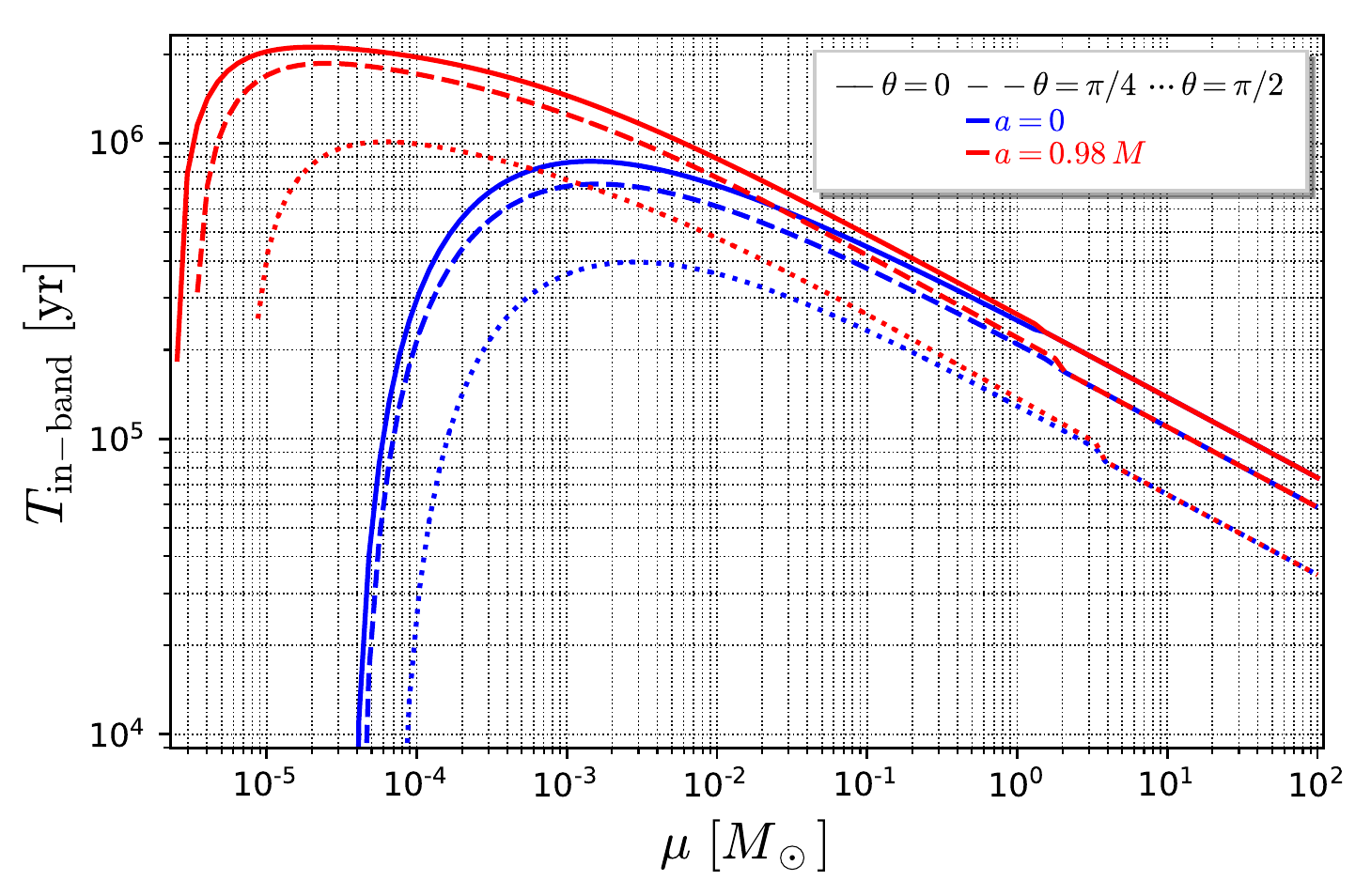}}
\caption{ \label{f:inspiral_time_inband}
Time elapsed between the entry in LISA band ($\SNRyr$ reaching 10)
and the ISCO for a compact object
inspiralling around Sgr~A*, as a function of the object's mass $\mu$.
}
\end{figure}

For a compact object, the time $T_{\mbox{\scriptsize in-band}} $ spent in LISA band is nothing but the inspiral
time from $r_{0,\rm max}$ to the ISCO:
\beq
    T_{\mbox{\scriptsize in-band}} \equiv T_{\rm ins}(r_{0,\rm max},r_{\rm ISCO})
    = T_{\rm life}(r_{0,\rm max}) ,
\eeq
where $T_{\rm ins}$ is given by Eq.~(\ref{e:inspiral_time}) and
$T_{\rm life}$ by Eq.~(\ref{e:life_time}).
The time in LISA band is depicted in Fig.~\ref{f:inspiral_time_inband}
and some selected values are given in Table~\ref{t:compact_obj}.
The trends in Fig.~\ref{f:inspiral_time_inband} can be understood by noticing that, at
fixed initial radius, the inspiral time is a decreasing function of $\mu$
[as $\mu^{-1}$, cf.~Eq.~(\ref{e:inspiral_time})], while it is an increasing
function of the initial radius [as $r_0^4$ at large distance, cf.~Eq.~(\ref{e:life_time_quadrupole})], the latter being larger for larger values of $\mu$,
since $r_0$ marks the point where $\SNRyr=10$, the S/N being an increasing
function of $\mu$ [cf.~Eq.~(\ref{e:SNR_sum_m})]. The behavior of the $T_{\mbox{\scriptsize in-band}}$ curves in
Fig.~\ref{f:inspiral_time_inband} results from the balance between these two competing
effects. The maximum is reached for masses around $10^{-3} M_\odot$ for $a=0$
($\mathrm{max}\; T_{\mbox{\scriptsize in-band}} \sim 9\times 10^5\; {\rm yr}$)
and around $10^{-5} M_\odot$
for $a=0.98 M$ ($\mathrm{max}\; T_{\mbox{\scriptsize in-band}} \sim 2\times 10^6\; {\rm yr}$) , which correspond to hypothetical primordial BHs.

\begin{table*}
\caption{
Orbital radius $r_{0,\rm max}$ at the entry in LISA band ($\SNRyr \geqslant 10$),
the corresponding gravitational wave frequency $f_{m=2}(r_{0,\rm max})$
and the time spent in LISA band until the ISCO, $T_{\mbox{\scriptsize in-band}} $, for various
compact objects orbiting Sgr~A*. The numbers outside (resp. inside) parentheses are for
Sgr~A* spin parameter $a=0$ (resp. $a=0.98M$).}
\label{t:compact_obj}
\centering
\begin{tabular}{lccccc}
\hline\hline
  & primordial & white & neutron & $10\, M_\odot$ & $30\, M_\odot$ \\[-0.5ex]
  & BH         & dwarf & star    & BH             & BH  \\
\hline
$\mu / M_\odot$                 & $10^{-3}$  & $0.5$  & $1.4$  & $10$   & $30$   \\
$r_{0,\rm max}/M\; (\theta=0)$    & $12.6\; (12.3)$   & $41.5\; (41.3)$
                                  & $50.3\; (50.3)$ & $72.0\; (72.0)$ & $88.0\; (88.0)$ \\
$r_{0,\rm max}/M\; (\theta=\pi/2)$& $10.7\; (10.4)$   & $35.3\; (35.1)$
                                  & $42.4\; (42.3)$ & $59.6\; (59.6)$ & $72.8\; (72.8)$ \\
$f_{m=2}(r_{0,\rm max}) \; (\theta=0) \ [{\rm mHz}]$
                                & $0.351\; (0.355)$    & $0.059\; (0.059)$
                                & $0.044\; (0.044)$ & $0.026\; (0.026)$ & $0.019\; (0.019)$ \\
$f_{m=2}(r_{0,\rm max}) \; (\theta=\pi/2) \ [{\rm mHz}]$
                                & $0.449\; (0.458)$    & $0.075\; (0.075)$
                                & $0.057\; (0.057)$ & $0.034\; (0.034)$ & $0.025\; (0.025)$ \\
$T_{\mbox{\scriptsize in-band}} \; (\theta=0) \ [10^5\; {\rm yr}]$
                                & $8.61\; (14.51)$ & $3.00\; (3.18)$
                                & $2.35\; (2.35)$ & $1.38\; (1.38)$ & $1.02\; (1.02)$ \\
$T_{\mbox{\scriptsize in-band}} \; (\theta=\pi/2) \ [10^5\; {\rm yr}]$
                                & $3.61\; (7.52)$ & $1.55\; (1.67)$
                                & $1.18\; (1.24)$ & $0.648\; (0.648)$ & $0.481\; (0.481)$ \\
\hline
\end{tabular}
\end{table*}

The key feature of Fig.~\ref{f:inspiral_time_inband} and Table~\ref{t:compact_obj}
is that the values of $T_{\mbox{\scriptsize in-band}}$
are very large, of the order of $10^5\; {\rm yr}$, except for very small values of
$\mu$ (below $10^{-4}\, M_\odot$). This contrasts with the time in LISA band
for extragalactic EMRIs, which is only $1$ to $10^2\; {\rm yr}$. This
is of course due to the much larger S/N resulting from the proximity of the
Galactic Center. This large time scale counter-balances the low event rate
for the capture of a compact object by Sgr~A* via the processes discussed in
Sect.~\ref{sec:formation}: even if only a single compact object is driven to the close
vicinity of Sgr~A* every $10^6\; {\rm yr}$, the fact that it
remains there in the LISA band for
$\sim 10^5\; {\rm yr}$ makes the probability of detection of order $0.1$. Given
the large uncertainty on the capture event rate, one can be reasonably optimistic.

One may stress as well that white dwarfs, which are
generally not considered as extragalactic EMRI sources for LISA
because of their low mass, have a larger value of $T_{\mbox{\scriptsize in-band}}$
than BHs (cf.~Table~\ref{t:compact_obj}). Given that they
are probably more numerous than BHs in the Galactic Center, despite mass segregation
(cf.~the discussion in Sect.~\ref{sec:formation} and \citet{Freitag03a}),
they appear to be good candidates for a detection by LISA.

\subsection{Main-sequence stars}
\label{sec:stars}

As discussed in Sect.~\ref{sec:tidal} (see Table~\ref{t:Roche_radius}),
main-sequence stars orbiting Sgr~A* have a Roche limit above the ISCO.
Away from the Roche limit, the evolution of a star on a quasi-circular orbit
is driven by the loss of energy and angular angular momentum via gravitational
radiation, as for the compact objects discussed above.
The orbit thus shrinks until the Roche limit is reached. At this point,
the star starts to loose mass through the Lagrange point $L_1$
\citep{HameuryKLA94,DaiB13} (standard accretion onto the MBH by
Roche lobe overflow) and possibly through the outer Lagrange point $L_2$
as well for stars of mass $\mu\gtrsim 1 M_\odot$ \citep{LinialS17}.
In any case, the mass loss is stable and proceeds on a secular time scale
(with respect to the orbital period).
The net effect on
the orbit is an increase of its radius \citep{HameuryKLA94,DaiB13,LinialS17},
at least for masses $\mu < 5.6 \, M_\odot$ \citep{LinialS17}.
Accordingly, instead of an EMRI, one may speak about
an extreme mass ratio \emph{outspiral} (EMRO) \citep{DaiB13},
or a \emph{reverse} chirp gravitational wave signal \citep{LinialS17,JaranowskiK92}
when describing the evolution of
such systems after they have reached the Roche limit.

For stars, let us denote by  $T_{\mbox{\scriptsize in-band}}^{\rm ins}$ the inspiral time from the
entry in LISA band
($r_0 = r_{0,\rm max}$, cf.~Fig.~\ref{f:max_detect_radius})
to the Roche limit ($r_0 = r_{\rm R}$, cf.~Table~\ref{t:Roche_radius}).
$T_{\mbox{\scriptsize in-band}}^{\rm ins}$ is
a lower bound for the total time spent in LISA band, the latter being
$T_{\mbox{\scriptsize in-band}}^{\rm ins}$ augmented by the mass-loss time at
the Roche limit, which can be quite large, of the order of $10^5\; {\rm yr}$
\citep{DaiB13}. The values of $T_{\mbox{\scriptsize in-band}}^{\rm ins}$
are given in Table~\ref{t:inspiral_time_Roche} for three typical main-sequence stars:
a Sun-like one,
a red dwarf ($\mu=0.2\, M_\odot$, same as in Table~\ref{t:Roche_radius}) and
a main-sequence star of mass $\mu = 2.4\, M_\odot$, which corresponds to
a spectral type A0V.
Let us mention that current observational data cannot rule out the presence of
such a rather luminous star in the vicinity of Sgr~A*: GRAVITY observations \citep{Gravity17}
have set the upper luminosity threshold to a B8V star, which is
a main-sequence star of mass $\mu=3.8\, M_\odot$.

$T_{\mbox{\scriptsize in-band}}^{\rm ins}$ appears to be very large,
of the order of $10^5\;  {\rm yr}$, except for the $2.4\, M_\odot$-star
for the inclination angle $\theta=\pi/2$, which has $r_{\rm R} > r_{0,\rm max}$,
i.e. it is not detectable by LISA.
As already argued for compact objects, this large value of the time spent in LISA
enhances the detection probability.

Regarding main-sequence stars, we note that the
recently claimed detection of a $149\; {\rm min}$ periodicity in the X-ray
flares from Sgr~A* \citep{Leibowitz18} has been interpreted as being caused
by a $\mu=0.18\, M_\odot$ star orbiting at $r_0 = 13.2 M$, where it is
filling its Roche lobe \citep{Leibowitz18}. If such a star exists, we can read from
Fig.~\ref{f:snr_radius_50} that LISA can detect it with a S/N equal to 76 (resp.
35) for $\theta=0$ (resp. $\theta=\pi/2$) in a single day of data.

\begin{table*}
\caption{
Inspiral time to the Roche limit in LISA band for a $\mu=0.062\, M_\odot$ brown dwarf and different types of main-sequence stars. The numbers outside (resp. inside) parentheses are for
Sgr~A* spin parameter $a=0$ (resp. $a=0.98M$).}
\label{t:inspiral_time_Roche}
\centering
\begin{tabular}{lcccc}
\hline\hline
  & brown dwarf & red dwarf & Sun & $2.4\,M_\odot$-star  \\
\hline
$\mu / M_\odot$   & $0.062$ & $0.20$ & $1$ & $2.40$ \\
$\rho/\rho_\odot$ & $131.$  & $18.8$ & $1$ & $0.367$ \\
$r_{0,\rm max}/M\; (\theta=0)$    & $28.2\; (28.0)$ & $35.0\; (34.9)$
                                  & $47.1\; (47.0)$ & $55.6\; (55.6)$ \\
$r_{0,\rm max}/M\; (\theta=\pi/2)$& $24.1\; (23.9)$ & $29.9\; (29.7)$
                                  & $40.0\; (39.8)$ & $46.8\; (46.6)$ \\
$f_{m=2}(r_{0,\rm max}) \; (\theta=0) \ [{\rm mHz}]$
                                & $0.105\; (0.106)$ & $0.076\; (0.076)$
                                & $0.049\; (0.049)$ & $0.038\; (0.038)$ \\
$f_{m=2}(r_{0,\rm max}) \; (\theta=\pi/2) \ [{\rm mHz}]$
                                & $0.133\; (0.134)$ & $0.097\; (0.097)$
                                & $0.062\; (0.063)$ & $0.049\; (0.049)$ \\
$r_{\rm R}/M$ ($\chi=0$) & $7.31\; (6.93)$ & $13.3\; (13.0)$
                         & $34.2\; (34.1)$ & $47.6\; (47.5)$ \\
$r_{\rm R}/M$ ($\chi=1$) & $7.86\; (7.56)$ & $14.5\; (14.3)$
                         & $37.7\; (37.6)$ & $52.5\; (52.5)$ \\
$T_{\mbox{\scriptsize in-band}}^{\rm ins} \; (\theta=0, \chi=0) \ [10^5\; {\rm yr}]$
                                & $4.98\; (5.55)$ & $3.72\; (3.99)$
                                & $1.83\; (1.89)$ & $0.938\; (0.945)$ \\
$T_{\mbox{\scriptsize in-band}}^{\rm ins} \; (\theta=0, \chi=1) \ [10^5\; {\rm yr}]$
                                & $4.98\; (5.54)$ & $3.67\; (3.96)$
                                & $1.49\; (1.54)$ & $0.409\; (0.418)$\\
$T_{\mbox{\scriptsize in-band}}^{\rm ins} \; (\theta=\pi/2, \chi=0) \ [10^5\; {\rm yr}]$
                                & $2.59\; (2.98)$ & $1.91\; (2.08)$
                                & $0.603\; (0.623)$ & $0\; (0)$ \\
$T_{\mbox{\scriptsize in-band}}^{\rm ins} \; (\theta=\pi/2, \chi=1) \ [10^5\; {\rm yr}]$
                                & $2.59\; (2.97)$ & $1.88\; (2.04)$
                                & $0.269\; (0.272)$ & $0\;  (0)$ \\
\hline
\end{tabular}
\end{table*}

\subsection{Brown dwarfs}
\label{s:brown_dwarfs}

Brown dwarfs are less massive than main-sequence stars, their mass
range being $\sim 10^{-2}$ to $\sim 0.08\, M_\odot$ \citep{ChabrierB00,Chabrier_al09}.
Accordingly, they enter
later (i.e. at smaller orbital radii) in the LISA band. However, they are
more dense than main-sequence stars, so that their Roche limit is closer
to the MBH, as already noticed in Sect.~\ref{sec:tidal}:
the $\mu = 0.062\, M_\odot$ brown dwarf of Table~\ref{t:Roche_radius}
has a Roche radius of order $7M$, i.e. quite close to the Schwarzschild ISCO.
In this region the S/N is quite high, despite the low value of $\mu$:
for $\mu = 0.062\, M_\odot$ and $\theta=0$, $\SNRyr = 7.4\times 10^3$
(resp. $\SNRyr = 5.4\times 10^3$) at the Roche limit with $\chi=0$
(resp. $\chi=1$). For $\theta=\pi/2$, these numbers become
$\SNRyr = 3.7\times 10^3$ ($\chi=0$) and $\SNRyr = 2.6\times 10^3$
($\chi=1$). Moreover,
brown dwarfs
stay longer in this region than compact objets since the inspiral time is
inversely proportional to the mass $\mu$ of the orbiting object
(cf.~Eq.~(\ref{e:inspiral_time})). As we can see from the values in
Table~\ref{t:inspiral_time_Roche}, the inspiral time in LISA band of
brown dwarfs is even larger than that of main-sequence stars:
$T_{\mbox{\scriptsize in-band}}^{\rm ins} \sim 5\times 10^5 \; {\rm yr}$
for $\theta=0$ and $T_{\mbox{\scriptsize in-band}}^{\rm ins} \sim 3\times 10^5 \; {\rm yr}$
for $\theta=\pi/2$.
These large values tends to make brown dwarfs good candidates for detection by LISA.
To conclude, one should know the capture rate of brown dwarfs by Sgr~A*.
It is highly uncertain but estimates have been provided very recently by
\citet{Amaro19}, which lead to a detection probability of one, with
$\sim 20$ brown dwarfs in LISA band at any moment, among which $\sim 5$
have almost circular orbits.


\subsection{Inner accretion flow}
\label{sec:accretion}

Sgr~A*'s accretion flow is known for generating particularly
low-luminosity radiation,
orders of magnitude below the Eddington limit,
and orders of magnitude below what
could be available from the gas supply at a Bondi radius~\citep{falcke13}.
This means that accretion models
should be very inefficient in converting viscously dissipated energy
into radiation. This energy will rather be stored in the disk
as heat, so that Sgr~A* accretion flow must be part of the
hot accretion flow family~\citep{yuan14}. Such systems are made of
a geometrically thick, optically thin, hot (i.e. close to the
virial temperature) accretion flow, probably accompanied by outflows.
A plethora of studies have been devoted to modeling the hot flow of
Sgr~A*, see \citet{falckemarkoff00,vincent15,broderick16,ressler17,
  davelaar18}, among many others, and references therein.

There is reasonable agreement between these different authors regarding
the typical number density and geometry of the geometrically thick hot
flow in the close vicinity of Sgr~A*. The electron maximum number density
is of order $10^8\,\mathrm{cm}^{-3}$ (to within one order of magnitude),
and the density maximum is located at a
Boyer-Lindquist radius of around $10\,M$ (to within a factor of a few).
It is thus straightforward
to give a very rough estimate of the mass of the flow, which is
of the order of $\approx 5\times 10^{-11}\,M_\odot$ (where we consider
a constant-density torus with circular cross section
of radius $4\,M$, such that its inner
radius is at the Schwarzschild ISCO).
This extremely small total mass of Sgr~A*'s accretion flow makes it
impossible to detect gravitational waves from orbiting inhomogeneities.
Figure \ref{f:snr_radius_50} shows that the LISA S/N would be vanishingly small,
assuming for instance an inhomogeneity of $10\%$ of the total mass.

\subsection{Dark matter}
\label{sec:DM}

The dark matter (DM) density profile in the inner regions of
galaxies is subject to debate. There is a controversy between observations
and cold-dark-matter simulations regarding the value of the DM
density power-law slope in the inner kpc,
observations advocating a cored profile
$\rho(r) \propto r^{0}$, while simulations predict
$\rho(r) \propto r^{-1}$~\citep{deblok10}.
The parsec-scale profile is even less well known.
\citet{gondolo99} have proposed a model
of the interaction of the central MBH
with the surrounding DM distribution for the Milky Way.
According to these authors, the presence of the MBH
should lead to an even more spiky inner profile,
with a scaling of $\rho(r) \propto r^{-2.3}$.
Such a dark matter spike can be constrained by
high-angular resolution observation at the Galactic Center~\citep{Lacroix18}.

Figure~1 of \citet{Lacroix18} shows the enclosed DM mass at the Galactic Center
as a function of radius, for various
DM models: either nonannihilating DM, or selfannihilating DM
(with particle mass equal to 1~TeV) for
various cross sections. Weakly-interacting DM
($\langle \sigma v \rangle < 10^{-30}\,\mathrm{cm}^3 \mathrm{s}^{-1}$)
leads to an enclosed mass higher than $10^{-4}\,M_\odot$
in the inner $10\,M$. Figure~\ref{f:snr_radius_50} shows
that this leads to $\SNRyr > 0.2$, assuming
that $10\%$ inhomogeneities would appear in the DM distribution
and orbit circularly around the MBH around $10\,M$.
For nonannihilating DM, the S/N values can be as high as $\SNRyr \sim 10^4$.
This makes a DM spike an interesting candidate for
a potential gravitational wave source at the Galactic Center, to be
studied in details in a forthcoming article \citep{LeTiec_al19}.


\subsection{Artificial sources}
\label{sec:alien}

The MBH Sgr~A* is indubitably a unique object in our Galaxy.
If\footnote{This is a very hypothetical "if".} an advanced civilization exists, or has existed, in the Galaxy, it would
seem unlikely that it has not shown any interest in Sgr~A*. On the contrary,
it would seem natural that such a civilization has put some material in close
orbit around Sgr~A*, for instance to extract energy from it via the Penrose
process. Whatever the reason for which the advanced
civilization acted so (it could be for purposes
that we humans simply cannot imagine), the orbital motion of this material
necessarily emits gravitational waves and if
the mass is large enough, these waves could be detected by LISA.
Given the S/N values obtained in Sect.~\ref{s:SNR} and assuming that
Sgr~A* is a fast rotator, an object of mass as low\footnote{Low is with
respect to an advanced civilization criterion.}
as the Earth mass orbiting close to the ISCO
is detectable by LISA.
This scenario is discussed further by \citet{AbramowiczBGS19}, who
consider a long lasting Jupiter-mass orbiter, left as a ``messenger''
by an advanced civilization, which possibly disappeared
billions of years ago.


\section{Discussion and conclusions}
\label{s:concl}

We have conducted a fully relativistic study of gravitational radiation
from bodies on circular orbits in the equatorial plane of the
$4.1\times 10^6\, M_\odot$ MBH at the Galactic Center, Sgr~A*.
We have performed detailed computations of the S/N in the LISA detector,
taking into account all the harmonics in the signal, whereas previous
studies \citep{Freitag03a,DaiB13,LinialS17,KuhnelMSF18}
were limited to the Newtonian quadrupole approximation, which yields only
the $m=2$ harmonic
for circular orbits. The Roche limits have been evaluated in a relativistic
framework as well, being based on the computation of the
Roche volume in the Kerr metric \citep{DaiB13}. This is specially important for
brown dwarfs, since their Roche limit occurs in the strong field region.

Setting the detection threshold to $\SNRyr=10$,
we have found that LISA has the capability to detect orbiting masses
close to Sgr~A*'s ISCO as small as ten Earth masses or even one Earth mass
if Sgr~A* is a fast rotator ($a \gtrsim 0.9 M$). Given the strong tidal forces
at the ISCO, these small bodies have to be compact objects, i.e. small BHs.
Planets and main-sequence stars have a Roche limit quite far from the ISCO:
$r_{\rm R} \sim 34\, M$ for a solar-type star (or Jupiter-type planet) and
$r_{\rm R} \sim 13\, M$ for a $0.2\, M_\odot$ star.
However, even at these distances, main-sequence stars
are still detectable by LISA, the entry in LISA band (defined by $\SNRyr=10$) being achieved for
$r_{0,\rm max} \sim 47\, M$ for a solar-type star and at $r_{0,\rm max} \sim 35\, M$
for a $0.2\, M_\odot$ main-sequence star, assuming an inclination angle $\theta=0$.
Because they are more dense, brown dwarfs have a Roche limit pretty close to
the ISCO, the minimal Roche radius being $r_{\rm R} \sim 7\, M$, which is achieved
for a $0.062\, M_\odot$ brown dwarf. For such an object, the entry in
LISA band occurs at $r_{0,\rm max} \sim 28\, M$.

Beside the S/N at a given orbit, a key parameter is the total time
spent in LISA band, i.e. the time $T_{\mbox{\scriptsize in-band}}$ during
which the source has $\SNRyr\geqslant 10$.
We have found that, once they have entered LISA band from the
low frequency side, all the considered objects, be they compact objects,
main-sequence stars or brown dwarfs, spend more than $10^5\; {\rm yr}$ in
LISA band\footnote{For high inclination angle $\theta\sim\pi/2$,
BHs and solar-type stars spend only one half of this value.}.
The minimal time in-band occurs for high-mass BHs ($\mu\sim 30\, M_\odot$),
for which $T_{\mbox{\scriptsize in-band}} \sim 1 \times 10^5\; {\rm yr}$ (assuming $\theta=0$)
and the maximal one, of the order of one million years, is achieved for a
Jupiter-mass BH ($\mu \sim 10^{-3}\, M_\odot$)
if Sgr~A* is a slow rotator ($a/M \ll 1$): $T_{\mbox{\scriptsize in-band}} \sim 9\times 10^5\; {\rm yr}$,
or for a $\mu\sim 10^{-5}\, M_\odot$ BH if Sgr~A* is a rapid rotator ($a/M \gtrsim 0.9$):
$T_{\mbox{\scriptsize in-band}} \sim 2\times 10^6\; {\rm yr}$. These small BH masses regard primordial
BHs. Among stars and stellar BHs, the maximum time spent in LISA band is achieved for
brown dwarfs: $T_{\mbox{\scriptsize in-band}} \geqslant 5\times 10^5\; {\rm yr}$, just followed
by low-mass main-sequence stars (red dwarfs) and white dwarfs, for which
$T_{\mbox{\scriptsize in-band}} \geqslant 3\times 10^5\; {\rm yr}$.
These large values of $T_{\mbox{\scriptsize in-band}}$ contrast with
those for extragalactic EMRIs, which are typically of the order of $1$ to $10^2\; {\rm yr}$.
This is of course due to the much larger S/N resulting from the proximity of Sgr~A*, which
allows one to catch compact objects at much larger orbital radii, where the
orbital decay is not too fast, and to catch
main-sequence stars above their Roche limit.

To predict some LISA detection rate from $T_{\mbox{\scriptsize in-band}}$, one shall
know the rate at which the considered objects are brought to close circular orbits around
Sgr~A* (``capture'' rate). While we have briefly described some scenarios proposed in the literature
in Sect.~\ref{sec:formation}, it is not the purpose of this work to make
precise estimates. Having those is probably very difficult, given the involved
uncertainties, both on the observational ground (strong absorption in the direction of the Galactic
Center) and the theoretical one (dynamics
of the tens of thousands of stars and BHs in the central parsec).
Some optimistic scenarios mentioned in Sect.~\ref{sec:formation} predict
a capture rate of the order of $10^{-6} \; {\rm yr}^{-1}$ for BHs. For
$T_{\mbox{\scriptsize in-band}} \sim 10^5\; {\rm yr}$, this would result in
a detection probability of $0.1$ by LISA. For white dwarfs, low mass main-sequence stars and brown dwarfs,
the capture rate could possibly be higher \citep{Freitag03a}, leading to a significant detection
probability by LISA, especially for brown dwarfs.
Instead of making any concrete prediction, we prefer an ``agnostic'' approach,
stating that Sgr~A* is definitely a target worth of attention for LISA, which
may reveal various bodies orbiting around it.

Let us point out that
\citet{Amaro19} has recently performed a study of gravitational radiation
from main-sequence stars and brown dwarfs orbiting Sgr~A*. He finds
results similar to ours regarding the S/N in LISA.  Also, he
derives the event rate for the Galactic Center taking into account the relativistic loss-cone and eccentric orbits, which are more typical in an astrophysical context.
The high event rate that he has obtained makes brown dwarfs promising
candidates for LISA.

In Appendix~\ref{s:M32}, we have considered bodies in close circular orbit around
the $2.5\times 10^6\, M_\odot$ MBH in the center of the nearby galaxy M32.
We find that main-sequence stars with $\mu \geqslant 0.2\, M_\odot$ are not
detectable by LISA in this case, while compact objects and brown dwarfs
are still detectable, with a lower probability: the time they are
spending in LISA band with $\SNRyr\geqslant 10$ is $10^3$ to $10^4$ years, that is
two orders of magnitude lower than for Sgr~A*.

A natural extension of the work presented here is towards noncircular orbits.
Gravitational waves from a compact body on eccentric, equatorial \citep{GlampedakisK02}, spherical \citep{Hughes00}, and generic bound \citep{Drasco:2005kz} geodesics have been studied before.
The application of these results to Sgr~A* including the calculation of the orbital decay for generic orbits, exploration of the inspiral parameter space, and the analysis of the tidal and Roche radii remains to be completed.
Another extension would be to study the gravitational emission from a (stochastic) ensemble
of small masses, such as brown dwarfs, in the case they are numerous
around Sgr~A*, or from dark matter clumps as mentioned in
Sect.~\ref{sec:DM} \citep{LeTiec_al19}.

\begin{acknowledgements}
We are grateful to Antoine Petiteau for having provided us with the LISA noise power spectral density curve and to Pau Amaro-Seoane, Micha\l{} Bejger, Christopher Berry,
Gilles Chabrier, Suzy Collin-Zahn,
and Thibaut Paumard for fruitful discussions. NW gratefully acknowledges support from a Royal Society - Science Foundation Ireland University Research Fellowship.
\end{acknowledgements}

\appendix
\section{The \texttt{kerrgeodesic\_gw} package} \label{s:kerrgeodesic_gw}

We have developed the open-source package \texttt{kerrgeodesic\_gw} for
the Python-based free mathematics software system SageMath\footnote{\url{http://www.sagemath.org/}}. This package implements all the computations
presented in this article. The installation of \texttt{kerrgeodesic\_gw} is very easy, since it
relies on the standard \texttt{pip} mechanism for Python packages.
One only needs to run
\begin{verbatim}
sage -pip install kerrgeodesic_gw
\end{verbatim}
to download and install the package in any working SageMath environment.
The sources of the package are available at the following \texttt{git} repository,
as part of the Black Hole Perturbation Toolkit\footnote{\url{http://bhptoolkit.org/}}:\\
\url{https://github.com/BlackHolePerturbationToolkit/kerrgeodesic_gw}

The reference manual of \texttt{kerrgeodesic\_gw} includes many examples and
is online at\\
\url{https://cocalc.com/share/2b3f8da9-6d53-4261-b5a5-ff27b5450abb/kerrgeodesic_gw/docs/build/html/index.html}

Various Jupyter notebooks making use of \texttt{kerrgeodesic\_gw}
are publicly available on the cloud platform CoCalc, including those used to generate all the
figures presented in the current article:\\
\url{https://cocalc.com/share/2b3f8da9-6d53-4261-b5a5-ff27b5450abb/PaperI/Notebooks?viewer=share/}\\
Other notebooks regard tests
of the package, like the comparison with the 1.5PN waveforms obtained
by \citet{Poisson93a} for $a=0$  and with the fully relativistic waveforms
obtained by \citet{Detweiler78} for $a=0.5M$ and $a=0.9M$:\\
\url{https://cocalc.com/share/2b3f8da9-6d53-4261-b5a5-ff27b5450abb/gw_single_particle.ipynb?viewer=share}


\section{Computation of the S/N integral} \label{s:comput_SNR}

In order to evaluate the S/N integral (\ref{e:SNR_eff}), we need
to compute the Fourier transforms $\tilde{h}_+(f)$ and $\tilde{h}_\times(f)$
over the observation time $T$ via Eq.~(\ref{e:FT_trunc}).
Let us focus first on $h_+(t)$ and rewrite its Fourier series (\ref{e:h_Fourier})
as
\beq
    h_{+}(t) = \frac{\mu}{r} \sum_{m=1}^{+\infty}
        H_m^+(\theta) \cos\left(2\pi m f_0 t+ \chi_m \right) ,
\eeq
where the amplitude $H_m^+(\theta)$ is defined by Eq.~(\ref{e:hm_def})
and the phase angle $\chi_m$ is defined by
(cf.~Eqs.~(\ref{e:h_Fourier}) and (\ref{e:def_psi}))
\beq
    \chi_m = m( \varphi_0 - \varphi - 2\pi f_0 r_* ) + \Phi_m \, ,
\eeq
with
\beq
\cos \Phi_m = \frac{A_m^{+}(\theta)}{H_m^{+}(\theta)}
\quad\mbox{and}\quad
\sin\Phi_m = - \frac{B_m^{+}(\theta)}{H_m^{+}(\theta)} \, .
\eeq
The Fourier transform (\ref{e:FT_trunc}) is then
\begin{align}
  \tilde{h}_{+}(f)
    &= \frac{\mu}{r} \sum_{m=1}^{+\infty} H_m^{+}(\theta)
        \int_{-T/2}^{T/2} \cos\left(2\pi m f_0 t+ \chi_m \right) \,
                     \mathrm{e}^{-2\pi \mathrm{i} f t} \, \mathrm{d}t \nonumber \\
  &= \frac{\mu}{2r} \sum_{m=1}^{+\infty} H_m^{+}(\theta)
  \int_{-T/2}^{T/2}  \left[ \mathrm{e}^{2\pi \mathrm{i} m f_0 t+
  \mathrm{i} \chi_m -2\pi \mathrm{i} f t}
  +  \mathrm{e}^{-2\pi \mathrm{i} m f_0 t - \mathrm{i}
        \chi_m -2\pi \mathrm{i} f t} \right]
  \, \mathrm{d}t \nonumber \\
  &= \frac{\mu}{2r} \sum_{m=1}^{+\infty} H_m^{+}(\theta) \left[
        \mathrm{e}^{\mathrm{i} \chi_m} \int_{-T/2}^{T/2}
        \mathrm{e}^{2\pi \mathrm{i} (m f_0 -f)t} \, \mathrm{d}t
      + \mathrm{e}^{-\mathrm{i} \chi_m} \int_{-T/2}^{T/2}
        \mathrm{e}^{-2\pi \mathrm{i} (m f_0 +f)t} \, \mathrm{d}t \right] \nonumber \\
  &= \frac{\mu}{2r} \sum_{m=1}^{+\infty} H_m^{+}(\theta) \left[
        \mathrm{e}^{\mathrm{i} \chi_m}
        \frac{2\mathrm{i} \sin(\pi (m f_0 - f) T)}{2\pi \mathrm{i} (m f_0 -f)}
        + \mathrm{e}^{-\mathrm{i} \chi_m}
        \frac{- 2\mathrm{i} \sin(\pi (m f_0 + f) T)}{- 2\pi \mathrm{i} (m f_0 +f)}
        \right] \nonumber \\
  &= \frac{\mu}{2r} T \sum_{m=1}^{+\infty} H_m^{+}(\theta) \left[
        \mathrm{e}^{\mathrm{i} \chi_m} \sinc\left( \pi  (f - m f_0) T \right)
       + \mathrm{e}^{-\mathrm{i} \chi_m} \sinc\left( \pi  (f + m f_0) T \right)
    \right] \, ,
\end{align}
where $\sinc$ stands for the cardinal sine function:
$\sinc(x) \equiv \sin x / x$.
The square of the modulus of $\tilde{h}_{+}(f)$, which appears in the
S/N formula (\ref{e:SNR_eff}), is then
\begin{align}
    |\tilde{h}_+(f)|^2 &= \tilde{h}_+(f) \tilde{h}_+(f)^*   \notag \\
    &= \left( \frac{\mu}{2r} T \right)^2
    \left( \sum_{m=1}^{+\infty} H_m^{+}(\theta) \left[
        \mathrm{e}^{\mathrm{i} \chi_m} \sinc\left( \pi  (f - m f_0) T \right)
       + \mathrm{e}^{-\mathrm{i} \chi_m} \sinc\left( \pi  (f + m f_0) T \right)
    \right] \right) \notag \\
    & \qquad \qquad\times \left( \sum_{n=1}^{+\infty} H_n^{+}(\theta) \left[
        \mathrm{e}^{-\mathrm{i} \chi_n} \sinc\left( \pi  (f - n f_0) T \right)
       + \mathrm{e}^{\mathrm{i} \chi_n} \sinc\left( \pi  (f + n f_0) T \right)
    \right] \right) \notag \\
  &= \left( \frac{\mu}{r} \right) ^2 \frac{T}{4}
   \sum_{m=1}^{+\infty} \sum_{n=1}^{+\infty}
   H_m^{+}(\theta) H_n^{+}(\theta)
    \Big[ \mathrm{e}^{\mathrm{i} (\chi_m - \chi_n)}
  \, T \sinc\left( \pi  (f - m f_0) T \right) \sinc\left( \pi  (f - n f_0) T \right)
    \notag \\
  & \quad\qquad\qquad\qquad\qquad + \mathrm{e}^{\mathrm{i} (\chi_m + \chi_n)}
  \, T \sinc\left( \pi  (f - m f_0) T \right) \sinc\left( \pi  (f + n f_0) T \right)
    \notag \\
  & \quad\qquad\qquad\qquad\qquad + \mathrm{e}^{-\mathrm{i} (\chi_m + \chi_n)}
  \, T \sinc\left( \pi  (f + m f_0) T \right) \sinc\left( \pi  (f - n f_0) T \right)
    \notag \\
  & \quad\qquad\qquad\qquad\qquad + \mathrm{e}^{\mathrm{i} (\chi_n - \chi_m)}
  \, T \sinc\left( \pi  (f + m f_0) T \right) \sinc\left( \pi  (f + n f_0) T \right)
  \Big]  \notag \\
  &= \left( \frac{\mu}{r} \right) ^2 \frac{T}{4}
   \sum_{m=1}^{+\infty} \sum_{n=1}^{+\infty}
   H_m^{+}(\theta) H_n^{+}(\theta)
    \Big[ \mathrm{e}^{\mathrm{i} (\chi_m - \chi_n)}
  \, \Delta_{T,mf_0}(f) \sinc\left( \pi  (f - n f_0) T \right)
    \notag \\
  & \quad\qquad\qquad\qquad\qquad + \mathrm{e}^{\mathrm{i} (\chi_m + \chi_n)}
  \, \Delta_{T,mf_0}(f)  \sinc\left( \pi  (f + n f_0) T \right)
    \notag \\
  & \quad\qquad\qquad\qquad\qquad + \mathrm{e}^{-\mathrm{i} (\chi_m + \chi_n)}
  \, \Delta_{T,-mf_0}(f)  \sinc\left( \pi  (f - n f_0) T \right)
    \notag \\
  & \quad\qquad\qquad\qquad\qquad + \mathrm{e}^{\mathrm{i} (\chi_n - \chi_m)}
  \, \Delta_{T,-mf_0}(f)  \sinc\left( \pi  (f + n f_0) T \right)
  \Big] \, ,   \label{e:hpf2}
\end{align}
where the functions $\Delta_{T,f_*}(f)$ are defined for any pair of real
parameters $(T,f_*)$ by
\beq
    \Delta_{T,f_*}(f) \equiv T \sinc\left( \pi  (f - f_*) T \right)  .
\eeq
For each value of $f_*$, the $\Delta_{T,f_*}$ constitute a family of
nascent delta functions, i.e. they obey\footnote{Eq.~(\ref{e:int_Delta_1})
immediately follows from the well known identity $\int_{-\infty}^{\infty} \sinc(\pi x) \, \mathrm{d}x = 1$.}
\begin{subequations}
\begin{gather}
    \int_{-\infty}^{+\infty} \Delta_{T,f_*}(f) \, \mathrm{d}f =  1 \label{e:int_Delta_1}\\
    \forall\; \delta f > 0,\quad
    \lim_{T\to+\infty} \int_{\mathbb{R}\setminus(f_*-\delta f, f_*+\delta f)} \Delta_{T,f_*}(f) \, \mathrm{d}f = 0 .  \label{e:int_Delta_0}
\end{gather}
\end{subequations}
These two properties imply that, for any integrable function $F$,
\beq
    \lim_{T\to+\infty} \int_{-\infty}^{\infty} F(f) \, \Delta_{T,f_*}(f) \, \mathrm{d}f
        = F(f_*) .
\eeq
In other words, when $T\to+\infty$, $\Delta_{T,f_*}$ tends to the Dirac delta distribution centered on $f_*$.
Considering successively the four terms that appear in Eq.~(\ref{e:hpf2}) and gathering
them two by two by means of $\pm$, we
have then
\begin{subequations}
\begin{align}
 & \int_0^{+\infty}
    \frac{\Delta_{T,mf_0}(f)  \sinc\left( \pi  (f \pm n f_0) T \right)}{S_{\rm n}(f)}
        \, \mathrm{d} f  \simeq \frac{\sinc\left( \pi  (m \pm n) f_0 T \right)}{S_{\rm n}(m f_0)} \quad \mbox{when}\quad T\to+\infty \, , \label{e:int_Dsinc_sinc}\\
 & \int_0^{+\infty}
    \frac{\Delta_{T, -mf_0}(f)  \sinc\left( \pi  (f \pm n f_0) T \right)}{S_{\rm n}(f)}
        \, \mathrm{d} f  \rightarrow 0 \quad \mbox{when}\quad T\to+\infty \, . \label{e:int_Dsinc_zero}
\end{align}
\end{subequations}
It should be noted that (\ref{e:int_Dsinc_zero}) readily follows from property (\ref{e:int_Delta_0})
since $-m f_0 < 0$. Regarding Eq.~(\ref{e:int_Dsinc_sinc}), we note that
\beq
    \lim_{T\to+\infty} \sinc\left( \pi  (m - n) f_0 T \right) =
    \begin{cases}
    1 & \text{if $n=m$} \\
    0 & \text{if $n\not= m$}
    \end{cases}
    \quad\mbox{and}\quad
    \lim_{T\to+\infty} \sinc\left( \pi  (m + n) f_0 T \right) = 0 , \label{e:lim_sinc}
\eeq
the last property resulting from $m+n \not =0$ for $m\geqslant 1$ and $n\geqslant 1$.
In view of Eqs.~(\ref{e:hpf2}) and (\ref{e:int_Dsinc_sinc})-(\ref{e:lim_sinc}),
we see that, when $T\rightarrow+\infty$, the only contribution to the S/N integral
(\ref{e:SNR_eff}) arises from the first term in Eq.~(\ref{e:hpf2}) with moreover
$n=m$, which implies $\mathrm{e}^{\mathrm{i} (\chi_m - \chi_n)}=1$. Hence we have
\beq
    \int_0^{+\infty}
    \frac{|\tilde{h}_+(f)|^2}{S_{\rm n}(f)} \, \mathrm{d} f \simeq
    \left( \frac{\mu}{r} \right) ^2 \frac{T}{4} \, \sum_{m=1}^{+\infty}
    \frac{H_m^{+}(\theta)^2}{S_{\rm n}(m f_0)}
    \quad\mbox{for}\quad T\rightarrow+\infty.
\eeq
The limit $T\rightarrow+\infty$, which arises from
Eqs.~(\ref{e:int_Dsinc_sinc}) and (\ref{e:lim_sinc}), can be translated
by $mf_0 T\gg 1$ for all $m$, i.e. by $f_0 T \gg 1$.
Obviously, we get a similar formula for the contribution of $|\tilde{h}_\times(f)|^2$
to the S/N, so that Eq.~(\ref{e:SNR_eff}) becomes
\beq
    \rho^2 = 4 \int_0^{+\infty}
    \frac{|\tilde{h}_+(f)|^2+|\tilde{h}_\times(f)|^2}{S_{\rm n}(f)}
        \, \mathrm{d} f
    \simeq \left( \frac{\mu}{r} \right) ^2 T \, \sum_{m=1}^{+\infty}
    \frac{H_m^{+}(\theta)^2 + H_m^{\times}(\theta)^2}{S_{\rm n}(m f_0)}
    \quad\mbox{for}\quad f_0 T \gg 1 ,
\eeq
hence the S/N value (\ref{e:SNR_sum_m}).


\section{Case of M32} \label{s:M32}

\begin{figure}
\centerline{\includegraphics[width=0.7\textwidth]{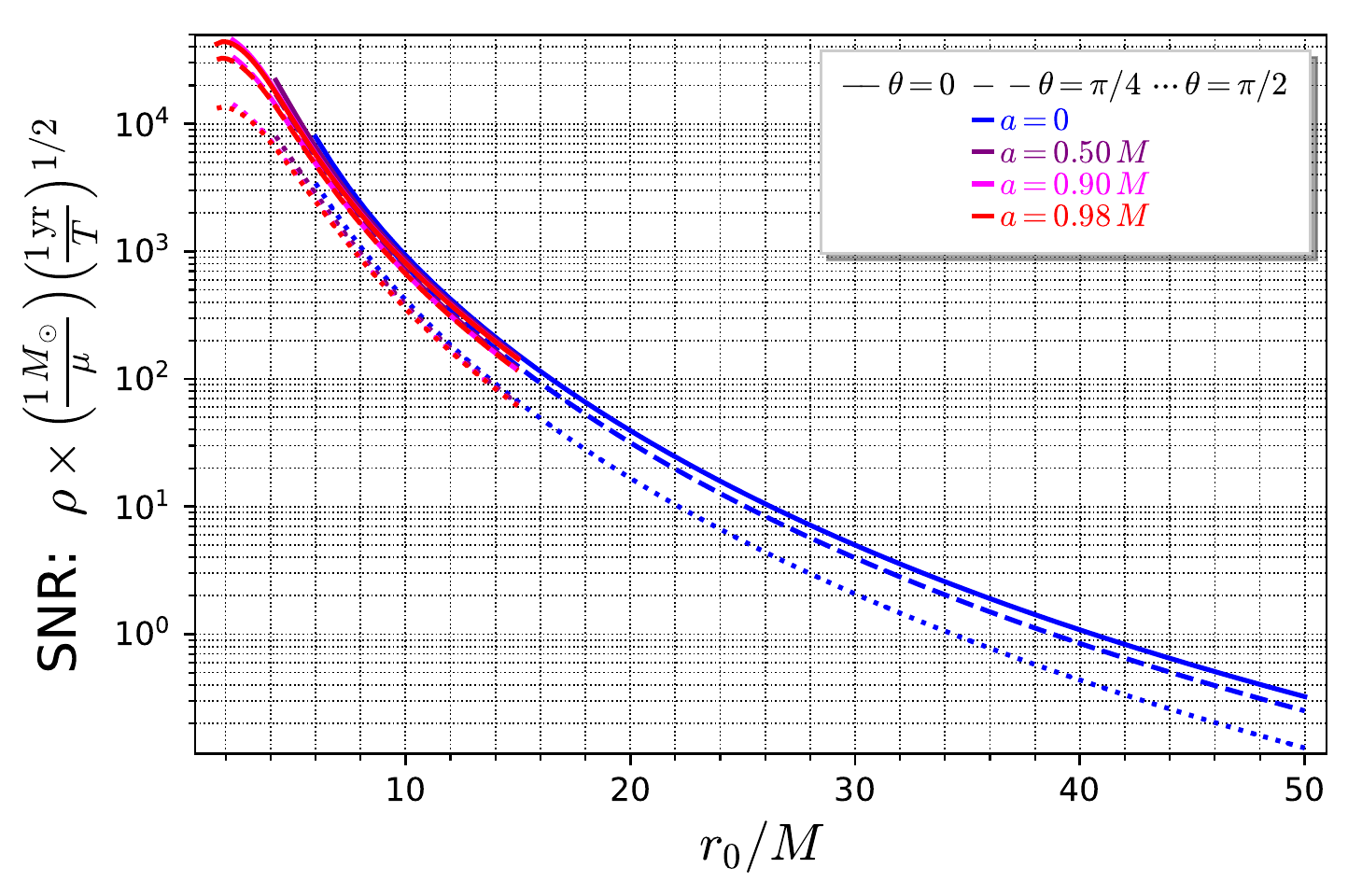}}
\caption{ \label{f:snr_radius_M32}
Effective (direction and polarization averaged) signal-to-noise in LISA
for a $T = 1\; {\rm yr}$ observation of an
object of mass $\mu=1\, M_\odot$ orbiting M32 MBH, as a function of
the orbital radius $r_0$ (in units of $M$, the mass of M32 MBH), and for
selected values of the MBH spin parameter $a$ as well
as selected values of the inclination angle $\theta$. Each curve starts at the
ISCO radius of the corresponding value of $a$. It should be noted that this figure
is scaled for $T = 1\; {\rm yr}$, while the equivalent figure for Sgr~A* (Fig.~\ref{f:snr_radius_50})
is scaled for $T = 1\; {\rm d}$.}
\end{figure}

\begin{figure}
\centerline{\includegraphics[width=0.7\textwidth]{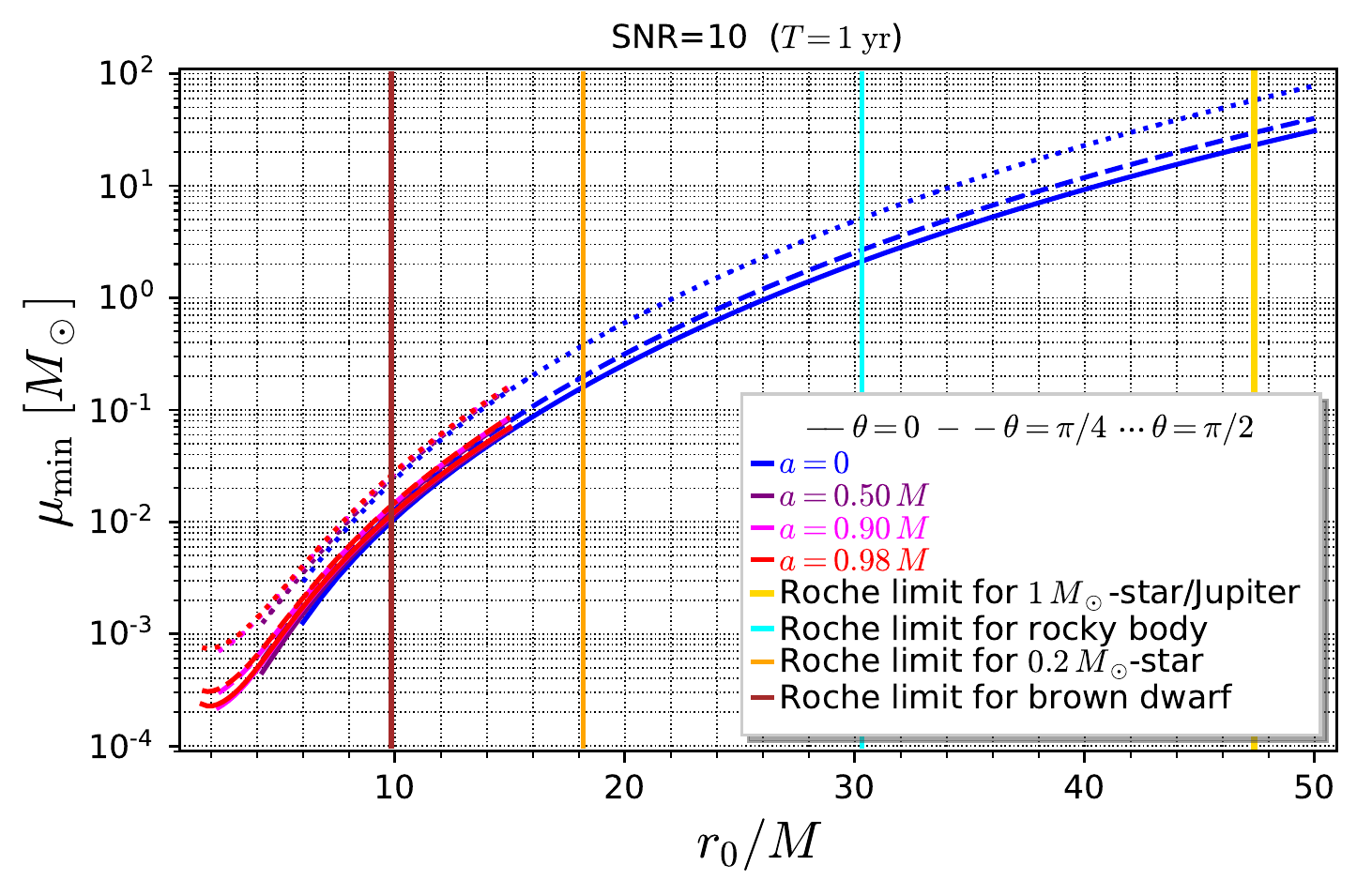}}
\caption{ \label{f:min_detect_mass_M32}
Minimal detectable mass with $\SNRyr\geqslant 10$ in LISA observations of
M32 center, as a function of the orbital radius $r_0$. The various Roche limits are
those considered in Sect.~\ref{sec:tidal}.
}
\end{figure}

\begin{figure}
\centerline{\includegraphics[width=0.7\textwidth]{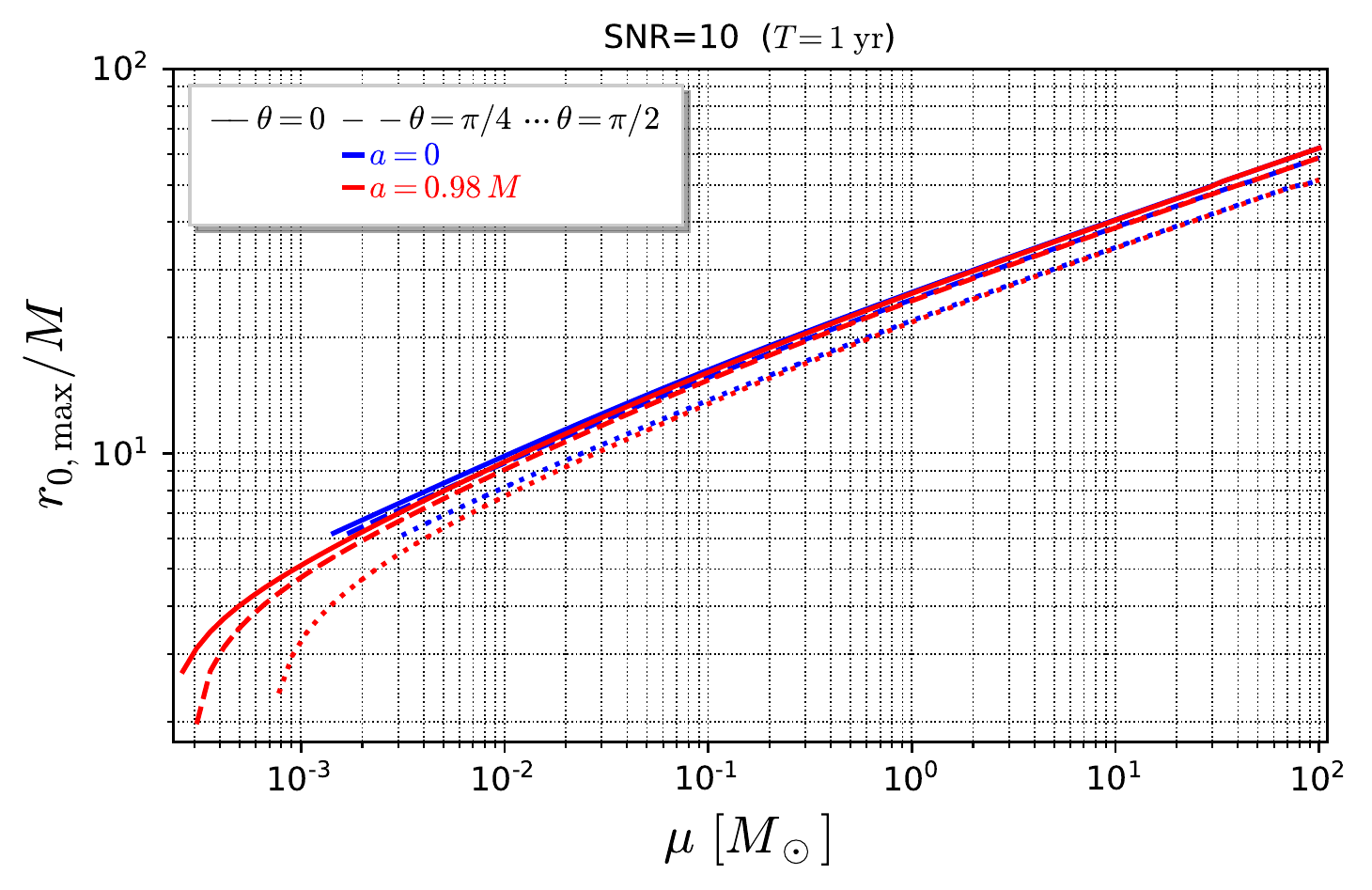}}
\caption{ \label{f:max_detect_radius_M32}
Maximum orbital radius $r_{0,\rm max}$ for a $\SNRyr=10$ detection by LISA, as a function of the mass $\mu$ of the object orbiting around M32 MBH.
}
\end{figure}

\begin{table*}
\caption{
Orbital radius $r_{0,\rm max}$ at the entry in LISA band ($\SNRyr$ reaching 10),
the corresponding gravitational wave frequency $f_{m=2}(r_{0,\rm max})$
and the time spent in LISA band until the ISCO, $T_{\mbox{\scriptsize in-band}} $, for various
compact objects orbiting M32 MBH. The numbers outside (resp. inside) parentheses are for
a MBH spin parameter $a=0$ (resp. $a=0.98M$). With respect to the equivalent table
for Sgr~A* (Table~\ref{t:compact_obj}), note that the primodial BH mass is chosen
to be $\mu=10^{-2}\, M_\odot$ and that the scale of $T_{\mbox{\scriptsize in-band}}$
is $10^3\; {\rm yr}$.}
\label{t:compact_obj_M32}
\centering
\begin{tabular}{lccccc}
\hline\hline
  & primordial & white & neutron & $10\, M_\odot$ & $30\, M_\odot$ \\[-0.5ex]
  & BH         & dwarf & star    & BH             & BH  \\
\hline
$\mu / M_\odot$                 & $10^{-2}$  & $0.5$  & $1.4$  & $10$   & $30$   \\
$r_{0,\rm max}/M\; (\theta=0)$    & $9.84\; (9.50)$   & $22.9\; (22.7)$
                                  & $28.0\; (27.8)$ & $40.6\; (40.4)$ & $49.8\; (49.6)$ \\
$r_{0,\rm max}/M\; (\theta=\pi/2)$& $8.16\; (7.73)$   & $19.3\; (19.1)$
                                  & $23.7\; (23.5)$ & $34.3\; (34.2)$ & $42.0\; (41.9)$ \\
$f_{m=2}(r_{0,\rm max}) \; (\theta=0) \ [{\rm mHz}]$
                                & $0.838\; (0.855)$    & $0.236\; (0.237)$
                                & $0.174\; (0.175)$ & $0.100\; (0.100)$ & $0.074\; (0.074)$ \\
$f_{m=2}(r_{0,\rm max}) \; (\theta=\pi/2) \ [{\rm mHz}]$
                                & $1.109\; (1.150)$    & $0.305\; (0.307)$
                                & $0.224\; (0.225)$ & $0.128\; (0.129)$ & $0.095\; (0.095)$ \\
$T_{\mbox{\scriptsize in-band}} \; (\theta=0) \ [10^3\; {\rm yr}]$
                                & $8.01\; (20.06)$ & $9.63\; (11.33)$
                                & $7.98\; (8.94)$ & $5.11\; (5.43)$ & $3.88\; (4.05)$ \\
$T_{\mbox{\scriptsize in-band}} \; (\theta=\pi/2) \ [10^3\; {\rm yr}]$
                                & $2.09\; (9.36)$ & $4.59\; (5.71)$
                                & $3.94\; (4.59)$ & $2.59\; (2.80)$ & $1.96\; (2.08)$ \\
\hline
\end{tabular}
\end{table*}

\begin{figure}
\centerline{\includegraphics[width=0.7\textwidth]{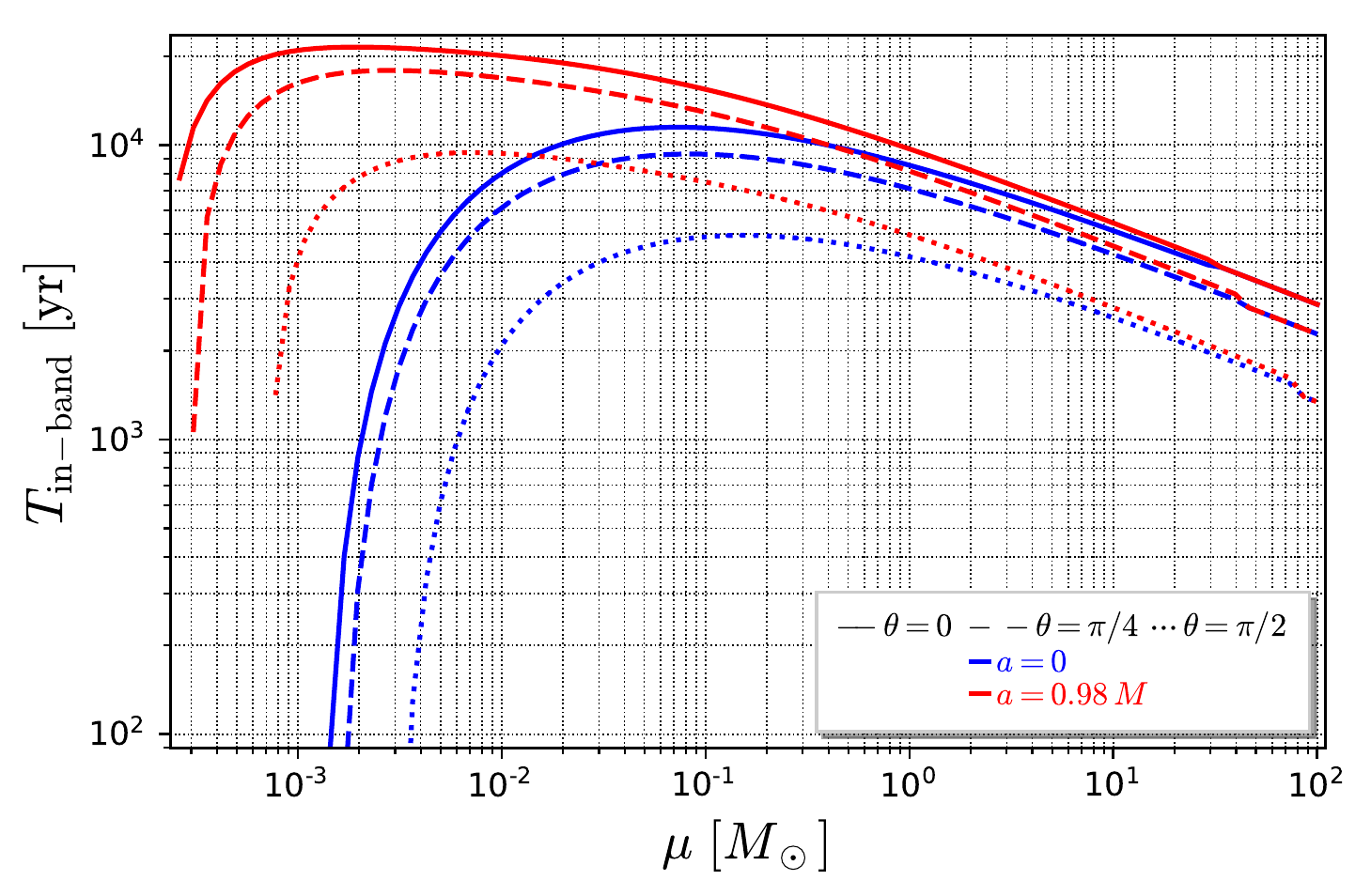}}
\caption{ \label{f:inspiral_time_inband_M32}
Time elapsed between the entry in LISA band ($\SNRyr\geqslant 10$)
and the ISCO for a compact object
inspiralling around M32 MBH, as a function of the object's mass $\mu$.
}
\end{figure}

\begin{table}
\caption{
Inspiral time to the Roche limit in LISA band ($\SNRyr\geqslant 10$)
for the brown dwarf and red dwarf
models considered in Sect.~\ref{sec:tidal}, when
orbiting M32 MBH. The numbers outside (resp. inside) parentheses are for
M32 MBH spin parameter $a=0$ (resp. $a=0.98M$).}
\label{t:inspiral_time_Roche_M32}
\centering
\begin{tabular}{lcc}
\hline\hline
  & brown dwarf & red dwarf  \\
\hline
$\mu / M_\odot$   & $0.062$ & $0.20$ \\
$\rho/\rho_\odot$ & $131.$  & $18.8$ \\
$r_{0,\rm max}/M\; (\theta=0)$    & $14.9\; (14.6)$ & $19.0\; (18.8)$ \\
$r_{0,\rm max}/M\; (\theta=\pi/2)$& $12.4\; (12.1)$ & $15.9\; (15.7)$ \\
$f_{m=2}(r_{0,\rm max}) \; (\theta=0) \ [{\rm mHz}]$
                                & $0.451\; (0.455)$ & $0.311\; (0.313)$\\
$f_{m=2}(r_{0,\rm max}) \; (\theta=\pi/2) \ [{\rm mHz}]$
                                & $0.594\; (0.603)$ & $0.406\; (0.410)$\\
$r_{\rm R}/M$ ($\chi=0$) & $9.85\; (9.55)$ & $18.2\; (18.0)$\\
$r_{\rm R}/M$ ($\chi=1$) & $10.7\; (10.5)$ & $19.9\; (19.8)$\\
$T_{\mbox{\scriptsize in-band}}^{\rm ins} \; (\theta=0, \chi=0) \ [10^3\; {\rm yr}]$
                                & $10.16\; (13.27)$ & $2.00\; (2.17)$\\
$T_{\mbox{\scriptsize in-band}}^{\rm ins} \; (\theta=0, \chi=1) \ [10^3\; {\rm yr}]$
                                & $9.34\; (11.92)$ & $0\; (0)$\\
$T_{\mbox{\scriptsize in-band}}^{\rm ins} \; (\theta=\pi/2, \chi=0) \ [10^3\; {\rm yr}]$
                                & $3.37\; (4.65)$ & $0\; (0)$\\
$T_{\mbox{\scriptsize in-band}}^{\rm ins} \; (\theta=\pi/2, \chi=1) \ [10^3\; {\rm yr}]$
                                & $2.55\; (3.29)$ & $0\; (0)$\\
\hline
\end{tabular}
\end{table}

Apart from Sgr~A*, the only MBH in the Local Group of galaxies whose mass fits
LISA band is the one in the center of M32 --- the compact elliptical galaxy
satellite of the Andromeda Galaxy M31\footnote{Andromeda Galaxy
itself harbors a MBH in its nucleus, but it has $M\sim 10^8\, M_\odot$ \citep{Bender_al05},
which is too massive for LISA band. Beyond the Local Group, nearby galaxies with a MBH in the LISA range have been considered by \citet{BerryG13b} in their study of
extreme mass ratio burts (cf. Sect.~\ref{s:intro}).}. Its mass is
$M = 2.5^{+0.6}_{-1.0}\times 10^6\, M_\odot$ \citep{Nguyen_al18}.
The distance to the Earth is $r \simeq 790 \; {\rm kpc}$ \citep{Nguyen_al18},
i.e. roughly a hundred time farther than Sgr~A*.

The LISA S/N for objects on circular equatorial orbits around M32 MBH
is depicted as a function of the orbital radius in Fig.~\ref{f:snr_radius_M32}.
The minimal mass $\mu_{\rm min}$ detectable with $\SNRyr\geqslant 10$
at a given orbital radius is shown in Fig.~\ref{f:min_detect_mass_M32}.
We note that the minimal detectable mass is $\sim 2\times 10^{-3} \, M_\odot$
(close to the ISCO) if M32 MBH is a slow rotator, down to $\sim 2\times 10^{-4} \, M_\odot$
in the case of a fast rotator. The Roche limits for the various kinds of stars
considered in Sect.~\ref{sec:tidal}, reevaluated to take into account
M32 MBH mass $M$, have been drawn in Fig.~\ref{f:min_detect_mass_M32}.
It appears then clearly that a solar-type star in circular orbit around M32 MBH cannot be detected by LISA and that a $0.2\, M_\odot$ red dwarf can be marginally detected,
while there is no issue in detecting a brown dwarf at its Roche limit.

Regarding the detection probability, the important parameter is the
time $T_{\mbox{\scriptsize in-band}}$ spent in LISA band, i.e. the time
elapsed between the orbit at which the object starts to be detectable
by LISA (cf.~Fig.~\ref{f:max_detect_radius_M32}) and either the
ISCO (for a compact object, cf.~Fig.~\ref{f:inspiral_time_inband_M32} and
Table~\ref{t:compact_obj_M32}) or the Roche limit
(brown dwarfs and red dwarfs, cf.~Table~\ref{t:inspiral_time_Roche_M32}).
From Fig.~\ref{f:inspiral_time_inband_M32}, the largest values of
$T_{\mbox{\scriptsize in-band}}$ are $T_{\mbox{\scriptsize in-band}}\sim 1.\times 10^4\;{\rm yr}$ (resp. $T_{\mbox{\scriptsize in-band}}\sim 2\times 10^4\;{\rm yr}$)
for $a=0$ (resp. $a=0.98\, M$) and are achieved for $\mu\sim 0.1\, M_\odot$
(resp. $\mu\sim 10^{-3}\, M_\odot$), which corresponds to hypothetical primordial
BHs. We note that for a $0.5\, M_\odot$ white dwarf,
$T_{\mbox{\scriptsize in-band}}\sim 1\times 10^4\;{\rm yr}$.
For stellar mass BHs, $T_{\mbox{\scriptsize in-band}}$ is of the order of
a few $10^3\;{\rm yr}$.

For the $0.2\, M_\odot$ red dwarf, we conclude from
Table~\ref{t:inspiral_time_Roche_M32} that it can be detected by LISA only
if the inclination angle $\theta$ is small and if it is not corotating
($|\chi|\ll 1$). One has then
$T_{\mbox{\scriptsize in-band}} > T_{\mbox{\scriptsize in-band}}^{\rm ins}\sim
2\times 10^3\;{\rm yr}$.

Regarding the $0.062\, M_\odot$ brown dwarf, we read in Table~\ref{t:inspiral_time_Roche_M32} that
$T_{\mbox{\scriptsize in-band}} > T_{\mbox{\scriptsize in-band}}^{\rm ins}\sim
1\times 10^4\;{\rm yr}$ for low inclinations and
$\sim 3\times 10^3\;{\rm yr}$ for large inclinations.

\bibliographystyle{aa}
\bibliography{references}

\end{document}